\documentclass[prr,aps,twocolumn,amsmath,amssymb,superscriptaddress,showpacs,showkeys,nofootinbib]{revtex4-2}
\usepackage{graphicx}
\usepackage{dcolumn}
\usepackage{bm}
\usepackage{float}
\usepackage{braket}
\usepackage[colorlinks=true, allcolors=blue, linktocpage=true]{hyperref}

\usepackage[dvipsnames]{xcolor}

\usepackage[markup=underlined]{changes}
\makeatletter
\@namedef{Changes@AuthorColor}{Red}
\colorlet{Changes@Color}{Red}
\makeatother
\usepackage[nodisplayskipstretch]{setspace}
\usepackage{xargs}
\newcommandx{\greencom}[2][1=]
{\todo[inline, color=green!40,#1]{#2}}
\newcommandx{\bluecom}[2][1=]
{\todo[inline, color=blue!40,#1]{#2}}
\newcommandx{\bluemargin}[2][1=]
{\todo[color=blue!40,#1]{#2}}
\newcommandx{\redcom}[2][1=]
{\todo[inline, color=red!40,#1]{#2}}
\newcommand{\RNum}[1]{\uppercase\expandafter{\romannumeral #1\relax}}
\begin{document}
	
	\title{Using the Autler-Townes and ac Stark effects to optically tune the frequency of indistinguishable single-photons from an on-demand source}
		
	\author{Chris Gustin}
	\email{cgustin@stanford.edu}
	\affiliation{Department of Applied Physics, Stanford University, Stanford, California 94305, USA}
	\affiliation{Department of Physics, Engineering Physics, and Astronomy, Queen's University, Kingston, Ontario K7L 3N6, Canada}
\author{\L{}ukasz Dusanowski}
%	\email{lukaszd@princeton.edu}
	\affiliation{Technische Physik and W\"{u}rzburg-Dresden Cluster of Excellence ct.qmat, Physikalisches Institut and Wilhelm-Conrad-R\"{o}ntgen-Research Center for Complex Material Systems, University of W\"{u}rzburg, Am Hubland, D-97074 W\"{u}rzburg, Germany}	
	\affiliation{Present address: Department of Electrical Engineering, Princeton University, Princeton, NJ 08544, USA}
	
	\author{Sven H\"{o}fling}
	\affiliation{Technische Physik and W\"{u}rzburg-Dresden Cluster of Excellence ct.qmat, Physikalisches Institut and Wilhelm-Conrad-R\"{o}ntgen-Research Center for Complex Material Systems, University of W\"{u}rzburg, Am Hubland, D-97074 W\"{u}rzburg, Germany}
		
	\author{Stephen Hughes}
	\affiliation{Department of Physics, Engineering Physics, and Astronomy, Queen's University, Kingston, Ontario K7L 3N6, Canada}
	
	\date{\today}
	
	\begin{abstract} 
We describe how a coherent optical drive that is near-resonant with the upper rungs of a three-level ladder system, in conjunction with a short pulse excitation, can be used to provide a frequency-tunable source of on-demand single photons. Using an intuitive master equation model, we identify two distinct regimes of device operation: (i) for a resonant drive, the source operates using the Autler-Townes effect, and (ii) for an off-resonant drive, the source exploits the ac Stark effect. The former regime allows for a large frequency tuning range but coherence suffers from timing jitter effects, while the latter allows for high indistinguishability and efficiency, but with a restricted tuning bandwidth due to high required drive strengths and detunings. We show how both these negative effects can be mitigated by using an optical cavity to increase the collection rate of the desired photons. We apply our general theory to semiconductor quantum dots, which have proven to be excellent single-photon sources, and find that scattering of acoustic phonons leads to excitation-induced dephasing and increased population of the higher energy level which limits the bandwidth of frequency tuning achievable while retaining high indistinguishability. Despite this, for realistic cavity and quantum dot parameters, indistinguishabilities of over $90\%$ are achievable for energy shifts of up to hundreds of $\mu$eV, and near-unity indistinguishabilities for energy shifts up to tens of $\mu$eV. Additionally, we clarify the often-overlooked differences between an idealized Hong-Ou-Mandel two-photon interference experiment and its usual implementation with an unbalanced Mach-Zehnder interferometer, pointing out the subtle differences in the single-photon visibility associated with these different setups.
	\end{abstract}
	\maketitle

\section{Introduction}\label{sec:intro}
The single-photon source (SPS) as a resource for quantum information technology has in recent years exhibited great progress in experimentally achieved efficiency and quantum state purity, pushing the technology towards practical near-term applications. Recent advances have enabled single photons to be generated on-demand with efficiencies exceeding 50\%~\cite{tomm21,thomas21,Wang2019Nov} and near-unity quantum indistinguishability~\cite{somaschi16} and purity~\cite{Hanschke2018,Sbresny2022Mar}, facilitating advances in boson sampling~\cite{wang2019} and quantum key distribution~\cite{francesco2021,kupko2020,kolodynski2020}, and even approaching minimum fidelities required for efficient linear optical quantum computation~\cite{Jennewein2011Feb,Varnava2008}.

For on-demand SPSs, these advances have largely been achieved using semiconductor quantum dots (QDs), where the dipole-active transition of an 
electron-hole pair (exciton) across the band gap in conjunction with the three-dimensional confinement afforded by the QD geometry provides an excellent quantum two-level system which, when inverted by excitation, emits a single photon  radiatively. %relaxes. 
Additional challenges to implementation of SPSs which have seen recent progress are the desirable criteria of scalability~\cite{uppu2020_2,uppu2020, Dusanowski2019,Ollivier2020Mar}, and frequency tuning~\cite{Lee2020}, as QDs are typically grown such that energy levels are stochastic in nature, but many applications require many SPSs with degenerate frequencies. Effective methods for frequency tuning QD SPSs (with a large variance in attainable bandwidth between methods) include electrical tuning~\cite{Schnauber2021, Nowak2014Feb}, strain tuning~\cite{dusanowska2020,Grim2019,ElShaari2018}, quantum frequency conversion via optical nonlinearity~\cite{Singh2019}, and multi-photon Raman transition processes utilizing multi-level systems~\cite{Breddermann2016, Gustin2017, Jonas2022Mar}. This last all-optical tuning process typically involves two sequential laser pulses, and uses the biexciton (two exciton) state, which extends the two-level structure of the QD to a cascade-type ladder system, and as such is applicable to any ladder system involving three or more energy levels, not just QDs. 

In a similar manner, we have shown recently---and demonstrated experimentally using the QD biexciton-exciton cascade---how  on-demand frequency-tunable single photons can be generated from such a ladder system with high efficiency, indistinguishability, and purity, by instead using a single pulse excitation under the presence of a cw laser {\it dressing} the exciton-biexciton transition~\cite{Dusanowski2022May}. Depending on whether the cw laser is resonant or detuned, this SPS then operates using either the Autler-Townes (AT) effect or ac Stark shift, respectively. Such an approach allows for the potential of all-optical frequency modulation of the emitter resonance~\cite{Lukin2020}, which has applications including creating high-dimensional entangled quantum states~\cite{Kues2017,Lukens2017}, and topological states~\cite{Silveri2017}. This optical frequency tuning may potentially also improve the performance of entangled photon pair sources in QDs~\cite{Schimpf2021Mar, Liu2019Jun, Olbrich2017Sep, Huber2014Nov, Zeuner2021Jul}, where the small fine structure splitting of polarized excitons can degrade entanglement fidelity.

While the possibilities of frequency-tuning a SPS at the level of the coherent optical system dynamics are interesting, single photon emitters for practical quantum information technology applications have very stringent requirements on efficiency, single-photon indistinguishability, and purity. It is thus an important question from a theoretical perspective what role the cw dressing laser plays in these SPS figures of merit, and how the source can be designed to minimize these effects. The analysis required to answer such a question would supplement and extend previous theoretical work that has helped elucidate the limits of SPS figures of merit in undressed systems, including the role of the pulse, cavity, and electron-phonon scattering~\cite{Gustin2018,Gustin2019, ilessmith17,somaschi16, Hanschke2018,tomm21,cosacchi19}.

In this work, we address this question in detail by studying theoretically 
%in detail 
a four-level ladder system, which can be physically realized using the QD biexciton-exciton cascade, as we have done in Ref.~\cite{Dusanowski2022May}. We find that the primary effects of source figure-of-merit degradation come from undesirable spontaneous emission from the higher energy state, and, in the case of semiconductor QDs, electron-phonon scattering induced by the cw laser causing excitation-induced dephasing during the emission process and, usually, increased population of the higher energy (biexciton) state. However, we find that incorporating an optical cavity resonant with the lower energy (exciton) state can mitigate these effects by accelerating emission into the preferred cavity mode. The ac Stark regime offers better SPS efficiency and indistinguishability at the cost of much reduced bandwidth of achievable frequency-tuning. We note that in this work we refer to the states of the system as QD (bi)excitons, but all results are also presented for the case of no phonons, which is generally applicable to any quantum four-level system, and the principles apply equally to a three-level system, with modified population dynamics due to the reduced number of decay channels.

The layout of the rest of the paper is as follows: in Sec.~\ref{sec:model}, we introduce our frequency-tunable SPS design and basic principles of operation, based on a four- (or three-) level quantum ladder system, of which the biexciton cascade in QDs is one physical realization. We present our quantum master equation (ME) model of the SPS, including for the case of QDs coupling to phonon reservoirs using the polaron master equation (PME) method. 

Next, we define in Sec.~\ref{sec:figs} the figures of merit we use to quantify the SPS fidelity. In particular, we include a quantum optical derivation of the two-photon interference visibility used to extract the single-photon indistinguishability of the source in a Mach-Zehnder (MZ) interferometer simulating a Hong-Ou-Mandel (HOM) interference experiment. The resulting expression is well known~\cite{santori2002}, and can be expressed in terms of the inteferometer properties, the single-photon purity, and the single-photon indistinguishability. However, most theoretical works on the subject to date assume an HOM inteferometer with two distinct SPSs; the MZ setup used in experiments only utilizes one physical SPS, which gives differing photon statistics. As a result, different expressions for the single-photon indistinguishability have existed in the literature---a discrepancy which becomes particularly important when the purity is non-ideal, as recent work has highlighted~\cite{Ollivier2021}. For the sake of comparing results directly to experiment, we expect this derivation will be of use in bridging the gap between theoretical and experimental works in the literature. 
%We also define a new figure-of-merit, unique to our SPS, called the cw error rate $\mathcal{E}_{\rm cw}$, which quantifies the amount of photons excited from the system in its ground state by the far off-resonant cw drive excitation.

In Sec.~\ref{sec:results}, we discuss our main results for the operation of the SPS in both AT and ac Stark regimes, and show how for the case of the QD SPS the phonon bath influences the performance of the device in both regimes and places limits on achievable figures of merit. We also show how an optical cavity can be used to significantly improve device performance by reducing timing jitter and phonon-related decoherence by means of selectively increasing the desired dipole transition rate. We then discuss aspects of the initial pulse excitation, including
%the error rate induced by the far off-resonant excitation of the system by the cw laser, and 
the effect of cw dressing on the source purity. Finally, in Sec.~\ref{sec:conc} we conclude. 

We also include five Appendices: in Appendix~\ref{app:sol}, we present a full analytical solution for the efficiency and indistinguishability for the case of resonant laser dressing in the absence of phonon effects. In Appendix~\ref{app:secular}, we show how a unitary transformation to the dressed state basis and secular approximation can be used to remove fast-oscillating terms in the ME, which drastically improves computational efficiency for most numerical calculations. In Appendix~\ref{app:weak}, we use a weak phonon coupling approximation, appropriate for the regimes studied in this work, to derive simple analytical expressions for the phonon interaction terms, and give an intuitive physical picture for the phonon processes. In Appendix~\ref{app:pulse}, we show how to extend the PME to include a time-dependent excitation pulse which we use to calculate the SPS purity. Lastly, we include in Appendix~\ref{app:cw} a study of the cw error rate induced by the far off-resonant excitation of the system by the cw drive (otherwise neglected for most of our analysis), which can typically be mitigated by spectral filtering.

\section{Theoretical model of a frequency-tunable SPS}\label{sec:model}
In this section, we present the main theoretical model we use to study the frequency-tunable SPS using the QD biexciton-exciton cascade, and describe its regimes of operation. In Sec.~\ref{subsec:laddermodel}, we describe the four-level cascade model and present the ME of the system Hamiltonian under cw driving with radiative emission. In Sec.'s~\ref{subsec:AT} and~\ref{subsec:ac} we describe the AT and ac Stark regimes of operation, respectively, and in Sec.~\ref{subsec:phonon} we describe how we model the electron-phonon interaction using the PME. Further detail and characterization of our SPS scheme, including emission spectra, can be found in Ref.~\cite{Dusanowski2022May}.

\subsection{Quantum ladder model}\label{subsec:laddermodel}
 We model the quantum ladder cascade system for the practical physical realization of the semiconductor QD as a four-level system with ground $\ket{G}$, excitons $\ket{X}$ and $\ket{Y}$ (with orthogonal linear polarizations), and biexciton $\ket{B}$ states, with the $\ket{B}$-$\ket{X}$ transition dressed by a coherent drive with strength $\Omega_{\rm cw}$. The coherent laser also weakly couples the $\ket{X}$-$\ket{G}$ transition, which we assume to be far-detuned due to the biexciton binding energy. 
 
\begin{figure}
		\centering
		\includegraphics[width=1\linewidth]{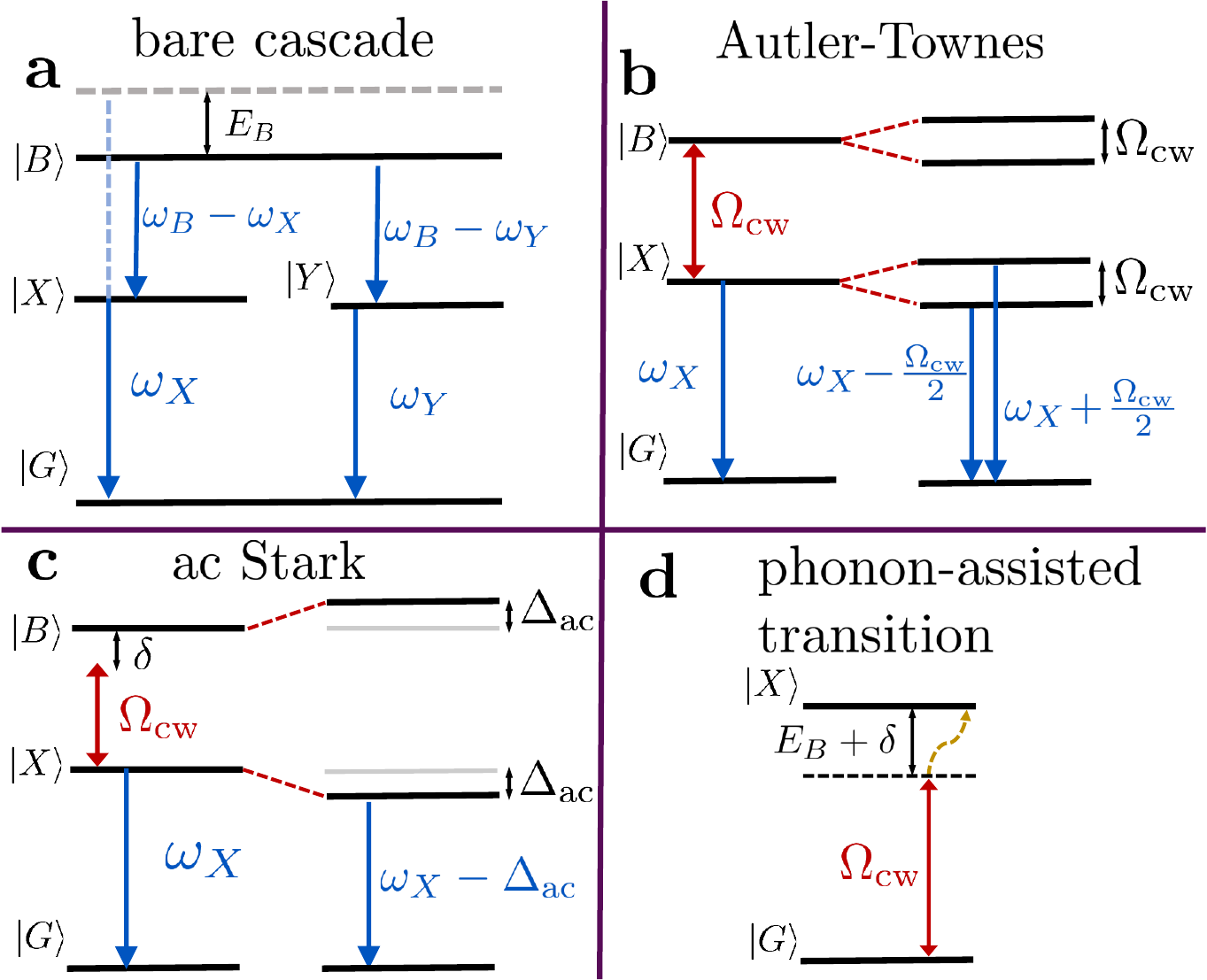}
		\caption{\label{fig:schematic}
		(a) Schematic of the quantum four-level system that we use to model the biexciton-exciton cascade for semiconductor QDs, and the fluorescence transitions in the absence of a dressing laser. (b) Dressed-state representation schematic of the SPS operating in the AT regime, with a detuning between the $\ket{B}$ and $\ket{G}$ states of $\delta=0$. (c) Schematic of the SPS operating in the ac Stark regime with $\delta/\Delta_{\rm ac} \gg1$. (d) Example schematic of a phonon-assisted excitation process; shown here is the process where the excitation from the ground $\ket{G}$ to excited $\ket{X}$ state, which is detuned by $E_B + \delta$, is assisted by the annihilation of a phonon in the phonon bath with energy $\sim E_B+\delta$. Note for (b,c) we have not shown the $\ket{Y}$ state which is involved in another decay channel as seen in (a).
		}
	\end{figure}

 The total Hamiltonian for this setup,  neglecting for now phonon coupling and radiative emission, is (letting $\hbar=1$ throughout)
	\begin{equation}\label{eq:H1}
	    H_{\rm tot} = \sum_{i}\limits  \omega_i \ket{i}\bra{i}  + \Omega_{\rm cw}\cos{(\omega_{\rm cw} t)}(\sigma_x^{B} + \sigma_x^X),
	\end{equation}
	where $\omega_i$ is the energy of the $i^{\rm th}$ state and $i \in \{X,Y, B\}$,  $\sigma_x^{B}=\sigma^+_{B} + \sigma^-_{B}$, $\sigma_x^X = \sigma^+_X + \sigma^-_X$, $\sigma^+_{B} = \ket{B}\bra{X}$, $\sigma^-_{B} = \ket{X}\bra{B}$, $\sigma^+_{X} = \ket{X}\bra{G}$, and $\sigma^-_X = \ket{G}\bra{X}$. The undressed system (i.e., with $\Omega_{\rm cw} =0$) gives rise to $X$-polarized fluorescence emission energies at  $\omega_{B} -\omega_{X}$ and $\omega_{X}$. In Fig.~\ref{fig:schematic}(a), we show a schematic of this undriven system in this bare state basis.

	The system is driven at (near-)resonant cw  frequency $\omega_{\rm cw} = \omega_{B} - \omega_{X} - \delta$ with an $X$ polarized laser,  such that $\delta$ is the laser detuning from the biexciton--X-exciton  transition. By moving into an interaction picture defined by $H_0 = (E_B +  \omega_{\rm cw})\sigma^+_X\sigma^-_X + \omega_{Y} \ket{Y}\bra{Y} + (E_B + 2 \omega_{\rm cw})\sigma_{B}^+ \sigma_{B}^- + E_B\sigma^-_X\sigma^+_X$, where the biexciton binding energy is defined as  $E_B =2\omega_{X}-\omega_{B}$, and performing the rotating wave approximation, we obtain the time-independent system Hamiltonian:
	\begin{align}\label{eq:r1}
	H_S &= 2\delta  \sigma^+_{B}\sigma^-_{B}  +\delta\sigma^+_X\sigma^-_X - E_B\sigma_X^- \sigma_X^+ \nonumber \\ &+  \frac{\Omega_{\rm cw}}{2}
	\left(\sigma_x^{B}+\sigma_x^X\right).
	\end{align}
	
	Equation \eqref{eq:r1} contains a far-detuned drive coupling the $\ket{X}$-$\ket{G}$ transition via the $\sigma_x^X$ term, and as such can model weak cw excitation of the exciton from the ground state. We shall assume for proper device operation that $E_B \gg | \delta|,  \Omega_{\rm cw}$ in all cases, such that this coupling can generally be neglected. However, we shall use Eq.~\eqref{eq:r1} for the simulations in Appendix.~\ref{app:cw} where we model the cw error rate. Neglecting this term, we can move into a different rotating frame instead defined by $H_0' = (\omega_X + \frac{\delta}{2})\sigma^+_X\sigma^-_X + (\omega_B - \frac{\delta}{2})\sigma_B^+\sigma_B^- + \omega_Y\ket{Y}\bra{Y}$, such that the system Hamiltonian can now be written as 
	\begin{equation}\label{eq:r2}
	H_S = \frac{\delta}{2} \sigma^B_z+ \frac{\Omega_{\rm cw}}{2}\sigma_x^{B},
	\end{equation} 
	where $\sigma_z^{B} = \sigma^+_{B}\sigma^-_{B} - \sigma^+_{X}\sigma^-_{X}$. This Hamiltonian has eigenenergies $E_{\pm} = \pm\eta /2$, where $\eta = \sqrt{\Omega_{\rm cw}^2 + \delta^2}$, and corresponding eigenstates
		\begin{equation}\label{eq:eigens}
\ket{\pm} = \frac{1}{\sqrt{2}}\left[\sqrt{1 \pm \frac{\delta}{\eta}} \ket{B} \pm \sqrt{1 \mp \frac{\delta}{\eta}}\ket{X}\right].
	\end{equation}
	The frequency splittings apparent in $E_{\pm}$ allow for frequency tuning of the source via radiative transitions between the dressed energy levels.
	
	Without yet considering phonon coupling, we can model spontaneous emission using a Lindblad ME for the reduced density operator of the system $\rho$:
	\begin{align}\label{eq:me1}
	\dot{\rho} =& -i[H_S,\rho] + \mathbb{L}_{\text{rad}}\rho,
	\end{align}
	where we have included radiative decay from both excitons with rate $\gamma_X$, and from the biexciton with \emph{total} rate $\gamma_{B}$ (i.e., we assume throughout orthogonal polarization channels have equal decay rates):
	\begin{align}
	\mathbb{L}_{\text{rad}}\rho&=\frac{\gamma_X}{2}\mathcal{L}[\sigma^-_X]\rho + \frac{\gamma_X}{2}\mathcal{L}\big[\ket{G}\bra{Y}\big]\rho \nonumber \\ &+ \frac{\gamma_{B}}{4}\mathcal{L}[\sigma^-_{B}]\rho + \frac{\gamma_{B}}{4}\mathcal{L}\big[\ket{Y}\bra{B}\big]\rho,
	\end{align}
	where $\mathcal{L}[A]\rho = 2A\rho A^{\dagger} - A^{\dagger}A\rho - \rho A^{\dagger}A$ is the Lindblad superoperator. 
	
	\subsection{Autler-Townes regime}\label{subsec:AT}
	For the case of no detuning (or small detuning relative to the drive strength $|\delta| \ll \Omega_{\rm cw}$), and a drive strength which exceeds the decay rates of the system (i.e., $\Omega_{\rm cw} \gg \gamma_X$), the SPS operates in the AT regime, where the $\ket{\pm}$ eigenstates of the Hamiltonian in Eq.~\eqref{eq:r2} become symmetric and antisymmetric superpositions of $\ket{X}$ and $\ket{B}$ states, with an AT energy splitting of $ \Omega_{\rm cw}$. The emission spectrum from the $\ket{X}$-$\ket{G}$ transition then consists of two peaks with energies $\omega_X \pm \Omega_{\rm cw}/2$. These peaks have nearly equal spectral weight (area), and as such the efficiency of a device operating in the AT regime is \emph{at most} $\sim 1/2$ if only one of the peaks is of interest. In Fig~\ref{fig:schematic}(b), we show a schematic of the four-level system QD model operating in this regime, and the associated energy splittings.
	
	\subsection{ac Stark regime}\label{subsec:ac}
	For larger detunings ($|\delta| \gg \Omega_{\rm cw})$, the $\ket{\pm}$ dressed eigenstates of the Hamiltonian in Eq.~\eqref{eq:r2} become unequal superpositions of $\ket{X}$ and $\ket{B}$ states, and as such, a system initialized in the $\ket{X}$ state will tend to emit photons preferentially from the eigenstate which contains a higher amplitude of the $\ket{X}$ state; 
	%and 
	the emission spectrum will consist of a dominant peak from the transition from this dressed state to the ground state and a subdominant peak from the other dressed state transition to the ground state. As the detuning is increased even further relative to the drive strength, the subdominant peak becomes negligible, and the spectrum consists of a single Stark shifted peak. In this limit, one also requires the detuning and drive rates to greatly exceed the damping.
	
	We then define a parameter equal to the undressed resonance $\omega_X$ minus the frequency of the dominant peak in the spectrum, which quantifies the frequency shift achieved in the SPS:
	\begin{equation}
	    \Delta_{\rm ac} \equiv \frac{\delta}{2}\left[\frac{\eta}{|\delta|} - 1\right].
	\end{equation}
	By construction, this quantity is positive (negative) when $\delta$ is positive (negative), corresponding to a red (blue) frequency shift. $\Delta_{\rm ac}$ can then be used to give the frequency shift of the dominant peak both in ac Stark shift and AT regimes, although $\Delta_{\rm ac}$ flips sign and thus changes discontinuously as $\delta \rightarrow 0$ as the dominant and subdominant peaks switch roles. Formally at $\delta =0$, $\Delta_{\rm ac}$ is undefined, as both peaks with frequency shifts $\pm \Omega_{\rm cw}/2$ are equally prominent in the AT regime (without considering, e.g., electron-phonon coupling). We can also solve for the drive strength in terms of this frequency shift and laser detuning,
	\begin{equation}
	    \Omega_{\rm cw} = 2\sqrt{\Delta_{\rm ac}^2 + \Delta_{\rm ac}\delta}.
	\end{equation}
	In the ac Stark regime, where $|\delta|/\Omega_{\rm cw} \gg 1$, $\Omega_{\rm cw} \approx 2\sqrt{\Delta_{\rm ac}\delta}$, and one also satisfies $\delta/\Delta_{\rm ac} \gg 1$. Thus,   we shall use both criteria interchangeably to denote the ac Stark regime. Also in this limit, $\Delta_{\rm ac} \approx \Omega_{\rm cw}^2/(4\delta)$, which is the usual ac Stark shift encountered in perturbation theory.
	
	In Fig.~\ref{fig:schematic}(c), we show a schematic of the QD model operating in the ac Stark regime and the associated frequency shift of the $\ket{X}$ state $\Delta_{\rm ac}$.
	
	For the sake of this work, we do not explicitly define for what values of $\delta/\Delta_{\rm ac}$ or $\delta/\Omega_{\rm cw}$ the system enters either regime, but rather seek to understand the two regimes as limiting cases associated with these parameters.
	
	\subsection{Exciton-phonon coupling and the polaron master equation}\label{subsec:phonon}
	It is well known that the coupling of excitons in semiconductor QDs to longitudinal acoustic (LA) phonon modes has important effects on their dynamics under optical driving, including excitation-induced dephasing, off-resonant feeding effects, Rabi frequency renormalization, and non-Markovian real phonon transitions which lead to the formation of a broad phonon sideband~\cite{Besombes2001Mar,Krummheuer2002May,Forstner2003,Ramsay10,mccutcheon10,roy11,Weiler2012,Hughes2011Apr,Quilter2015,Ulrich2011Jun}. Using a spherical QD wavefunction model, the phonon coupling can be characterized by a super-Ohmic spectral function
	\begin{equation}
	J(\omega) = \alpha \omega^3 e^{-\frac{\omega^2}{2\omega_b^2}}, 
	\end{equation}
	where $\alpha$ is the phonon coupling constant and $\omega_b$ is a cutoff frequency which scales inversely with the size of the QD~\cite{nazir16}. The Hamiltonian that couples excitons with phonons takes the form of the independent Boson model, which is exactly diagonalizable~\cite{mahan,Wilson-Rae2002May}. Employing a unitary ``polaron'' transform to a frame in which this interaction is diagonalized thus allows one to construct a perturbative expansion in the optical drive strength; this approach, under the Born-Markov approximation, yields the PME~\cite{mccutcheon10}. 
	%\sh{check PME is defined} 
	%defined in introduction - chris
	
	Assuming the different transitions in the cascade have equal dipole moments~\cite{hargart16}, the result is that to incorporate phonon coupling, we add to the ME in Eq.~\eqref{eq:me1} the following term
\begin{align}\label{eq:pme}
& \mathbb{L}_{\rm PME}\rho = \nonumber \\ & \sum\limits_{m = x, y} \int_0^{\infty}\!\!  \text{d}\tau  G_m(\tau)   
[\widetilde{X}_m(-\tau) \rho(t),X_m] + {\rm H.c.},\end{align}
where $G_x(\tau) =  \cosh{[\phi(\tau)]} -1$, $G_y(\tau) =  \sinh{[\phi(\tau)]}$, $X_m = \frac{\Omega_{\rm cw}}{2\langle B \rangle}(\sigma_m^B+\sigma_m^X)$, with  $\sigma_y^j = i[\sigma_j^- - \sigma_j^+]$, and the time-dependent complex phase term is defined through 
\begin{align}
    &\phi(\tau) = \nonumber \\ & \int\limits_{0}^{\infty} \! \! \text{d}\omega \frac{J(\omega)}{\omega^2}\left[\coth{\left(\frac{ \omega}{2 k_B T}\right)}\cos{(\omega\tau)} - i\sin{(\omega\tau)}\right],
\end{align}
and $\langle B \rangle = e^{-\phi(0)/2}$. We have absorbed a coherent attenuation factor $\langle B \rangle$ from the PME into our definition of $\Omega_{\rm cw}$ for easy comparison with the no-phonon case, as well as a polaron shift in exciton resonance frequencies. Except for when we model the cw error rate, we can neglect the $\sigma_m^X$ term in $X_m$, which is consistent with using the approximate Hamiltonian in Eq.~\eqref{eq:r2}.
The operators $\widetilde{X}_m(-\tau)=U(\tau)X_mU^{\dagger}(\tau)$ are calculated using $U(\tau) = \exp{[-iH_S\tau]}$. Note this unitary transform can be simplified analytically when using Eq.~\eqref{eq:r2} as $H_S$~\cite{ross16,mccutcheon10}, and we do this in Appendix~\ref{app:secular} in the dressed state frame. 

It is worth noting that the $X_m$ terms in Eq.~\eqref{eq:pme} lead to an overall scaling factor of $\sim\Omega_{\rm cw}^2$ in $\mathbb{L}_{\rm PME}\rho$, which dominates for small effective drive $\eta$ relative to $\omega_b$ and $k_B T$, although the full functional dependence of the phonon scattering on the drive strength will also depend on the interplay between the phonon function $\phi(\tau)$ and coherent dynamics induced by $H_S$ in the $\widetilde{X}_m(-\tau)$ functions; this  leads to rich and highly nonlinear features in the phonon decoherence rates as a function of $\Omega_{\rm cw}$~\cite{ross16}. In Appendix~\ref{app:weak}, we derive simplified expressions for the phonon coupling rates valid for strong driving and weak phonon coupling strengths.

Throughout our calculations, we shall use for our calculations two different sets of phonon parameters, denoted \RNum{1} and \RNum{2}, such that $\alpha_{\RNum{1}}= 0.04 \ \text{ps}^2$, and $\omega_{b, \RNum{1}} = 0.9 \ \text{meV}$, while $\alpha_{\RNum{2}}= 0.006 \ \text{ps}^2$, and $\omega_{b, \RNum{2}} = 5.5 \ \text{meV}$. For the most part (a notable exception being the calculation of the cw error rate),  set \RNum{1} corresponds to a ``stronger'' phonon coupling strength, and is similar to what has been extracted from measurements with InAs/GaAs QDs~\cite{Weiler2012,Gustin2021Jan,Quilter2015,Hughes2011Apr,Ota2009Aug,Ulhaq2013Feb}, while  set \RNum{2} gives a (in most cases) weaker phonon coupling strength, and is more similar to numbers consistent with experimental results in some waveguide structures~\cite{Reigue2017Jun}, including our own~\cite{Dusanowski2022May}. For our study of the source purity, we also use a set \RNum{3} with intermediate values $\alpha_{\RNum{3}} = 0.025 \ \text{ps}^2$  and $\omega_{b,\RNum{3}} = 2.5 $ meV. In all cases, we use a phonon bath temperature of $T = 4$ K.

One of the main consequences of phonon coupling is the formation of a broad phonon sideband, which arises from non-Markovian real phonon transitions concurrent with photon emission. Photons emitted into this sideband have poor indistinguishability~\cite{ilessmith17,Grange2017Jun}, and as such this sideband is usually filtered out for HOM interference measurements, leaving only the zero-phonon--line (ZPL), which has much better coherence properties due to the fact that phonon dephasing of the ZPL for bulk phonons vanishes very rapidly at low temperatures~\cite{Tighineanu2018Jun} (and this is a higher-order process, not captured by our PME). The PME is in fact capable of capturing this non-Markovian effect by means of the exponential factor which arises upon transformation back to the lab frame from the polaron frame~\cite{mahan,ilessmith17}. We can, for the sake of this work, approximate the filtering process that we assume to occur to remove this phonon sideband by simply neglecting this factor, and calculating all observable quantities directly in the polaron frame. In doing so, we miss an efficiency cut that arises from neglecting the sideband contribution to the emission. However, we can analytically approximate this contribution using the factor $\langle B \rangle$, and we quantify this efficiency reduction in Sec.~\ref{sec:results}.  

In addition to this phonon sideband in the emission spectrum, it is also important in this work to consider the phonon sideband in the \emph{absorption} spectrum---specifically, the potential for phonon-assisted excitation of energy levels under detuned driving. An example of this process is shown in Fig.~\ref{fig:schematic}(d): in this example, we show that the far detuned excitation of the $\ket{X}$ state by the cw laser (due to the $\sigma_x^X$ term in the full Hamiltonian of Eq.~\eqref{eq:r1}) can be assisted by the absorption of a phonon in the phonon bath with energy $~E_B +\delta$---if the phonon spectral function $J(\omega)$ is appreciable over this frequency range, this process may become significant. In the case of Fig.~\ref{fig:schematic}(d), the process is suppressed at low temperatures due to the small thermal occupation of phonons in the bath (although it still plays a potentially significant role as we show in Sec.~\ref{sec:results}), but the corresponding process for a cw drive with frequency exceeding the energy transition of interest is highly significant even at low temperatures, as it involves phonon creation. The dynamics of the driven $\ket{B}$-$\ket{X}$ transition are also subject to similar considerations, and Appendix~\ref{app:weak} gives simplified analytical rates and a schematic picture of these phonon processes.

Finally, we note that the coherent (unitary) part of the phonon effects also leads to small frequency shifts in the emission spectrum; for the sake of this work we neglect these and focus on the nominal frequency splitting given by $\Delta_{\rm ac}$; if desired, the analytical simplifications in Appendices~\ref{app:secular} and~\ref{app:weak} can be used to calculate these small shifts explicitly.

\section{SPS figures of merit and two-photon interference experiments}\label{sec:figs}

In experiment, the two-photon interference (TPI) visibility of the source is typically measured by simulating an HOM interferometery setup using an unbalanced MZ interferometer, excited with two photon pulses separated in time by $T_0$, and with overall repetition time of the laser $T_{\rm rep}$. In contrast to this setup, theoretical analyses of the single-photon indistinguishability (which is often conflated with---or defined to be equal to---the TPI visibility) often derive this parameter by assuming an HOM setup with two identical but distinct SPSs described by the same density operator~\cite{PhysRevA.69.032305,Gustin2018,woolley13}. While both approaches should lead to perfect TPI for perfectly indistinguishable single photons with unity purity (no multiphoton probability from a source excitation), the photon statistics of the two scenarios are different, leading to different normalizations. This can make direct comparison of experiment and theory difficult, particularly in the case of non-unity single-photon purity. In the work of Kiraz \emph{et al.}~\cite{PhysRevA.69.032305}, a HOM-type experiment was analyzed, but the authors omitted terms corresponding to the second order correlation function of the source field; as this correlation function separates into a product of photon flux expectation values at large delay times, this omission erroneously led to a definition of the TPI visibility which in fact corresponds to an MZ-type experiment---although only in the limit of an ideal interferometer and zero multiphoton emission probability per excitation.

Furthermore, some authors use the single-photon indistinguishability interchangeable with the ``corrected'' TPI visibility, after accounting for the finite multiphoton probability of the source, imbalance of the beam splitters, and deviation from perfect interference fringe contrast of the interferometer~\cite{Wei2014Oct}. This is, however, potentially ambiguous, as the latter two effects are considerations arising from the experimental detection process, whereas the multiphoton probability of the source is a {\it fundamental} and physical limitation on the degree of TPI achievable with the source. Additionally, there exists an alternative definition of the TPI visibility which involves normalizing by a  cross-polarized cross-coincidence histogram peak at zero delay, which gives a different value for the observed visibility for nonzero two-photon emission probability. The difference in photon statistics from HOM vs. MZ inteferometry experiments was correctly pointed out by Fischer \emph{et al.}~\cite{fischer16}, although the metric they propose to quantify the TPI visibility differs from those typically used in experimental works.

To help clarify the matter, and for the sake of one-to-one comparison of experiment and theory, we present in Sec.~\ref{subsec:derivation} a quantum mechanical derivation based on field correlation functions of the TPI visibility for a real MZ interferometer (the expression for which is already well-known from photon counting arguments~\cite{santori2002,loredo2016,fischer16}) in the spirit of previous studies of the corresponding quantity for an HOM interferometer~\cite{PhysRevA.69.032305,woolley13,Schofield2022Jan}. We explicitly define the indistinguishability to a measure of the first-order degree of coherence of the source, and $g^{(2)}[0]$ to be a measure of the second-order degree of coherence of the source (purity, or lack of multiphoton emission events). We define the raw TPI visibility to be what is measured in experiment, and the corrected TPI visibility what would be measured in an idealized experiment with perfect fringe contrast and balanced beam splitters; this latter metric is the most important single parameter to characterize the fidelity of the SPS as it encompasses both first and second order coherence of the source.

In Sec.~\ref{subsec:fom}, we then relate these experimentally observable quantities to the theoretical figures of merit for our frequency-tunable SPS, by means of conventional quantum optics input-output theory~\cite{gardiner_quantum_2004}. In particular, we show how the dressed state basis of Eq.~\eqref{eq:eigens} can be used to derive figures of merit for each sidepeak of the spectrum separately. 
%As well, we define the additional figure-of-merit---the cw error rate $\mathcal{E}_{\rm cw}$---which is unique to our cw dressed SPS.

\subsection{Derivation of TPI visibility}\label{subsec:derivation}

\begin{figure*}
		\centering
		\includegraphics[width=0.9\linewidth]{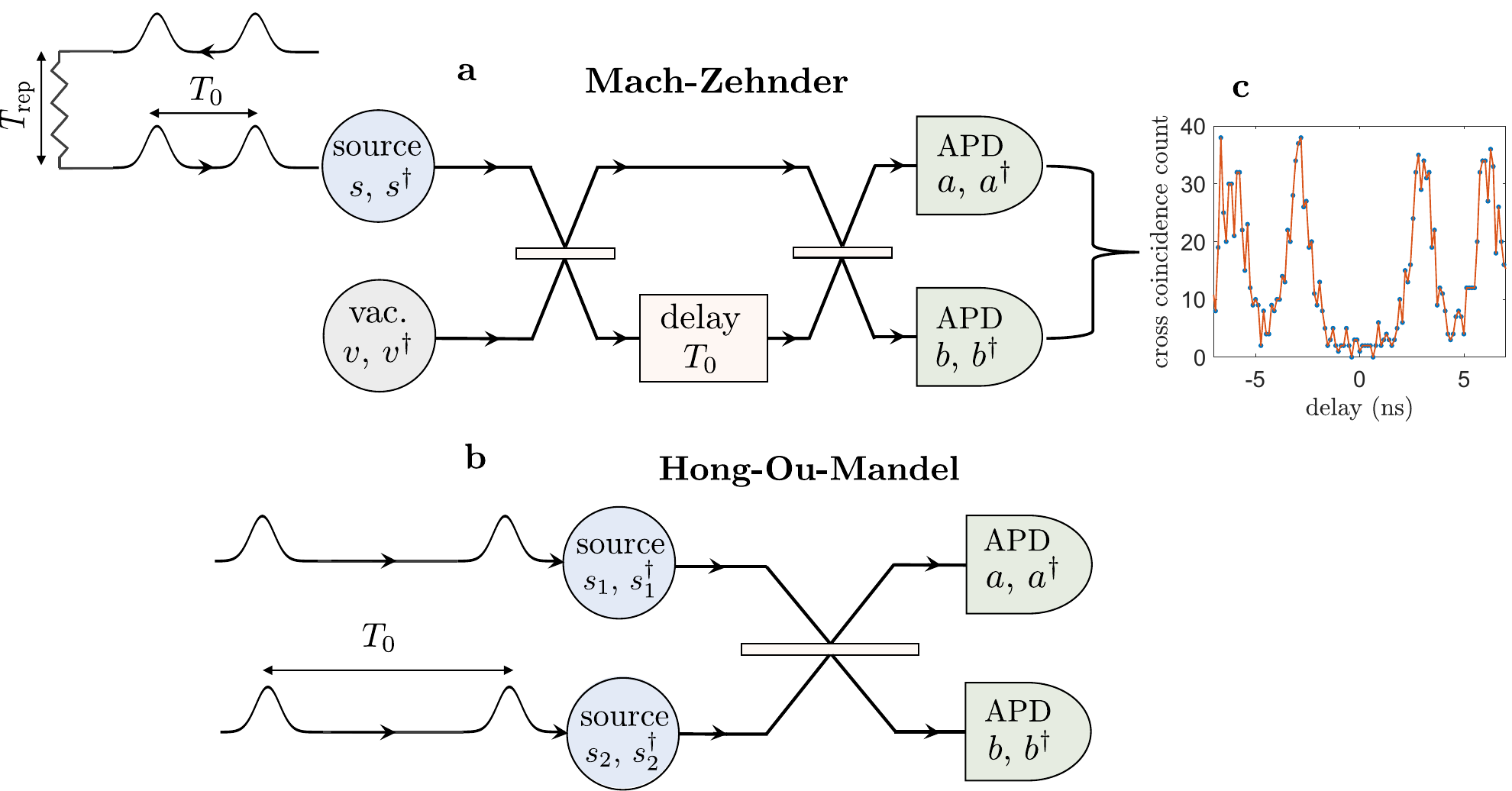}
		\caption{\label{fig:MZ}
			Schematics of (a) an unbalanced MZ simulating a HOM two-photon interferometry experiment with (c) sample cross-coincidence count data  taken for an undressed QD in Ref.~\cite{Dusanowski2022May}, and (b) an idealized HOM experiment with distinct but identical sources. 
		}
	\end{figure*}

A simplified schematic of the experimental procedure for extracting the TPI visibility of photons emitted sequentially from a SPS using a MZ interferometer is shown in Fig.~\ref{fig:MZ}(a). The SPS, excited every $T_{\rm rep}$ with two pulse excitations separated in time by $T_0$ (assumed to be much greater than the relaxation time of the SPS, to ensure independent excitation events), emits photons into a decay channel mode.  These photons then pass through two sequential beam splitters, where one of the transmission channels between the beam splitters is subject to a time delay $T_0$. The outputs of the second beam splitter then propagate to photodetectors from which a cross-correlation HOM coincidence signal can be constructed as a histogram of detection events; for perfect quantum single-photon interference, this signal vanishes at zero time delay~\cite{Hong1987Nov}. We also show in Fig.~\ref{fig:MZ}(c) an experimental example of this cross-coincidence function taken from the data in Ref.~\cite{Dusanowski2022May}. Note that in this analysis we assume a purely pulsed SPS; any residual cw contribution which arises due to (for example) the small excitation of the ground-excited state transition from the far off-resonant cw laser is assumed to be filtered out of the emission spectrum, although it is possible to extend the analysis to also account for the cw background~\cite{Ollivier2021,Kirsanske2017Oct}.

In experiment, the raw visibility of TPI is often defined as (sometimes without the factor of 2)~\cite{santori2002,Dusanowski2019,somaschi16}
\begin{equation}\label{eq:v_def}
    \mathcal{V}_{\rm raw} = 1 - \frac{A_0}{(A_{+ }+A_{-})/2},
\end{equation}
where $A_0$ denotes the area of the peak in the cross-correlation coincidence histogram at $\tau = 0$, and $A_{\pm }$ denote the area of the neighbouring peaks. To theoretically calculate this value, we include here a quantum optics derivation of the cross-correlation signal that is detected by an unbalanced MZ interferometer setup with delay $T_0$ as show schematically in Fig.~\ref{fig:MZ}(a). We assume a signal mode with annihilation operator $s$,
%\sh{slight overlap with the source one, but distinguishable ...}
which we will relate to the system modes of the QD via standard input-output theory~\cite{PhysRevA.31.3761}, as well as a vacuum mode with annihilation operator $v$. We assume that the timescale of decay for the system dynamics $T_{\rm lifetime} \sim \gamma_X^{-1}$ is much smaller than $T_0$. 

Assuming for simplicity the two beam splitters to be identical and lossless, we can express the modes detected by the photodetectors as 
\begin{equation}\label{eq:a_eq}
a(t) = \mathcal{R}^2 s(t) + \mathcal{RT} v(t) + \mathcal{T}^2 s(t\!-\!T_0) + \mathcal{RT} v(t\!-\!T_0),
\end{equation}
and
\begin{equation}\label{eq:b_eq}
    b(t) = \mathcal{R}^2 v(t\!-\!T_0) + \mathcal{RT} s(t\!-\!T_0) + \mathcal{T}^2 v(t) + \mathcal{RT} s(t),
\end{equation}
where the reflectivity and transmissivity of the beam splitters satisfy $|\mathcal{T}|^2+|\mathcal{R}|^2 =1$ and $\mathcal{R}\mathcal{T}^* + \mathcal{R}^*\mathcal{T}= 0$, enforcing unitarity. The vacuum terms do not contribute to any normal-ordered expectation values and are dropped henceforth.
%
%\begin{widetext}
%\end{widetext}
%

For convenience, we define quantities associated with the source \emph{excited only with a single pulse excitation}. These constitute the main single-photon figures of merit for the \emph{total} spectrum emitted from the source (i.e., including both peaks of the split spectrum in the AT regime). These are  (i), the number of photons emitted by the source:
\begin{equation}\label{eq:N}
    \mathcal{N} = \int_0^{\infty} \text{d}t \langle s(t)^{\dagger} s(t)\rangle,
\end{equation}
which we shall sometimes refer to as the ``efficiency'' (or ``brightness'', although note there are other sources of end-to-end efficiency degradation not captured by this metric); (ii)
the normalized Hanbury-Brown-Twiss (HBT) $g^{(2)}[0]$---a measure related to the two-photon emission probability of the source, which can be measured by blocking one arm of the MZ interferometer and normalizing the $t$-integrated peak (time-averaged) in the cross-correlation signal around $\tau\sim0$ to any other peak:
\begin{equation}\label{eq:g2_0}
    g^{(2)}[0] = \frac{1}{\mathcal{N}^2/2}\int_0^{\infty} \! \! \! \text{d}t \! \int_0^{\infty} \! \! \! \text{d}\tau \langle s^{\dagger}(t)s^{\dagger}(t\!+\! \tau)s(t\!+\!\tau)s(t)\rangle;
\end{equation}
and (iii), the single-photon indistinguishability:
\begin{equation}\label{eq:In}
    \mathcal{I} = \frac{1}{\mathcal{N}^2/2}\int_0^{\infty} \! \! \!  \text{d}t \! \int_0^{\infty} \! \! \!  \text{d} \tau |\langle s^{\dagger}(t)s(t\!+\!\tau)\rangle|^2,
\end{equation}
which is a measure of the first order degree of coherence of the SPS.

The cross-correlation signal from the `a' and `b' photodetectors is proportional to the probability that if one detector detects a photon at time $t$, the other will detect a photon at time $t+ \tau$, and is proportional to (assuming identical detectors)
\begin{align}\label{eq:G2MZ}
    G^{(2)}_{\rm MZ}(t,\tau) &=\langle a^{\dagger}(t)b^{\dagger}(t\!+\!\tau)b(t\!+\!\tau)a(t)\rangle \nonumber \\ &+\langle b^{\dagger}(t)a^{\dagger}(t\!+\!\tau)a(t\!+\!\tau)b(t)\rangle .
\end{align}
The vanishing of $G^{(2)}_{\rm MZ}(t,\tau)$ around $\tau=0$ for times around $t$ where one might expect classically a cross-correlation photon detection event due to two wave-packets hitting the (second) beam splitter at the same time and travelling down different channels  is, similar to the HOM setup, a hallmark of TPI and a signature of single-photon indistinguishability. However, the photon statistics of the MZ interferometer differ from that of an HOM interferometer, and the calculation of the TPI visibility is modified accordingly. Note that we could instead define $G^{(2)}_{\rm MZ}(t,\tau)$ using only one of the two cross-correlation functions, which will, in the presence of an unbalanced beam splitter change the various peak weights of the time-integrated cross correlation function~\cite{santori2002}, but our metric of two-photon visibility is independent of this choice. Using Eq.'s~\eqref{eq:a_eq},~\eqref{eq:b_eq}, and~\eqref{eq:G2MZ}, we find:
\begin{align}\label{eq:Gbig}
    & \frac{G^{(2)}_{\rm MZ}(t,\tau)}{|\mathcal{RT}|^2} =  2|\mathcal{R}|^4 \langle s^{\dagger}(t) s^{\dagger}(t\!+\!\tau) s(t\!+\!\tau) s(t)\rangle 
    \nonumber \\ &+ 2|\mathcal{T}|^4 \langle s^{\dagger}(t\!-\!T_0)s^{\dagger}(t\!+\!\tau\!-\!T_0)s(t\!+\!\tau\!-\!T_0)s(t\!-\!T_0)\rangle \nonumber \\ &-2|\mathcal{RT}|^2\Big[ \langle s^{\dagger}(t) s^{\dagger}(t\!+\!\tau\!-\!T_0)s(t\!+\!\tau)s(t\!-\!T_0)\rangle + \text{H.c.} \Big] \nonumber \\
    & + (|\mathcal{T}|^4+|\mathcal{R}|^4)\Big[\langle s^{\dagger}(t)s^{\dagger}(t\!+\!\tau\!-\!T_0)s(t\!+\!\tau\!-\!T_0)s(t)\rangle \nonumber \\ & \phantom{testetst}+ \langle s^{\dagger}(t\!-\!T_0) s^{\dagger}(t\!+\!\tau) s(t\!+\!\tau) s(t\!-\!T_0)\rangle \Big] .
\end{align}

In the above, we have made note that since we have assumed that $T_0 \gg T_{\rm lifetime}$, the dynamics of operators separated in time by $T_0$ are uncorrelated. Thus we have  dropped terms where for any values of $t,\tau$, using this property, we can decompose the correlation function as a product of multiple expectation values involving a phase oscillation (i.e., unequal numbers of annihilation and creation operators within a correlation function). These terms are small but nonzero for finite $g^{(2)}[0]$---however, they are phase-sensitive and thus will average out to zero in an experimental setting where data is collected over a timescale longer than the coherence of the system~\cite{santori2002, fischer16}.

In the TPI experiment, the MZ interferometer is fed two photons from the same source separated in time by $T_0$, and this process is repeated every $T_{\rm rep}$. For the following derivation, we shall assume that $T_{\rm rep} - 4T_0 \gg T_{\rm lifetime}$, such that each laser repetition is an independent event containing only two QD excitation events. However, as the figures of merit we derive only involve peaks at $\tau$ delays of zero and $\sim \pm T_0$, the final results are valid for the weaker condition $T_{\rm rep} - 3T_0\gg T_{\rm lifetime}$.

Under the assumption that $T_{\rm rep} \gg 4T_0$, we only need to consider from the perspective of a theoretical analysis the function $G_{\rm MZ}^{(2)}(t,\tau)$ for a source excitation $s(t)$ that is excited only at times $t = 0$ and $t = T_0$ (i.e., one laser pulse cycle). Thus, any correlation functions involving operators with time arguments that are not within $\sim T_{\rm lifetime}$ of $0$ or $T_0$ vanish. Furthermore, $G_{\rm MZ}^{(2)}(t,\tau)$ is nonzero around times $t$ and delays $\tau$ of  $\sim 0$, $ \sim T_0$, and $\sim 2T_0$ only.

In the TPI experiment, photon detection events are integrated over time, and thus we will integrate $G^{(2)}_{\rm MZ}(t,\tau)$ over the time bounds $t \in [0, T_{\rm rep}/2]$, or $t \in [0, \infty)$ as we are only considering a single laser pulse cycle. However, due to the periodicity of the system in time over multiple excitations separated in time by more than the relaxation time of the system, the system density operator is prepared in the same excited state from both QD pulse excitations, and thus certain correlation functions will be equal around time arguments $\sim 0$ and $\sim T_0$. 
Using this knowledge of the MZ cross-coincidence correlation function, we can simplify the time integration:
\begin{align}\label{eq:tint}
   & \int_0^{\infty}\! \! \text{d}t G_{\rm MZ}^{(2)}(t,\tau) = \left(\int_0^{T_0} \!\!+\!\! \int_{T_0}^{2T_0} \!\!+\!\! \int_{2T_0}^{3T_0}\right)\text{d}t G_{\rm MZ}^{(2)}(t,\tau) \nonumber \\ &= \int_0^{T_0} \! \! \text{d}t\big[G_{\rm MZ}^{(2)}(t,\tau) + G_{\rm MZ}^{(2)}(t\!+\!T_0,\tau) + G_{\rm MZ}^{(2)}(t\!+\!2T_0,\tau)\big].
\end{align}

The function $\int_0^{\infty} \text{d}t G^{(2)}_{\rm MZ}(t,\tau)$ corresponds to the coincidence count histogram as shown schematically in Fig.~\ref{fig:MZ}(c)  and has a characteristic 5-peak structure (experimentally, peaks at larger delays also occur due to the laser repetition rate).
With this function in mind, we define the TPI visibility by normalizing the peak around $\tau = 0$ to its neighbouring peaks:
\begin{equation}\label{eq:Vraw1}
    \mathcal{V}_{\rm raw} = 1 -  \frac{ \int_0^{\infty}\text{d}t \int \text{d}\tau G_{\rm MZ}^{(2)}(t,\tau)}{ \int_0^{\infty}\! \!\text{d}t \int \! \! \text{d}\tau \frac{1}{2}\Big( G_{\rm MZ}^{(2)}(t,\tau\!+\!T_0)+ G_{\rm MZ}^{(2)}(t,\tau\!-\!T_0)\Big)},
\end{equation}
where the integration bounds of $\tau$ are chosen to capture the peak of interest alone (much larger than $T_{\rm lifetime}$ to capture the entire peak, but not so large as to integrate over neighbouring peaks).

Using Eq.'s~\eqref{eq:tint} and~\eqref{eq:Vraw1}, we localize the integration bounds for $t$ and $\tau$ to occur to around $t, \tau \sim 0$. The expression for $\mathcal{V}_{\rm raw}$ can then be highly simplified by noting that any correlation functions containing operators $s(t)$, $s^{\dagger}(t)$ where $t$ is not in the vicinity of $\sim 0$ or $\sim T_0$ vanish, recalling that operators separated in time by $\sim T_0$ become uncorrelated, and finally noting that correlation functions evaluated around $T_0$ are equivalent to to those evaluated around $0$.

As an example of how this works, consider the integration around $\tau \approx 0$:
$\int_0^{\infty} \text{d}t \int \text{d}\tau G^{(2)}_{\rm MZ}(t,\tau)$.
We have three terms coming from Eq.~\eqref{eq:tint}, all given by Eq.~\eqref{eq:Gbig} at different time arguments. Considering just the third term $G^{(2)}(t+2T_0,\tau)$ as an example, this term only has one nonzero correlation function, physically corresponding to a multiphoton detection event from the output given from the second pulse excitation of the source, having travelled down the longer arm of the MZ interferometer, and is given by the second term in Eq.~\eqref{eq:Gbig},
\begin{align}
   & \int_0^{T_0} \! \text{d}t \!  \int \! \text{d}\tau G^{(2)}_{\rm MZ}(t+2T_0,\tau) = \nonumber \\ & \int_0^{T_0} \!\!\! \text{d}t  \!\! \int \!  \text{d}\tau 2RT^3 \langle s^{\dagger}(t\!\!+\!\!T_0)s^{\dagger}(t\!\!+\!\!T_0\!\!+\!\! \tau)s(t\!\!+\!\!T_0\!\!+\!\!\tau)s(t\!\!+\!\! T_0)\rangle \nonumber \\ 
    & = \int_0^{T_0} \!  \text{d}t \! \int \! \text{d}\tau 2RT^3 \langle s^{\dagger}(t)s^{\dagger}(t\!+\! \tau)s(t\!+\!\tau)s(t)\rangle \nonumber \\ & = \mathcal{N}^2 RT^3 g^{(2)}[0],
\end{align}
where $R = |\mathcal{R}|^2$ and $T = |\mathcal{T}|^2$.

Applying this procedure to all terms, we ultimately find
\begin{equation}
    \frac{A_0}{N_0} = RT\mathcal{N}^2\left[\frac{R^2+T^2}{2}(1+2g^{(2)}[0]) - RT(1-\epsilon)^2\mathcal{I}\right], 
\end{equation}
\begin{equation}
    \frac{A_{\pm}}{N_0} = RT\mathcal{N}^2\left(\frac{R^2+T^2}{2}\right)(1+g^{(2)}[0]),
\end{equation}
where $N_0$ is the number of laser pulse cycles over which data is collected, multiplied by an overall efficiency factor which contains, for example, both extraction and detector efficiencies and is assumed equal for both detectors. 

From this, we find the visibility, recovering known results~\cite{santori2002,fischer16},
\begin{equation}\label{eq:Vraw2}
\mathcal{V}_{\rm raw} = \frac{\mathcal{I}_{s}/\chi_{\rm cor} - g^{(2)}[0]}{1 + g^{(2)}[0]},
\end{equation}
where 
\begin{equation}
    \chi_{\rm cor} = \frac{R^2+T^2}{2RT(1-\epsilon)^2},
\end{equation}
is a correction factor related to imperfections associated with the interferometry setup;
 we have added a factor of the interferometer fringe contrast $(1-\epsilon)$ wherever first order degree of coherence correlation functions appear to account for optical surface imperfections reducing interference visibility.
Clearly, in the limit of no multiphotons ($g^{(2)}[0]=0$), the visibility of TPI in a perfect MZ interferometer ($1-\epsilon =1$) with balanced beam splitters ($R = T = 1/2$) and the single-photon indistinguishability are equivalent.
We can find a corrected TPI visibility (which would occur in a perfect MZ interferometer with balanced beam splitters) using
\begin{align}
    \mathcal{V} &= \frac{\chi_{\rm cor}\left[(1+g^{(2)}[0])\mathcal{V}_{\rm raw} + g^{(2)}[0]\right] - g^{(2)}[0]}{1 + g^{(2)}[0]} \nonumber \\ & \approx 
    \chi_{\rm cor}\mathcal{V}_{\rm raw} + g^{(2)}[0](\chi_{\rm cor} -1) + \mathcal{O}\left(\left[g^{(2)}[0]\right]^2\right),
\end{align}
where the second line is appropriate in the usual high purity case that $g^{(2)}[0] \ll 1$.
We can also solve for the single-photon indistinguishability:
\begin{equation}
\mathcal{I} = \chi_{\rm cor}\left[(1+g^{(2)}[0])\mathcal{V}_{\rm raw} + g^{(2)}[0]\right].
\end{equation}

It is useful to contrast this result to that obtained from a HOM interferometry experiment using two distinct SPSs with identical expectation values, as shown in Fig.~\ref{fig:MZ}(b). In this case, we can consider two sources with bosonic operators $s_1(t)$ and $s_2(t)$ incident upon a single beam splitter, and compute the cross-correlation function of the detectors. Following a completely analogous derivation as in the MZ case,  we can find the area of the peaks that appear around $\tau \sim 0$:

\begin{equation}\label{eq:a0HOM}
    \frac{A_0^{\rm (HOM)}}{N_0} = \mathcal{N}^2\left[R^2+T^2-2RT\left(\mathcal{I}(1-\epsilon)^2 - g^{(2)}[0]\right)\right],
\end{equation}
\begin{equation}
    \frac{A_{\pm}^{\rm (HOM)}}{N_0} = \mathcal{N}^2,
\end{equation}
which leads to 
\begin{equation}\label{eq:Vhom}
    \mathcal{V}_{\rm raw}^{\rm (HOM)} = 2RT\left[1+ \mathcal{I}(1-\epsilon)^2 - g^{(2)}[0]\right].
\end{equation}
Equation~\eqref{eq:a0HOM} is a direct generalization of a result initially found by Hong, Ou, and Mandel~\cite{Hong1987Nov} to allow for nonzero $g^{(2)}[0]$.

In the case of an ideal inteferometer, Eq.~\eqref{eq:Vhom} reduces to $\mathcal{V}_{\rm raw}^{\rm (HOM)} = \frac{1}{2}\left[1+ \mathcal{I} - g^{(2)}[0]\right]$. This definition has appeared in some theoretical works~\cite{Raghunathan2009Mar, Pathak2010Jul,ross16,Gustin2018,Hughes2019Aug}, sometimes referred to therein as the indistinguishability. Most notably, this definition leads to a TPI visibility of $\sim 1/2$ in the limit of distinguishable single photons, in contrast to the MZ setup. It is worth noting however, that there also exist definitions of the visibility which differ from Eq.~\eqref{eq:v_def} by a factor of two, such that the visibility in the MZ setup also goes to $\sim 1/2$ for distinguishable photons~\cite{santori2002}. Such a definition is still slightly different than the HOM setup in the case of an imperfect interferometer or nonzero $g^{(2)}[0]$.

One should also be aware that a different convention for the visibility is sometimes encountered, where the raw visiblity is instead defined as $1 - A_0/A_{0,\text{cross}}$, where $A_{0,\text{cross}}$ is the zero delay peak area obtained after rotating the polarization of one of the MZ arms prior to the second beam splitter. $A_{0,\text{cross}}$ thus differs from $A_0$ by the indistinguishability term, which vanishes for cross-polarized photons. From Eq.~\eqref{eq:a0}, it can be seen that the resultant expression for the raw visibility in this case differs slightly from the definition used in this work for nonzero $g^{(2)}[0]$, and this is important to keep in mind when comparing visibilities calculated using different methods.

\subsection{Input-output relations and SPS figures of merit}\label{subsec:fom}
In the previous subsection, we derived the brightness $\mathcal{N}$, HBT visibility $g^{(2)}[0]$, indistinguishability $\mathcal{I}$, and TPI visibility $\mathcal{V}$ for photons emitted from a SPS in terms of the output channel operators $s$, $s^{\dagger}$. Using input-output theory, these can be related to QD exciton operators for the desired $\ket{X}$-$\ket{G}$ transition in the Heisenberg picture as $s(t) = \sqrt{\gamma_X}\sigma^-_X(t)$ (neglecting vacuum input noise terms which do not contribute to any expectation values) for the total emission spectrum, neglecting any extraction efficiency loss from photons emitted into undesired modes~\cite{gardiner_quantum_2004}. However, in the AT regime (and also in the ac Stark regime to some extent), there exist two spectral components emitted from the exciton transition due to the cw laser dressing inducing energy splittings. When the difference in frequency between these peaks is greater than their spectral widths, input-output theory can be applied to each of the transitions separately~\cite{gardiner_quantum_2004}, and figures of merit can be expressed for these peaks separately~\cite{Dusanowski2022May}. This is done using the eigenstates of Eq.~\eqref{eq:eigens}.

For example, consider the total emitted photon number (brightness):
	\begin{align}\label{eq:npm}
	    \mathcal{N} &= \gamma_X \int_0^{\infty} \text{d}t \langle \sigma^+_{X}\sigma^-_{X}\rangle(t)  \nonumber \\ & = \frac{\gamma_X}{2}\!\left (1\!-\!\frac{\delta}{\eta}\right)\!\int_0^{\infty} \!\!\text{d}t \rho_+(t) + \frac{\gamma_X}{2}\!\left (1\!+\!\frac{\delta}{\eta}\right)\!\int_0^{\infty}\!\! \text{d}t \rho_-(t) \nonumber \\ &  \ \ \ \ \ \ \ \ \ \ \  - \frac{\gamma_X \Omega_{\rm cw}}{2\eta}\int_0^{\infty} \text{d}t \text{Re}\big\{\rho_{+-}(t)\big\} \nonumber \\ & \approx \mathcal{N}^+ + N^-,
	\end{align}
		where $\rho_{\pm}(t) = \bra{\pm} \rho(t) \ket{\pm}$, and $\rho_{+-}(t) = \bra{+}\rho(t)\ket{-}$. In the last line, we dropped an integration over a coherence term, as in the limit of well-separated peaks (i.e., with center-frequencies separated by $\gg \gamma_X$), the integrand is highly oscillatory and contributes negligibly; such is in the same spirit as the secular approximation made in Appendix~\ref{app:secular}.
		
		One can also define
		indistinguishability 
		for the sidepeaks using the dressed operators:
    	\begin{equation}\label{eq:ipm}
	    \mathcal{I}^{\pm} =  \frac{\int_0^{\infty} \text{d}t \int_0^{\infty} \text{d}\tau |g^{(1)}_{\pm}(t,\tau)|^2}{\frac{1}{2}\left[\int_0^{\infty} \text{d}t \rho_{\pm}(t)\right]^2},
	\end{equation}
		where $g^{(1)}_{\pm}(t,\tau) =\langle \sigma^+_{\pm}(t+\tau)\sigma^-_{\pm}(t)\rangle$, $\sigma^-_{\pm} = \ket{G}\bra{\pm}$, $\sigma^+_{\pm} = \ket{\pm}\bra{G}$, and $\rho_{\pm} = \bra{\pm} \rho\ket{\pm}$. While it is possible to define an HBT $g_{\pm}^{(2)}[0]$ for the sidepeaks, this quantity is likely strongly dependent on the filter width used to isolate the peak of interest (and thus can not be unambiguously defined without reference to the filter width), as the short excitation pulse leads to broad two-photon emission tails in the spectrum. As for a pulse shorter than the cw dressing timescale, the first emitted photon during the pulse is centered around the undressed $\omega_X$ energy, we can expect most of this to be filtered out of the isolated sidepeak, and generally we expect the $g_{\pm}^{(2)}[0]$ to be much smaller than the total $g^{(2)}[0]$. In fact, this purity-enhancing effect has been predicted even with unshifted emission frequencies in the context of pulse excitation of QDs in cavities, which play the role of spectral filtering~\cite{Gustin2018}.

			%Additionally, the presence of the $\sigma_x^X$ term in Eq.~\eqref{eq:r1} leads to weak cw excitation of the $\ket{X}$ state, by means of the far off-resonant drive. As a result, in addition to the (pulse-wise) emitted photon number $\mathcal{N}$, which is calculated in absence of this term, using instead Eq.~\eqref{eq:r2} for the system Hamiltonian, there is a small, constant in time, photon emission flux, which in some cases may be removable by frequency filtering. To quantify this \emph{cw error rate}, we define a quantity $\mathcal{E}_{\rm cw}$ to be the ratio of the average number of photons emitted in the absence of any pulse excitation over a duration equal to the laser repetition rate $T_{\rm rep}$, divided by the number of photons emitted by the source with a pulse excitation. 
			%In Sec.~\ref{subsec:cwerror}, we show how this rate can be calculated using the model of Sec.~\ref{sec:model}.
			
			%To calculate $\mathcal{E}_{\rm cw}$, we assume as a first approximation that the pulse initializes the system in the $\ket{X}$ state. We then can simulate, using the Hamiltonian of Eq.~\eqref{eq:r1}, the photons emitted $\mathcal{N}_0$ over a duration $T_{\rm rep}$ starting from the initial condition $\rho_0$, which we choose to be the steady-state condition of the ME, and then divide this quantity by the number of photons emitted using the same Hamiltonian with instead initial condition $\ket{X}$, which we denote $\mathcal{N}_X$. Then,
			%\begin{equation}\label{eq:cwerr}
			%\mathcal{E}_{\rm cw} = %\frac{\mathcal{N}_0}{\mathcal{N}_X}.
			%\end{equation}
			
			In summary, the figures of merit we use to quantify the SPS are the emitted photon number (or efficiency/brightness) $\mathcal{N}$, the HBT $g^{(2)}[0]$, and the single-photon indistinguishability $\mathcal{I}$, given by Eq.'s~\eqref{eq:N},~\eqref{eq:g2_0}, and~~\eqref{eq:In}, respectively (all with $s(t) = \sqrt{\gamma_X}\sigma_X(t)$).
			%, and the cw error rate $\mathcal{E}_{\rm cw}$ given by Eq.~\eqref{eq:cwerr}.
			We also have the corresponding quantities defined for the sidepeaks of the dressed system, $\mathcal{N}^{\pm}$ and $\mathcal{I}^{\pm}$, given by Eq.'s~\eqref{eq:npm} and~\eqref{eq:ipm}, respectively. In Appendix~\ref{app:cw}, we define and study an additional figure-of-merit, the \emph{cw error rate} $\mathcal{E}_{\rm cw}$, which quantifies the proportion of photons emitted due to the weak off-resonant excitation of the ground-exciton transition by the dressing laser. This effect has not been taken into account in the calculation of the figures of merit in the main text, as we assume this contribution can be removed from the emitted spectrum (and also is usually very small) by spectral filtering, in typical cases.

\section{SPS operation and figures of merit for Autler-Townes and ac Stark regimes}\label{sec:results}
In this section, we analyze the operation and figures of merit of our SPS source in both AT and ac Stark regimes. In Sec.~\ref{subsec:adiabatic}, we derive approximate formulae for the emitted photon number and indistinguishability in the ac Stark regime by adiabatic elimination of the $\ket{B}$ state, assuming the source to be excited in the $\ket{X}$ excited state by a short pulse at time $t = 0$. In Sec.~\ref{subsec:num}, we assume the same initial condition, and calculate the evolution of the reduced density operator using the full ME Eq.~\eqref{eq:me1} with the Hamiltonian (and associated rotating frame) of Eq.~\eqref{eq:r2}. 

Throughout this section, unless otherwise stated, we let $\gamma_X = 1.32 \ \mu \text{eV}$ as in Ref.~\cite{Dusanowski2022May}. For the case relevant to QDs where all radiative transitions have the same rate, such that the $\ket{B}$ state has half the lifetime of the $\ket{X}$ state, we have $\gamma_B = 2 \gamma_X$, and the solution to the ME with initial condition $\ket{X}$ is analytically calculable for the AT regime $\delta = 0$ (neglecting any phonon effects). This situation corresponds closely to the biexciton cascade realization of our SPS source, where for example in Ref.~\cite{Dusanowski2022May}, $\gamma_B/\gamma_X = 1.92$. The solution is given in full in Appendix~\ref{app:sol}, but in the well-dressed limit $\Omega_{\rm cw}/\gamma_X \gg 1$, the emitted photon number is $\mathcal{N}=1/2$, and the indistinguishability is a very poor $\mathcal{I} = 11/21$, whereas for each sidepeak the emitted photon number is $\mathcal{N}^{\pm}=1/4$ and $\mathcal{I}^{\pm}=2/3$.

In Sec.~\ref{subsec:cav}, we show how a cavity mode can be used to increase the spontaneous emission rate of the $\ket{X}$-$\ket{G}$ transition, improving the SPS figures of merit at the cost of larger emission linewidths relative to the frequency shifts. We assume that the only significant effect of the cavity in the regimes studied is to change the ratio $\gamma_X/\gamma_B$, so the results of this section are also applicable to non-QD systems where the decay rates of each transition may be quite different.

In Sec.~\ref{subsec:ph_eff}, we discuss the efficiency loss that arises from the filtering of the phonon sideband, which can be captured using the polaron transform, and in
%Sec.~\ref{subsec:cwerror} we discuss the cw error rate associated with the far off-resonant driving of the $\ket{X}$-$\ket{G}$ transition, and in 
Sec.~\ref{subsec:g2}, we discuss the role of the pump pulse to initialize the system in the $\ket{X}$ state at time $t= 0$, and what role a pulse with a nonzero duration plays in the HBT $g^{(2)}[0]$.

	Well into the ac Stark and/or AT regime, we have a Hamiltonian which oscillates rapidly compared to the dissipation rates of the system (spontaneous emission and phonon decoherence rates). To simplify our numerical calculations by avoiding having to resolve these rapid oscillations, in Appendix~\ref{app:secular}, we perform a secular approximation by moving into an interaction frame defined by the system Hamiltonian Eq.~\eqref{eq:r2} and dropping these rapidly-oscillating terms. For all numerical calculations presented in this work, we have checked that the secular approximation gives the same results (i.e., visually indistinguishable on any plots) as the full ME presented in the main text as the driving rates are increased into regimes where the secular approximation is expected to asymptotically recover the full solution, thus ensuring the accuracy of our simulations.

Finally, we note that if the effective cw drive Rabi oscillation period $\sim \eta^{-1}$ is not much larger than the excitation pulse width, the QD will experience Rabi oscillations between the $\ket{B}$ and $\ket{X}$ state during the process of the pulse excitation, if in the AT regime, or become off-resonant with the Stark shifted $\ket{X}$ state during the pulse excitation, if in the ac Stark regime. Such a process will highly degrade the inversion efficiency of the pulse, and an initial condition of $\ket{X}$ will no longer be applicable. Better inversion efficiency could perhaps be achieved using different excitation techniques, such as a non-$\pi$ pulse, off-resonant phonon-assisted excitation~\cite{Quilter2015,Weiler2012,Gustin2019,Barth2016Jul}, or adiabatic rapid passage~\cite{Wu2011Feb,Wei2014Oct}; however, we leave a full study of this to future work. For reference, assuming a Gaussian pulse with full width at half maximum in \emph{intensity} of 2 ps, proper inversion is achieved for $\eta/\gamma_X \ll 249$.

	\subsection{Adiabatic elimination under far off-resonant driving}	\label{subsec:adiabatic}
	
	For $|\delta| \gg \Omega_{\rm cw}$,
	the biexciton state remains largely unpopulated as it is driven far off resonance. In this regime, for the case of no phonon coupling, we can approximate the system as a two-level system under radiative decay via adiabatic elimination of the biexciton state. We consider the dynamics of the $\sigma_{B}^-$ operator under the Heisenberg-Langevin equation (neglecting noise fluctuation terms):
\begin{equation}\label{eq:hl}
\dot{\sigma}_{B}^- = -\left(\frac{1}{2}(\gamma_X+\gamma_{B})+ i \delta\right) \sigma_{B}^- + i\frac{\Omega_{\rm cw}}{2}\sigma_z^{B}.
\end{equation}

As the dynamics of $\sigma_{B}^-$ are fast oscillating, we approximate $\dot{\sigma}_{B}^- \approx 0$, and solve for $\sigma_{B}^-$:
\begin{equation}
\sigma_{B}^- \approx \mathcal{A}\sigma_z^{B},
\end{equation}
where 
\begin{align}\label{eq:a0}
    \mathcal{A} &= \frac{i\Omega_{\rm cw}}{\gamma_X+\gamma_{B}+ 2i\delta} \nonumber \\ &\approx \frac{\Omega_{\rm cw}}{2\delta},
\end{align}
and the approximation in the second line is justified on the grounds that for the adiabatic elimination procedure to be valid the detuning should greatly exceed the linewidths.
We note that in the ac Stark regime, $\mathcal{A} \ll 1$, and thus from Eq.~\eqref{eq:a0} we can also find $\sigma^+_B\sigma^-_B \approx \mathcal{A}^2\sigma^+_X\sigma^-_X + \mathcal{O}(\mathcal{A}^4)$. 

Substituting this result into the ME and expanding to second order in $\mathcal{A}$, we find that the dynamics for a QD initialized in the $\ket{X}$ state can be described with a simple ME for an effective two-level system,
\begin{align}\label{eq:mead}
    \dot{\rho} = i[\Delta_{\rm ac}\sigma^+_X\sigma^-_X,\rho]&+ \frac{\gamma_X}{2}\mathcal{L}[\sigma^-_X]\rho + \mathcal{A}^2\frac{\gamma_{B}}{4}\mathcal{L}[\sigma^+_X\sigma^-_X]\rho \nonumber \\ & + \mathcal{A}^2\frac{\gamma_B}{4}\mathcal{L}
    \left [\ket{Y}\bra{X}\right]\rho,
\end{align}
where here we have used that to second order in $\mathcal{A}$, $\Delta_{\rm ac} \approx \delta \mathcal{A}^2$. Equation \eqref{eq:mead}, when considering the radiative decay of the $X$ exciton, is an ME for spontaneous emission with effective decay rate $\gamma_X + \mathcal{A}^2\gamma_B/2 $ and pure dephasing with rate $ \gamma_{\rm eff}' = \mathcal{A}^2\gamma_{B}/2 \approx \frac{\gamma_B}{2}\frac{\Delta_{\rm ac}}{\delta}$. Note that this result can be derived as well in a more rigorous manner by using the effective operator formalism, which utilizes a Feschbach projection and perturbation theory to separate slow and fast subspaces~\cite{Reiter2012Mar}. The single-photon indistinguishability can be easily calculated for this system: again with accuracy up to order $\mathcal{A}^2$,
\begin{align}
    \mathcal{I} &= \frac{1}{1+\frac{\gamma_{B}}{2\gamma_X}\mathcal{A}^2}.
\end{align}

Also to order $\mathcal{A}^2$, we have $\mathcal{N} \approx \mathcal{I}$.
For $\gamma_{B} = 2\gamma_X$, we have  $\mathcal{I} = 4\delta^2/(4\delta^2+\Omega_{\rm cw}^2)$. We can express this in terms of the ac Stark shift $\Delta_{\rm ac}$, so that
\begin{align}\label{eq:Ixg}
    \mathcal{I} &= \frac{\delta^2}{\delta^2+\Delta^2_{\rm ac} + \delta \Delta_{\rm ac}} 
    \nonumber \\ & 
    \approx \frac{\delta}{\delta+\Delta_{\rm ac}} %\nonumber \\ &
    = \frac{\Omega_{\rm cw}^2}{\Omega_{\rm cw}^2  +4 \Delta_{\rm ac}^2}.
\end{align}
In the absence of sources of additional dephasing or decoherence, Eq.~\eqref{eq:Ixg} gives an estimate of the highest achievable indistinguishability for a given $\Delta_{\rm ac}$ in terms of the maximum detuning $\delta$ that can be introduced without exciting other unwanted energy levels, with $\Omega_{\rm cw} \approx 2\sqrt{\delta \Delta_{\rm ac}}$, or in terms of the maximum drive strength without introducing additional decoherence.
%In Fig. 4b of the main text, this solution is compared with the full numerical solution, showing excellent agreement in the large detuning regime.

\subsection{Numerical results for SPS efficiency and indistinguishability}\label{subsec:num}

In this subsection, we present the results of our numerical (and analytical) solutions of the ME for the single-photon emitted photon number and indistinguishability, without yet considering any cavity coupling (for $\gamma_X = \gamma_B/2$). For all plots here and in subsequent subsections, unless otherwise stated, we show results without any phonon coupling as solid lines, phonon parameter set \RNum{1} as dashed lines, and phonon parameter set \RNum{2} as dashed-dotted lines. Red (lower energy) sidepeaks are shown in red, blue (higher energy) sidepeaks are shown in blue, and the total spectrum results are shown in black. To calculate the two-time correlation functions that appear in the definition of the indistinguishability and HBT $g^{(2)}[0]$, we use the quantum regression theorem~\cite{carmichael}.

	\begin{figure}
		\centering
		\includegraphics[width=1\linewidth]{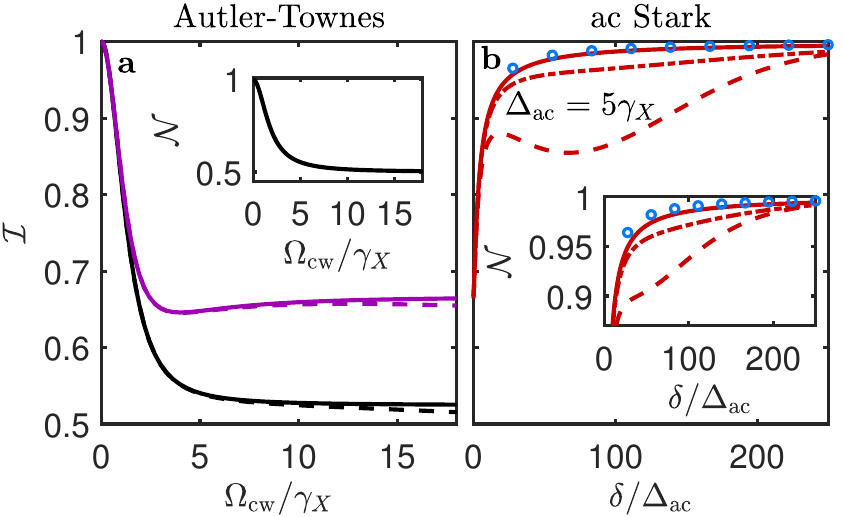}
		\caption{\label{fig:R1}
		Basic operation figures of merit for SPS indistinguishability and (insets) emitted photon number for (a) the AT regime with $\delta =0$, and (b) the ac Stark regime with $\Delta_{\rm ac} = 5\gamma_X$.  The purple lines are used to indicate that the red and blue sidepeaks have visually indistinguishable figures of merit in the regimes plotted. In (a), the curves corresponding to the case of no phonon coupling and phonon parameter set \RNum{2} are visually indistinguishable. In (b), the analytical solution under adiabatic elimination as given in Sec.~\ref{subsec:adiabatic} is shown as blue circles.
		}
	\end{figure}

In Fig.~\ref{fig:R1}(a), we show the analytical solution of Appendix~\ref{app:sol}, as well as the numerical solution with phonon coupling in the AT regime with $\delta=0$ for the SPS indistinguishability and emitted photon number. As the AT regime requires weaker drive strengths to achieve equivalent energy splittings as compared with the ac Stark regime, the influence of phonon scattering is quite weak here, only being perceptible for the phonon parameter set \RNum{1}. 

In Fig.~\ref{fig:R1}(b), we show the numerical solutions to the ME for the indistinguishability and emitted phonon number for a device operating the ac Stark regime with a fixed frequency shift of $\Delta_{\rm ac} = 5\gamma_X$, as well as the approximate solution derived under adiabatic elimination of the higher energy state $\ket{B}$ in Sec.~\ref{subsec:adiabatic}. Note that we only present results here for the red (lower-energy) sidepeak, but well into the ac Stark regime as $\delta/\Delta_{\rm ac} \gg 1$, the spectral weight of the other sidepeak rapidly goes to zero, and the red sidepeak becomes nearly equal to the total spectrum, as we show explicitly later.

As $\delta/\Delta_{\rm ac} \gg 1$, the full numerical solution without phonon coupling asymptotically approaches the approximation solution under adiabatic elimination. For phonon parameter set \RNum{1}, the effect of phonon coupling initially increases with increasing detuning, before ultimately decreasing again at very high detunings.

To understand this observation, it is useful to refer to the approximate simplification of the PME which is presented in Appendices~\ref{app:secular} and~\ref{app:weak} as the weak phonon coupling ME under the secular approximation. While we use the full PME (with the secular approximation) for all numerical calculations, the expressions derived in Appendix~\ref{app:weak} allow for insight into the physics of the phonon interaction and its effect on the source figures of merit. We show that for the phonon parameters studied in this work, the dominant effect of the phonon interaction is to induce transitions from the $\ket{+}$ to the $\ket{-}$ state with rate $\tilde{\Gamma}'_0\left[n_{\rm ph}(\eta,T) + 1\right]$, which corresponds to a phonon creation process, as well as transitions from the $\ket{-}$ to the $\ket{+}$ state with rate $\tilde{\Gamma}'_0n_{\rm ph}(\eta,T)$, which is a phonon absorption process, where $\tilde{\Gamma}'_0$ is given by Eq.~\eqref{eq:gamma0}, and $n_{\rm ph}(\omega,T) = [e^{\omega/(k_B T)} - 1]^{-1}$ is the thermal phonon occupation number. A schematic of these processes in the dressed state frame is shown in Fig.~\ref{fig:AppSchem}. For $\eta \ll k_B T$\footnote{Note we can compare $\eta$ (a frequency, or rate) with energy here ($k_BT$) as we are using units with $\hbar=1$.}, the phonon creation and absorption processes occur at similar rates, and the phonon coupling leads to an incoherent dephasing-like effect which scales with $\sim \eta^2$. At higher effective drive strengths $\eta \gg k_B T$, only phonon creation becomes probable, and the $\ket{+}$ to $\ket{-}$ transition becomes driven with rate proportional to $\sim \eta^3$.

In light of this, the initial increase in the role of phonons can be understood as a consequence of the concurrent increase in the drive strength $\Omega_{\rm cw} \approx 2\sqrt{\Delta_{\rm ac}\delta}$, which, for small $\eta$ relative to $\omega_b$ and $k_B  T$ (for $T = 4$ K, $k_B T = 345 \ \mu \text{eV} = 178\gamma_X$), leads to a roughly linear increase in the phonon decoherence rates as given by Eq.~\eqref{eq:pme} as a function of $\delta$ (see Appendix~\ref{app:weak}) when holding $\Delta_{\rm ac}$ fixed. This manifests in an increased population of the higher lying biexciton state, which reduces the efficiency (as seen in the inset), and the indistinguishability via timing jitter. In fact, for phonon parameter set \RNum{1}, this increased decoherence is sufficient to outweigh the increased indistinguishability afforded by moving further into the ac Stark regime by increasing the detuning, leading to \emph{non-monotonic} behavior of the indistinguishability as a function of $\delta/\Delta_{\rm ac}$, and an initial local maximum of the indistinguishability as the detuning is increased. In addition, this behavior  allows us to deduce that excitation-induced--dephasing also reduces the coherence of the emitted photons, as over the range of detunings where the indistinguishability decreases, the emitted photon number continues to increase, suggesting that the reduction of coherence can not be entirely attributed to timing jitter. 

Moving into even higher detuning regimes, we see
the indistinguishability and efficiency increase
again for phonon parameter set \RNum{1}, which is due to the phonon absorption process which takes states from $\ket{-}$ to $\ket{+}$ becoming improbable as $\eta \gg k_B T$, as the number of phonons in the bath with the required energy becomes small. 
It is worth noting that this effect
is only present for rather large detunings (of
order $\sim$meV) and drive strengths
($\sim$hundreds of $\mu$eV), where the $\pi$-pulse inversion is likely very inefficient.

			\begin{figure}
		\centering
		\includegraphics[width=1\linewidth]{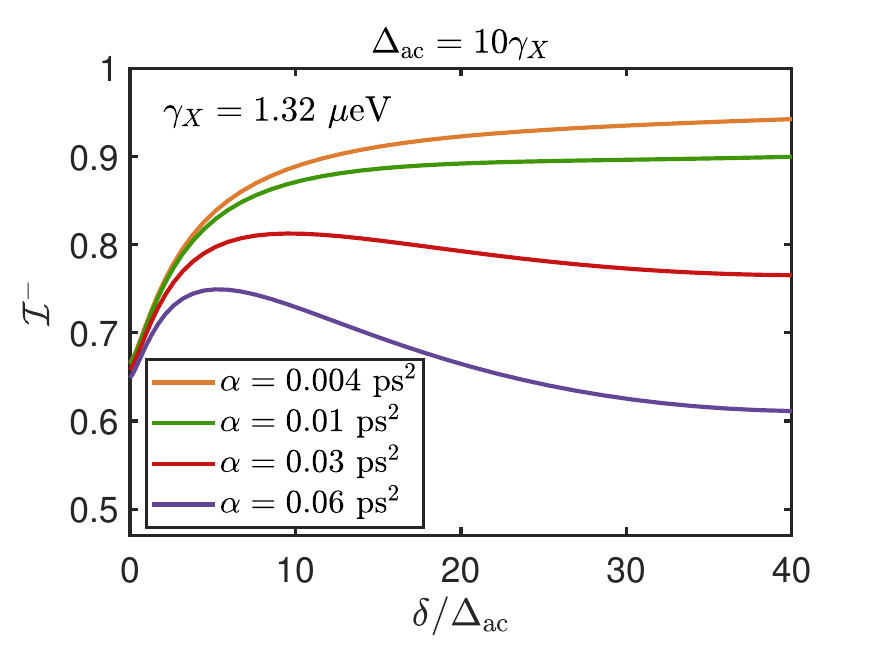}
		\caption{\label{fig:R5}
			SPS indistinguishability for the dominant red sidepeak as a function of detuning for frequency shift $\Delta_{\rm ac} = 10\gamma_X$, and for varying values of the phonon coupling strength $\alpha$, at fixed $\omega_{b} =2$ meV (intermediate value between $\omega_{b,\RNum{1}}$ and $\omega_{b,\RNum{2}}$), showing the formation of a local maximum for sufficiently large $\alpha$.
		}
	\end{figure}

	In Fig.~\ref{fig:R5}, we show the formation of the local maximum for sufficiently large phonon coupling strengths by plotting the indistinguishability of the dominant red sidepeak as a function of detuning for a fixed frequency shift $\Delta_{\rm ac}$. 

	\begin{figure}
		\centering
		\includegraphics[width=1\linewidth]{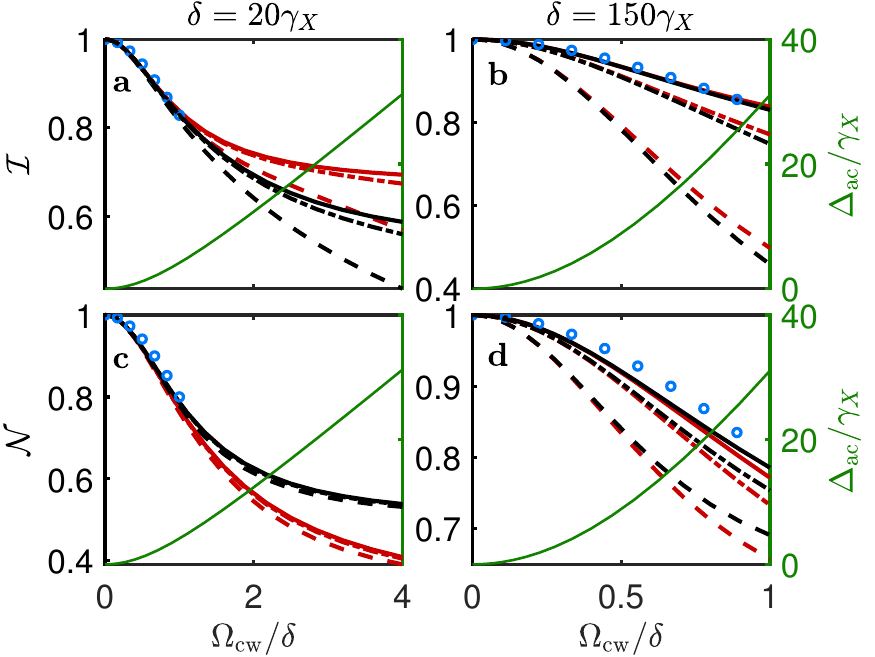}
		\caption{\label{fig:R2}
			SPS indistinguishability (a,b) and emitted photon number (c,d) with fixed detuning $\delta = 20\gamma_X$ (a,c) and $\delta = 150 \gamma_X$ (b,d) as a function of drive strength $\Omega_{\rm cw}/\delta$, showing the transition from ac Stark regime to AT regime. On the right green axis, the frequency shift of the dominant peak $\Delta_{\rm ac}$ is shown. The analytical solution under adiabatic elimination as given in Sec.~\ref{subsec:adiabatic} is shown as blue circles.
		}
	\end{figure}

While holding the ac Stark shift $\Delta_{\rm ac}$ fixed and varying the detuning and drive strengths to achieve this shift (as we have done in Fig.~\ref{fig:R1}(b) and Fig~\ref{fig:R5}) is useful from a theoretical perspective to reveal the achievable figures of merit for a given Stark shift, as well as to show the transition from the AT regime at $\delta =0$ to the ac Stark regime as $\delta/\Delta_{\rm ac} \gg 1$, it is less suited to what would be directly observed in an experiment, as the drive strength $\Omega_{\rm cw}$ must also be varied simultaneously with the detuning to achieve a constant Stark shift. Thus, in Fig.~\ref{fig:R2} we plot the indistinguishability and emitted photon numbers for a \emph{fixed} detuning, and instead vary the drive strength $\Omega_{\rm cw}$. Here, we can see that for a fixed detuning, the figures of merit vary inversely to the effective frequency shift; as we increase the drive strength, we move further away from the ac Stark regime with weak frequency shifts and good figures of merit (due to minimal excitation of higher energy states), towards the AT regime where $\delta/\Delta_{\rm ac}$ is small, and the energy splittings are larger, but with increased decoherence mostly due to increased timing jitter.

\begin{figure}
		\centering
		\includegraphics[width=1\linewidth]{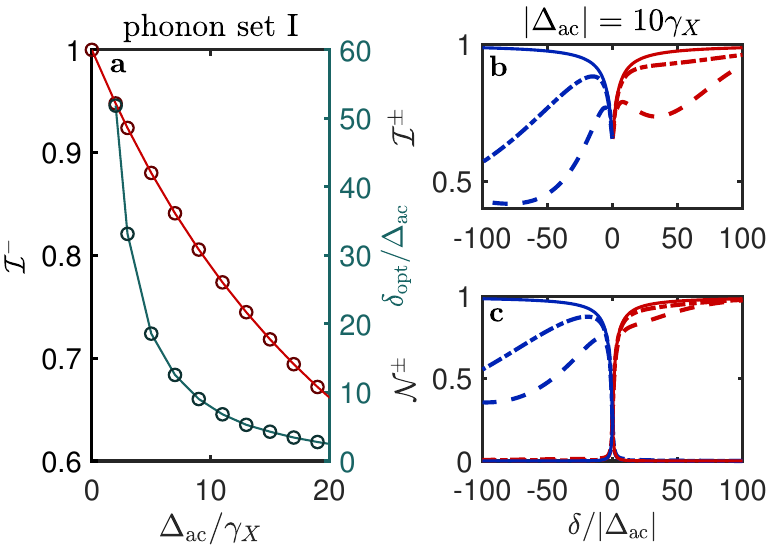}
		\caption{\label{fig:R3}
			(a) Indistinguishability  (red line and circles) of red sidepeak photon as a function of frequency shift $\Delta_{\rm ac}/\gamma_X$ at the detuning $\delta_{\rm opt}/\Delta_{\rm ac}$ (blue line and circles) of local maximum for phonon parameter set \RNum{1}. (b) indistinguishability and (c) emitted photon number for dominant sidepeak as a function of detuning for a frequency shift of $|\Delta_{\rm ac}| = 10\gamma_X$, revealing the asymmetry in figures of merit associated with the phonon bath at low temperatures.
		}
	\end{figure}

In Fig.~\ref{fig:R3}, we show 
more details on some aspects associated with the phonon bath interaction
for  a QD SPS. As visible in Fig.~\ref{fig:R1}(b), for phonon parameter set \RNum{1}, there occurs a local maximum of the indistinguishability as a function of the detuning in the ac Stark regime, for a given Stark shift $\Delta_{\rm ac}$. This local maximum occurs for modest drive strengths and detunings, and as such is important to consider in light of practical experimental considerations. In Fig.~\ref{fig:R3}(a), we plot the indistinguishability of the red sidepeak as a function of shift $\Delta_{\rm ac}$, where for each value of $\Delta_{\rm ac}$ we sweep the detuning $\delta$ to find the value $\delta_{\rm opt}/{\Delta_{\rm ac}}$ where this local maximum occurs, and plot this as well as the corresponding indistinguishability at this detuning value. We can clearly see here that for phonon parameter set \RNum{1}, the maximum achievable indistinguishablity drops off rapidly as the splitting is increased. By frequency shifts of $\Delta_{\rm ac} \sim 20 \gamma_X$, the optimal indistinguishability occurs rather close to the AT regime, with small detunings $\delta/\Delta_{\rm ac}$ and indistinguishabilities not much higher than the AT regime value of (without phonons) $\mathcal{I}^{-} = 2/3$.

In Fig.~\ref{fig:R3}(b,c), we compare the performance of the device for positive (red) and negative (blue) dominant peak frequency shifts by plotting the source figures of merit as a function of detuning $\delta/|\Delta_{\rm ac}|$ for both positive and negative detunings. In the absence of phonon coupling (and, as with all of the calculations in this subsection, neglecting the far off resonant $\ket{X}$-$\ket{G}$ drive term in the Hamiltonian), as expected, the device performance is perfectly symmetric with respect to the sign of the detuning. 

Upon introducing phonon coupling, however, the device operation becomes highly asymmetric with respect to the detuning (and thus the sign of the dominant peak frequency shift $\Delta_{\rm ac}$). In all cases, the figures of merit for the SPS are better for positive (red) frequency shifts $\Delta_{\rm ac}$ in the ac Stark regime. This is because at low temperatures, there are few phonons present in the phonon bath (i.e., in the sense of a thermal distribution of bosons); for negative detunings, the laser frequency is larger than the transition frequency between the $X$ exciton and biexciton states, and a resonant process can occur where the difference in energy between the laser and transition can be absorbed by the creation of a real phonon in the bath with this energy, leading to phonon-assisted absorption. 
The corresponding process, where for positive detunings the energy difference is made up for by annihilation of a phonon with energy near equal to the energy gap (Fig.~\ref{fig:schematic}(d)), is more strongly suppressed due to the small number of phonons present in the equilibrium bath with this energy range at $T = 4$ K. Thus, for positive detunings (red energy shifts), the population of the undesired higher energy biexciton state is less, and as such, the efficiency and timing jitter is reduced. 

In terms of the simplified model presented in Appendix~\ref{app:weak}, the process associated with phonon emission becomes more dominant as $\eta$ is increased, which leads to increased transitions from the $\ket{+}$ state to the $\ket{-}$ state; for positive $\delta$, $\ket{-}$ is more $\ket{X}$ like, whereas for negative detunings, it is more $\ket{B}$ like. For this reason we mostly focus on positive detunings and frequency shifts throughout this work. As we show later in Appendix~\ref{app:cw}, the usual case of positive biexciton binding energies can also lead one to favor positive detunings and frequency shifts.

Note we also see in Fig.~\ref{fig:R3}(c), that the non-dominant peak weight (emitted photon number) goes very rapidly to zero as the detuning is increased, indicating that the device is operating in the ac Stark regime.

		\begin{figure}
		\centering
		\includegraphics[width=1\linewidth]{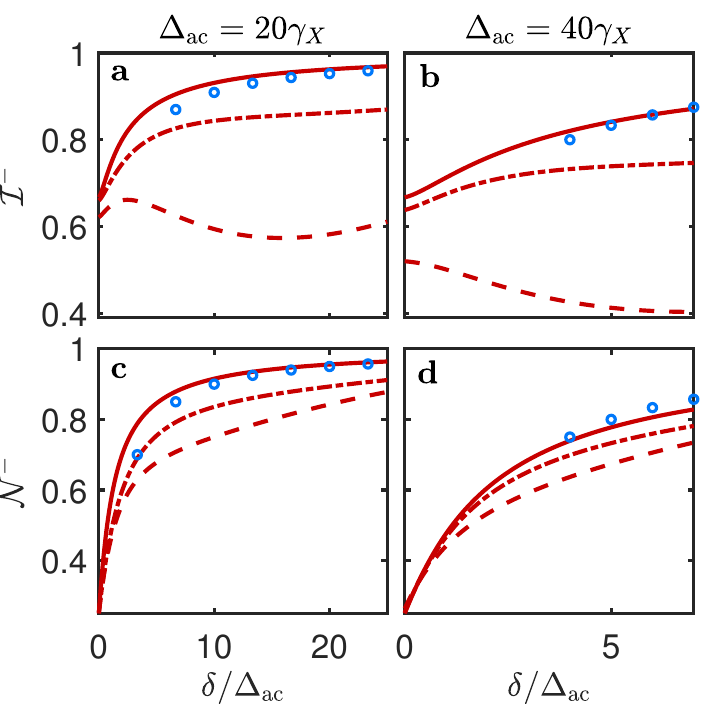}
		\caption{\label{fig:R4}
			SPS indistinguishability (a,b) and emitted photon number (c,d) in the ac Stark regime for (a,c) $\Delta_{\rm ac} = 20\gamma_X$ and (b,d) $\Delta_{\rm ac} = 40\gamma_X$.  The analytical solution under adiabatic elimination as given in Sec.~\ref{subsec:adiabatic} is shown as blue circles.
		}
	\end{figure}

In Fig.~\ref{fig:R4}, we plot the indistinguishability and emitted photon for the dominant red sidepeak as a function of detuning, as in Fig.~\ref{fig:R1}(b), but for larger frequency shifts of $\Delta_{\rm ac} = 20\gamma_X$ and $\Delta_{\rm ac} = 40 \gamma_X$. Here the plots span a similar range of drive strengths $\Omega_{\rm cw}$, showing how larger drive strengths are required to reach the ac Stark regime for larger frequency shifts, which thus increases phonon related decoherence.

\subsection{Use of a cavity to improve device performance via the Purcell effect}\label{subsec:cav}

In our results thus far, we have assumed that the higher lying state $\ket{B}$ has a spontaneous emission rate $\gamma_B$ which is twice that of the desired state $\ket{X}$ such that $\gamma_B = 2\gamma_X$, which is close to the case in QDs where $\ket{B}$ corresponds to the biexciton. However, it is the simultaneous dipole radiation of the higher energy state $\ket{B}$ and the desired state $\ket{X}$ which leads to timing jitter and reduced coherence of the spontaneously emitted (and frequency shifted) photons. Thus, it is intuitively reasonable that should we increase the radiation rate $\gamma_X$ relative to the decay rate $\gamma_B$, we should expect to see improved figures of merit for both the indistinguishability (for reasons of timing jitter mentioned above) and emitted photon number (as emission into the $X$-polarized decay channel becomes accelerated relative to the $Y$-polarized one). 

One way to increase the effective radiation rate $\gamma_X$ relative to $\gamma_B$ is to use the Purcell effect afforded by a cavity with an enhanced density of optical states (near)-resonant with the $\ket{X}$-$\ket{G}$ transition. To be concrete, we can consider a single-mode cavity with bosonic operators $[a,a^{\dagger}]=1$, QD-cavity coupling rate $g$, which couples to the $\ket{X}$ exciton with a Hamiltonian term 
\begin{equation}\label{eq:cav}
    H_{\rm cav}=g(a\sigma^+_X + a^{\dagger}\sigma^-_X),
\end{equation}
and has photon decay rate (spectral full width at half maximum) $\kappa$, which should satisfy $\kappa \ll E_B$ to ensure the biexciton transition is not also broadened. Then, in the bad cavity (weak coupling) limit that $g/\kappa \ll 1$, the cavity mode can be adiabatically eliminated, and the result is that the spontaneous emission rate $\gamma_X$ is increased by a factor $\gamma_X \rightarrow (1+F_P)\gamma_X$, where 
\begin{equation}\label{eq:purcell}
F_P = \frac{4g^2}{\kappa\gamma_0},
\end{equation}
and $\gamma_0$ is the bare $\gamma_X$ before any cavity enhancement.

In principle, to determine the quantitative influence of incorporating a cavity mode on the SPS figures of merit, the Hamiltonian $H_{\rm cav}$ in Eq.~\eqref{eq:cav} should be included in the system Hamiltonian $H_S$, and the PME should be modified to reflect this change (as in, e.g., Ref.'s~\cite{roy11,hargart16,Gustin2018,Gustin2019}). Output observables of the system should then be calculated in terms of the cavity operators $a$ and $a^{\dagger}$, as input-output theory tells one that the scattered fields of the reservoir in the Heisenberg picture (which are ultimately detected) differ from their input by $\sqrt{\kappa}a(t)$~\cite{gardiner_quantum_2004,Franke2020Sep}. This approach leads to a correct description of some of the subtleties involved with cavity coupling, including filtering of the output spectrum which occurs due to finite cavity width $\kappa$, the emitted photon numbers in each respective channel (i.e., the cavity mode collects a factor $F_P$ greater photons than background emission channels, in the weak-coupling limit), and cavity-induced dephasing. 

For simplicity, however, we shall for the results in this section assume that the influence of the cavity is solely to increase the exciton decay rate $\gamma_X$. This approach has, for one, the advantage of generality, as it can apply to any (e.g., atomic) system with decay rates which do not satisfy $\gamma_B = 2\gamma_X$, with or without any cavity coupling. Furthermore, in the weak coupling limit $g \ll \kappa$, which is the ideal regime of operation for SPSs~\cite{Gustin2018}, the phonon decoherence rates associated with the cavity-QD interaction become insignificant~\cite{ilessmith17,Gustin2018}, and the dominant effect of the cavity on the QD dynamics is enhanced spontaneous emission. Thus, we simply use a variable $\gamma_X$ in our simulations with the model of Sec.~\ref{sec:model}, with the understanding that Eq.~\eqref{eq:purcell} can be used to get an estimate of the cavity parameters required to observe the corresponding figures of merit as a function of $\gamma_X$. Neglecting the background spontaneous emission channels, we can thus let $F_P \approx \gamma_X/\gamma_0$, where $\gamma_0 = 1.32 \ \mu$eV, and $F_P$ is given by Eq.~\eqref{eq:purcell}.
	
				\begin{figure}
		\centering
		\includegraphics[width=1\linewidth]{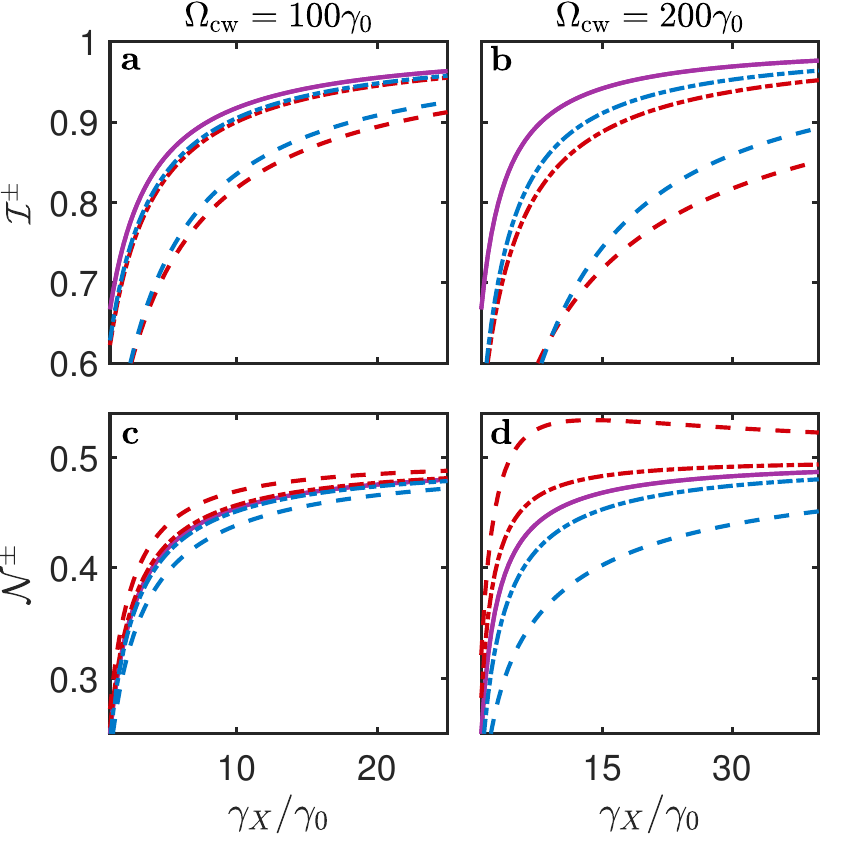}
		\caption{\label{fig:R6}
			figures of merit for SPS operating in AT regime with $\delta=0$ as a function of the variable decay rate $\gamma_X/\gamma_0$, with $\gamma_0 = 1.32 \ \mu$eV. The purple lines are used to indicate that the red and blue sidepeaks have the same figures of merit in the absence of phonons. 
		}
	\end{figure}
	
	In Fig.~\ref{fig:R6}, we plot the figures of merit of the SPS operating in the AT regime with $\delta = 0$ as a function of the variable decay rate $\gamma_X$. We see in the case of QD SPSs that, provided the enhancement is given by a cavity with a sufficiently broad linewidth to capture the frequency splittings and operate in the weak-coupling regime, that the source can emit with near $1/2$ efficiency (the best case scenario in the AT regime) and high ($>90-95\%$) indistinguishability for AT splittings on the order of hundreds of $\mu$eV (but with highly broadened linewidths).
	
					\begin{figure}
		\centering
		\includegraphics[width=1\linewidth]{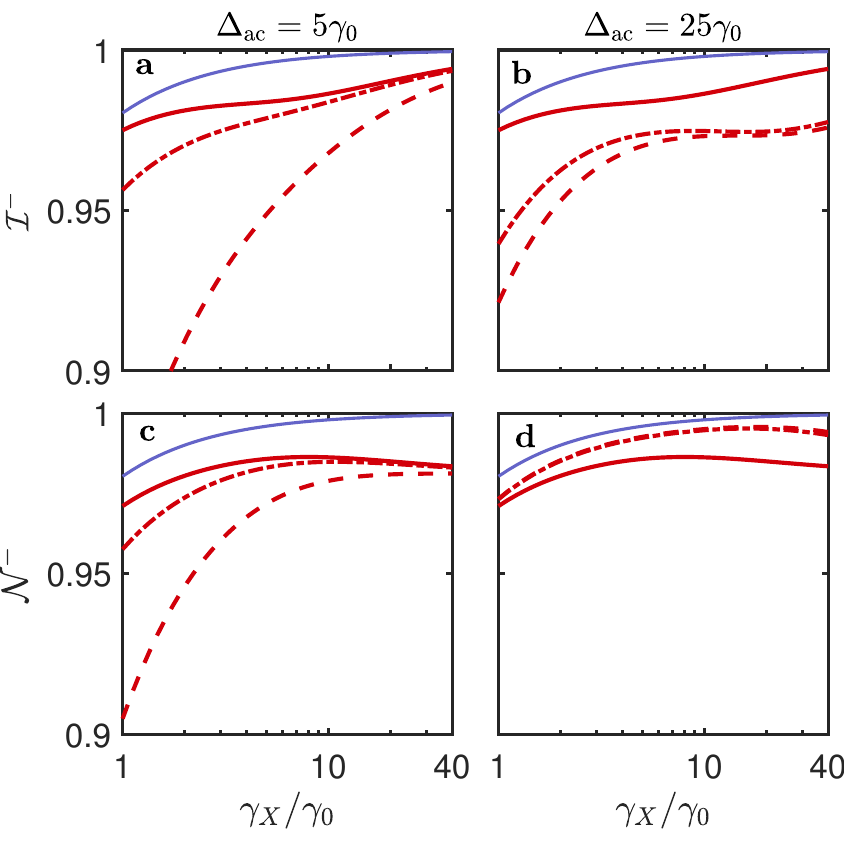}
		\caption{\label{fig:R7}
			figures of merit for a SPS operating in the ac Stark regime with fixed detuning $\delta_{\rm ac} = 50 \Delta_{\rm ac}$ for (a,c) Stark shift $\Delta_{\rm ac} = 5 \gamma_0$ and (b,d) $\Delta_{\rm ac} = 25 \gamma_0$. The analytical solution under adiabatic elimination as given in Sec.~\ref{subsec:adiabatic} is shown as blue lines.
		}
	\end{figure}

	In Fig.~\ref{fig:R7}, we explore the ac Stark regime of operation with a variable decay rate $\gamma_X/\gamma_0$. Here, we see that as the decay rate is increased, the emitted photon number increases as losses due to $Y$-polarized emission channels are decreased, up to a point at which the emitted photon number begins to decrease again; this decline can be attributed to the laser detuning becoming smaller relative to the effective decay rate, which moves the operation regime towards the AT regime, and thus slightly increases the population of the non-dominant peak at the cost of the dominant peak.
	Nonetheless, for large emission rates $\gamma_X/\gamma_0$, the emission exceeds the phonon scattering rates and the figures of merit remain quite high; for the case of QDs, we see that near-unity indistinguishability ($>97\%$ for both phonon cases a $\Delta_{\rm ac}  = 25\gamma_0$) for frequency shifts on the order of tens of $\mu$eV. 
	
	Interestingly, in Fig.~\ref{fig:R7}(d), the emitted photon number $\mathcal{N}^-$ actually becomes larger with phonon coupling. This can be understood in the context of the weak phonon coupling approximation of Appendix~\ref{app:weak}.  In particular, the ratio of the rate that takes states from $\ket{+}$ to $\ket{-}$ to the rate that takes states from $\ket{-}$ to $\ket{+}$ is simply $\exp{[\eta/k_B T]}$--- a direct consequence of the thermal occupation distribution of phonons in the bath. Thus, for $\eta \gg k_B T $ (in Fig.~\ref{fig:R7}(d), and positive detunings $\delta >0$, we expect phonons to \emph{increase} the proportion of photons emitted into the dominant peak. Indeed, for Fig.~\ref{fig:R7}(d), we have $\eta/k_B T \approx 5$. In contrast, Fig.~\ref{fig:R7}(c) has $\eta/k_B T \approx 1$, and this effect is not seen. The fact that the indistinguishability in Fig.~\ref{fig:R7}(b) is not also improved relative to the no-phonon case is indicative that the excitation-induced dephasing of the $\ket{X}$-$\ket{G}$ transition dominates over the reduced timing jitter afforded by this phonon-induced transition. 
	
	We stress that the numbers presented here can be achieved with cavity and QD parameters which have been demonstrated in the literature with semiconductor microcavities~\cite{somaschi16,Gazzano2013Feb,Wang2019Nov,Ding2016Jan,Reitzenstein2010Jan,Kolatschek2021Sep,Najer2019Nov,Liu2018Sep,Hepp2020Dec,Dusanowski2020Jul,Moczala-Dusanowska2019Jun,Madsen2014Oct};
the range of Purcell factors shown here (less than $40$; cf.~a value achieved with a photonic crystal cavity of 43~\cite{Liu2018Sep}) can be achieved with (e.g., using dielectric micropillar resonators~\cite{somaschi16,Gazzano2013Feb,Wang2019Nov,Ding2016Jan,Reitzenstein2010Jan,Wang2020Sep}) a linewidth $\kappa$ of a few hundred $\mu$eV (corresponding to $Q$ factors $\sim 10^3\!-\!10^4$), and a coupling $g$ on the order of (at most) tens of $\mu$eV, and the phonon parameter sets we use reflect measured values as discussed in Sec.~\ref{sec:model}.
	
	\subsection{Efficiency loss due to phonon sideband coupling}\label{subsec:ph_eff}
	As mentioned in Sec.~\ref{sec:model}, the presence of the non-Markovian broad phonon sideband due to scattering with LA phonons leads to much lower photon indistinguishability if it is not filtered out of the detected spectrum (retaining only the ZPL---in this case the frequency shifted peak(s)). Since we are assuming in this work that the sideband is removed by frequency filtering after emission, we would like to quantify this efficiency cut in terms of the phonon coupling parameter sets \RNum{1} and \RNum{2} we use for the presented simulations.
	
	Without any cavity coupling (i.e., assuming a post-emission filtered ZPL), the fraction of total photons emitted that remain unfiltered is $\eta_{\rm eff} = \langle B \rangle^2$, whereas with efficient cavity filtering, it is $\eta_{{\rm eff}, \rm{cav}} = \langle B \rangle^2 F_P/(1+\langle B \rangle^2F_P)$~\cite{ilessmith17}. In Table~\ref{tab:numbers}, we show these filter efficiencies for phonon parameter sets \RNum{1}, \RNum{2}, and \RNum{3} at $T = 4$ K. We can also find low-temperature analytical expressions for these efficiencies by noting that
	\begin{align}
	    \langle B \rangle ^2 &= \exp{\left[-\alpha\int_0^{\infty}\text{d}\omega \omega(1+2n_{\rm ph}(\omega,T))e^{-\frac{\omega^2}{2\omega_b^2}}\right]} \nonumber \\ 
	    & = \exp{\left[-\alpha\omega_b^2\left(1+\frac{1}{3}\tilde{T}^2 - \frac{1}{15}\tilde{T}^4 + \mathcal{O}(\tilde{T}^6)\right)\right]},
	\end{align}
	where $\tilde{T} = \pi k_B T/\omega_b$, and in the second line we have, for the term containing $n_{\rm ph}(\omega,T)$,  expanded the exponential cutoff as a power series and evaluated the resulting Bose-Einstein integrals\footnote{That is, using the relation $\int_0^{\infty} \text{d} x \frac{x^{n-1}}{e^x-1} = \Gamma(n)\zeta(n)$, where $\zeta(n)$ is the Riemann zeta function.}; at $T = 4$ K, retaining only terms up to order $\tilde{T}^4$ is an excellent approximation for $\omega_b \gtrsim 1.5$ meV, and qualitatively accurate for $\omega_{b,\RNum{1}} = 0.9$ meV as well.
	
	Note that the efficiency cut has \emph{not} already been taken into account in our simulations for emitted photon numbers $\mathcal{N}^{(\pm)}$, and the efficiency given here must also ultimately be considered to yield the total SPS efficiency (in addition to other experimental considerations, e.g., output fiber coupling efficiencies, etc.)

	\begin{table*}[htp]
    \centering
    \begin{tabular}{l | l l l l l}
 ~ & $\alpha$ ($\text{ps}^2$) & $\omega_b$ (meV) & $\langle B \rangle$ & $\eta_{\rm eff}$ (\%) & $\eta_{{\rm eff},\rm{cav}}$ (\%) \\ \hline phonon parameter set \RNum{1} & 0.04 & 0.9 & 0.949 & 90.1 & 90.0 \\ phonon parameter set \RNum{2} & 0.006 & 5.5 & 0.809 & 65.4 & 86.7 \\ phonon parameter set \RNum{3} & 0.025 & 2.5 & 0.826 & 68.2 & 87.2
    \end{tabular}
    \caption{Phonon parameters and efficiency cut due to phonon sideband post-emission spectral filtering or cavity filtering for $F_P = 10$. }
    \label{tab:numbers}
\end{table*}

	\subsection{Single photon purity $g^{(2)}[0]$ associated with pulse excitation}\label{subsec:g2}
In our simulations presented in Sec.'s~\ref{subsec:num} and~\ref{subsec:cav}, we have simply assumed the SPS to be initialized in the $\ket{X}$ state, and we have not explicitly modelled the pulse excitation. As we have also neglected the $\sigma_x^X$ term in $H_S$ that can give rise to (far off-resonant)  excitation of the $\ket{X}$-$\ket{G}$ transition, we have not included in our model any mechanism for more than one photon to be emitted from the transition of interest to our SPS. Formally, then, all of the above simulations are for $g^{(2)}[0] =0$. 

To improve on this approximation, we include the pulse excitation directly in the system Hamiltonian, and use a time-dependent PME which incorporates the pulse-induced phonon scattering~\cite{ross16, Gustin2019,Gustin2018}, described in Appendix~\ref{app:pulse}. However, the re-excitation probability (and corresponding two-photon emission probability) is not strongly affected by the dressing field so long as $\eta^{-1}$ is much larger than the pulse width in time, as the pulse occurs much quicker than the period of Rabi oscillations between the $\ket{X}$ and $\ket{B}$ states. Thus, the statistics of two-photon emission are very closely related to results known for the simple two-level system pulse excitation~\cite{Gustin2018, Hanschke2018,Fischer2018,Dusanowski2022May}. The primary modification to the $g^{(2)}[0]$ comes from the fact that in a two-photon emission event, the first photon will, for $\eta^{-1}$ much larger than the pulse duration, be emitted in the desired $\ket{X}$-$\ket{G}$ transition. The sequential photon, however, is emitted with probability $\mathcal{N}$ (i.e., the efficiency). 

Specifically, consider a Gaussian pulse with full width at half maximum in \emph{intensity} $\tau_{p}$, and area in amplitude $\pi$ (i.e., after the coherent attenuation factor from phonons $\langle B \rangle$ is applied, as with the rest of this paper). Then, for a short pulse that satisfies $\eta \tau_p \ll 1$ and $\gamma_X \tau_p \ll 1$, the results of Fischer~\emph{et al.}~\cite{Fischer2018} give  $g^{(2)}[0] \approx \eta_G \gamma_X \tau_p/\mathcal{N}^2$ for the case of no cw dressing laser, where $\eta_G = [2\text{ln}(2)]^{-1/2}\int_0^{\infty}\text{d}x \cos^2{[\pi \text{erf}(x)/2]} \approx 0.4376$ is a factor associated with the Gaussian pulse. For the reasons outlined previously, with cw dressing, this value is reduced by a factor of $\mathcal{N}$ such that \begin{equation}\label{eq:g2_al}
g^{(2)}[0] \approx \eta_G \frac{\gamma_X \tau_p}{\mathcal{N}}.
\end{equation}
Next, using Eq.~\eqref{eq:Vraw2}, we find
\begin{equation}\label{eq:V_a}
    \mathcal{V}_{\rm raw} \approx \frac{\mathcal{I}}{\chi_{\rm cor}} - \frac{\eta_G\gamma_X \tau_p}{\mathcal{N}}\left(1+\frac{\mathcal{I}}{\chi_{\rm cor}}\right) ,
\end{equation}
where we neglect small terms of order $(\gamma_X \tau_p)^2$. Note if we multiply the second term on the right hand side of Eq.~\eqref{eq:V_a} by a factor of $1/\mathcal{N}$, this equation also applies to the usual scheme of QD SPSs with resonant pulse excitation and no dressing, and is thus broadly useful in characterizing the relationship between indistinguishability and TPI visibility as measured in an MZ interferometry experiment. It is important to keep in mind, however, that the indistinguishability (first-order coherence) is also degraded due to pulse excitation in a manner which scales similarly to the $g^{(2)}[0]$ behavior~\cite{Gustin2018,Hughes2019Aug}.

	\begin{figure}
		\centering
		\includegraphics[width=1\linewidth]{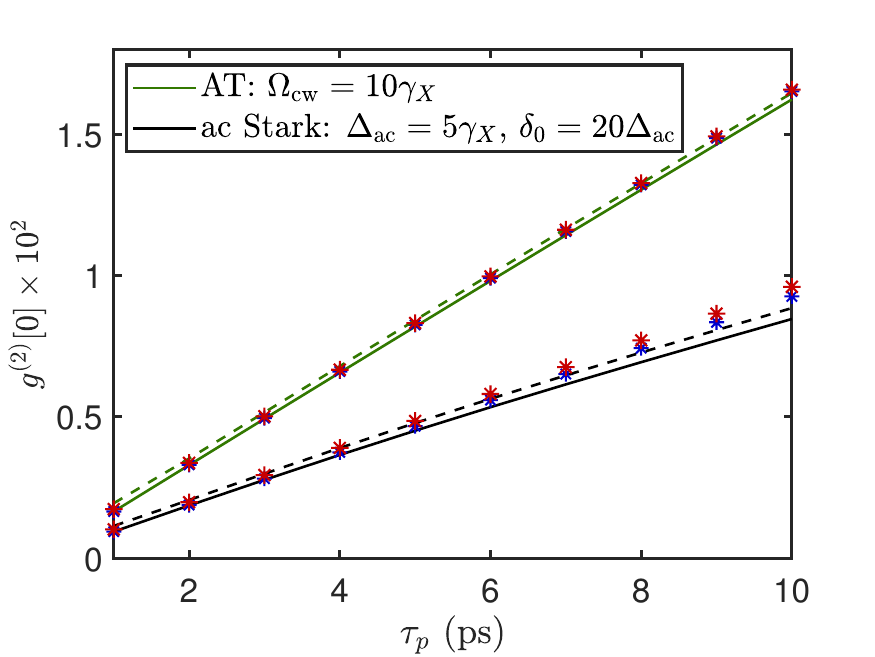}
		\caption{\label{fig:G2}
		HBT $g^{(2)}[0]$ as a function of pulse width for AT (green lines) and ac Stark (blue lines) regimes, without phonons (solid lines) and with phonons using parameter set \RNum{3} (dashed lines). Also plotted as stars is the semi-analytical approximate solution given in Eq.~\eqref{eq:g2_al} with (red) and without (blue) phonons.
		}
	\end{figure}
	
	In Fig.~\ref{fig:G2}, we study the full $g^{(2)}[0]$ of the source, using the PME model of the excitation pulse described in Appendix~\ref{app:pulse}, as well as the approximate solution in Eq.~\eqref{eq:g2_al}. For ps-duration pulses, the approximate formula is very accurate, agreeing almost exactly with the full calculation (without phonons) for $\tau_p \approx 2$ ps, and deviating only slightly in the case of phonons. With phonon coupling, the $g^{(2)}[0]$ is uniformly larger than without, which is likely due to the reduction in inversion efficiency from pulsed-excitation--induced dephasing~\cite{Gustin2018}. For long pulses in the ac Stark regime, the approximate solution overestimates the $g^{(2)}[0]$. This may be because in this scenario  $\eta\tau_p \ll 1$ is no longer satisfied even for a short pulse, and the Rabi oscillations between the $\ket{X}$ and $\ket{B}$ states lead to a small reduction in the re-excitation probability and thus $g^{(2)}[0]$.

As such, we can conclude that to maximize the purity of this source and minimize $g^{(2)}[0]$, very short pulse excitation should be used to minimize re-excitation probability (but not so short as to excite unwanted energy levels). Additionally, the use of a high-$Q$ factor cavity, which already is beneficial to the SPS figures of merit for reasons of collection efficiency and minimization of timing jitter (see discussion in Sec~\ref{subsec:cav}), also suppresses $g^{2}[0]$ strongly by means of a dynamical decoupling effect leading to a pulse-induced time-dependent Purcell factor, as found in Ref.~\cite{Gustin2018}; this effect is \emph{not} captured by our simple model which involves merely changing the relative decay rates.

There is, in addition to the contribution to $g^{(2)}[0]$ from the pulse excitation, a potential contribution which arises from the cw drive directly via the $\sigma_x^X$ term in Eq.~\eqref{eq:H1}.  Generally we assume these photons can be filtered out of the collected spectrum, however one can expect this contribution to the $g^{(2)}[0]$ to remain small relative to the pulse so long as $\mathcal{E}_{\rm cw}$ (discussed in Appendix~\ref{app:cw}) is sufficiently small, and  Ref.'s~\cite{Ollivier2021,Kirsanske2017Oct} contain information on how to incorporate a cw contribution to the $g^{(2)}[0]$ into the analysis if desired.

\section{Conclusions}\label{sec:conc}
In conclusion, we have theoretically analyzed the important factors that affect the figures of merit of our experimentally realized SPS~\cite{Dusanowski2022May}, which allows for optical frequency tuning of emitted single photons, using a polaron transform ME method to incorporate rigorously the effects of electron-phonon scattering. 
Our detailed study is also relevant to wide range of 
semiconductor QD platforms.

In the AT regime, we have shown that (without any cavity coupling and equal dipole transition rates) large frequency shifts can be achieved, but at the cost of poor indistinguishability ($\sim$66\%) and efficiency ($\sim$25\% at best). In the ac Stark regime, with a detuned laser drive, we have shown that much higher indistinguishabilities can be achieved ($>$90\% for frequency shifts of tens of $\mu$eV), but with the need for much higher cw drive strengths leading to increased phonon-related decoherence, including increased population of the higher energy state and thus increased timing jitter and reduced efficiency, as well as excitation-induced dephasing of the photon emission channel. This phonon-related degradation of the source figures of merit increases rapidly as the frequency shift $\Delta_{\rm ac}$ is increased. For large enough phonon coupling rates, phonon scattering also leads to, for a fixed frequency shift $\Delta_{\rm ac}$, a local maximum in the indistinguishability as a function of detuning, whereas without phonon coupling the indistinguishability continues to increase as the detuning increases further according to the approximate relation derived by adiabatic elimination of the higher energy state $\mathcal{I} \approx \mathcal{N} \approx \delta/(\delta+\Delta_{\rm ac})$. In this case, the maximum achievable indistinguishability is set by the presence of other energy levels neglected in this analysis, other sources of excitation-induced decoherence, or other experimental limitations.

In addition, we have shown how the low-temperature asymmetry associated with the phonon bath due to low phonon occupation probabilities leads to preferential operation of the SPS in the ac Stark regime with positive (red detuned) detunings (and thus red Stark shifts of the emitted photons), as this leads to reduced phonon-induced excitation of the higher energy state and thus reduced timing jitter and efficiency loss.

We have also elucidated how cavity coupling (or more generally, application of the model to systems with different transition rates between levels) can be used to improve the source figures of merit via the Purcell effect, at the cost of broader emission linewidths relative to the frequency shift. In this case, using the AT regime to generate indistinguishable photons becomes more practical, with $>$90\% indistinguishability achievable with $\mathcal{N} \approx 1/2$ for frequency shifts up to hundreds of $\mu$eV for realistic QD and cavity parameters which have been widely achieved in the literature for dielectric resonators. The presence of selectively enhanced spontaneous emission rates also was shown to benefit the ac Stark regime, enabling near unity indistinguishabilities with high efficiency for Stark shifts up to tens of $\mu$eV.

Finally, we analyzed the multiphoton statistics of the source, including the $g^{(2)}[0]$ parameter, which was shown to follow closely related trends as previous studies on simple undressed two-level system models have predicted~\cite{Gustin2018,Fischer2018}.
%We also introduced a new figure-of-merit unique to our proposed dressed source, the cw error rate $\mathcal{E}_{\rm cw}$, which quantifies the ratio of photons emitted due to the cw excitation weakly dressing the ground state to first excited state transition to the total number of photons emitted over a laser pulse cycle; we then  showed how this can, if not filtered out of the collected spectrum, place limitations on the biexciton binding energies for which the source will operate effectively.

We 
%would like to 
reiterate that while the results in this paper have been presented for the specific realization of the source with the semiconductor QD biexciton cascade four-level system model (which necessitates an analysis of LA phonon coupling), the principles of operation only require a quantum 3 (or more) level ladder system, and many of the results of our analysis for the case of no phonons should apply qualitatively to these systems as well.

Furthermore, we expect that the specific implementation of the frequency tuning mechanism we have illustrated in this work could be modified, expanded upon, or optimized further by the engineering or implementation of different energy levels or radiative decay rates. For example, the QD biexciton cascade also involves a $Y$-polarized exciton, which could be used to create Stark shifted photons of the $X$-cascade by instead dressing the $\ket{Y}$-$\ket{G}$ transition with an orthogonally polarized cw laser. This would remove the issue of the cw background by employing polarization filtering. In this case, the indistinguishability follows similar trends as in the case of biexciton-exciton dressing. For the sake of this work, however, we have restricted our detailed analysis to a cascade type dressing involving the biexciton state, as this is what we have reported in experiment.

Overall, our results indicate that the use of the AT and ac Stark effects to produce optical frequency shifts in a quantum ladder system can be effective in generating indistinguishable single photons with high efficiency. While the achievable figures of merit are ultimately limited by phonon effects in the case of the QD cascade system studied here, frequency shifts of up to tens-hundreds of $\mu$eV are achievable with realistic cavity and QD parameters which have regularly appeared in the literature, while maintaining high indistinguishabilities, efficiencies, and purities, and this analysis is consistent with experimental results we have reported in Ref.~\cite{Dusanowski2022May}.

	\begin{acknowledgments}
	\L{}.D. and Stephen H. acknowledge financial support from the Alexander von Humboldt Foundation. We are also grateful for the support by the State of Bavaria, the Natural Sciences and Engineering Research Council of Canada, and the Canadian Foundation for Innovation. We would like to thank Christian Schneider for useful discussions and work in fabrication of the sample used to obtain the data in Fig.~\ref{fig:MZ}(c).
	\end{acknowledgments}

\appendix
\section{Exact solution for case of equal dipole transition rates in the Autler-Townes regime}\label{app:sol}
	For $\gamma_{B} = 2\gamma_X$, and $\delta=0$ (AT regime), neglecting phonons, the solution to the ME of Eq.~\eqref{eq:me1} with the Hamiltonian (and rotating frame) of Eq.~\eqref{eq:r2} can be expressed analytically. Considering the system to be in the $\ket{X}$ state at $t=0$, the single-photon indistinguishability is, up to an integral,
	\begin{equation}
	\mathcal{I}= \frac{2\gamma_X^2}{\mathcal{N}^2}\int_0^\infty \text{d}t T(t),
	\end{equation}
	where
		\begin{equation}
	    \mathcal{N} = \frac{W_0}{2} + \frac{5\gamma_X^2\Omega_{\rm cw}^2/4}{(\Omega_{\rm cw}^2 + \frac{\gamma_X^2}{2})(\Omega_{\rm cw}^2  + 3\gamma_X^2)},
	\end{equation}
	and
	\begin{align}
	T(t) =& \frac{|c_+(t)|^2}{\frac{3\gamma_X}{2} - \frac{i}{2}(\widetilde{\Omega} - \widetilde{\Omega}^*)} + \frac{|c_-(t)|^2}{\frac{3\gamma_X}{2} + \frac{i}{2}(\widetilde{\Omega} - \widetilde{\Omega}^*)} + \nonumber \\ & \frac{c_+(t)c_-(t)^*}{\frac{3\gamma_X}{2} - \frac{i}{2}(\widetilde{\Omega} + \widetilde{\Omega}^*)} + \frac{c_-(t)c_+(t)^*}{\frac{3\gamma_X}{2} + \frac{i}{2}(\widetilde{\Omega} + \widetilde{\Omega}^*)},
	\end{align}
	with
		\begin{equation}
	W_0 = \frac{\Omega_{\rm cw}^2 + \gamma_X^2}{\Omega_{\rm cw}^2 + \frac{\gamma_X^2}{2}},
	\end{equation}
	and
	\begin{equation}
	c_{\pm}(t) = \frac{1}{2}\left( \rho_X(t)\left(1 \mp i \tan{\phi}\right)  \mp \frac{1}{\cos{\phi}}\rho_{B-X}(t) \right),
	\end{equation}
	where $\cos{\phi} = \widetilde{\Omega}/\Omega_{\rm cw}$, $\sin{\phi} = \gamma_X/(2\Omega_{\rm cw})$, and $\widetilde{\Omega} = \sqrt{\Omega_{\rm cw}^2 - \frac{\gamma_X^2}{4}}$.
	
	For the density matrix elements, we obtain
	\begin{equation}
	    \rho_{X}(t)= \bra{X}\rho\ket{X} = \frac{e^{-\gamma_X t}}{2}\left[W_0 + a_+(t) + a_-(t)\right],
	\end{equation}
	\begin{align}
	    \rho_{B-X}(t) &= \bra{B}\rho\ket{X} \nonumber \\ &=\frac{i e^{-\gamma_X t}}{2\Omega_{\rm cw}}\Bigg( \left[\frac{\gamma_X}{4} + i\widetilde{\Omega}'\right]a_+(t) \nonumber \\ & \ + \left[\frac{\gamma_X}{4} - i\widetilde{\Omega}'\right]a_-(t) - \frac{\Omega_{\rm cw}^2 \gamma_X}{\Omega_{\rm cw}^2+\gamma_X^2/2}\Bigg),
	\end{align}
	and
	\begin{equation}
		\rho_{B}(t) = e^{-\gamma_X t} - \rho_{X}(t),
	\end{equation}
	where
	\begin{equation}
	a_{\pm}(t) = \frac{\frac{\Omega_{\rm cw}^2}{2}}{\Omega_{\rm cw}^2+\frac{\gamma_X^2}{2}}\left(1 \mp i\frac{3\gamma_X}{4\widetilde{\Omega}'}\right)\exp\left({-\frac{3\gamma_X}{4}t \pm i \widetilde{\Omega}' t}\right),
	\end{equation}
	with
	$\widetilde{\Omega}' = \sqrt{\Omega_{\rm cw}^2-\frac{\gamma_X^2}{16}}$. 
	
	In the dressed state basis (using Eq.~\eqref{eq:eigens}), we have simply $\rho_+(t) = \rho_-(t) = e^{-\gamma_X t}/2$, and so $\mathcal{N}^{\pm} =1/4$.
 The first-order two time correlation function is
	\begin{equation}
	g^{(1)}(t,\tau)= e^{-\frac{3}{4}\gamma_X \tau}\left( c_+(t) e^{i \widetilde{\Omega} \tau/2} + c_-(t) e^{-i \widetilde{\Omega} \tau/2}\right).
	\end{equation}
	Notably, in the well-dressed limit ($\gamma_X/\Omega_{\rm cw} \rightarrow 0$), we have $\mathcal{N} = 1/2$ (as the dressed system has equal decay rates to both polarization channels), and $\mathcal{I} = 11/21$.
	
	We also consider the indistinguishability of a photon emitted from one of the sidepeaks, $\mathcal{I}^{\pm}$. By symmetry, for $\delta=0$ both sidepeaks give the same indistinguishability such that $\mathcal{I}^+ = \mathcal{I}^-$.
	The result for this $\mathcal{I}^{\pm}$ is (assuming $\widetilde{\Omega}$ is real, as these observables are only well-defined for dressing exceeding the damping)
	\begin{equation}
	\mathcal{I}^{\pm} = 8\gamma_X^2\int_0^{\infty}\text{d}t S(t),
	\end{equation}
	where 
	\begin{align}
	S(t) &=  \frac{2\Omega_{\rm cw}}{3 \gamma_X}\left[|d_+|^2(\Omega_{\rm cw} - \widetilde{\Omega}) + |d_-|^2(\Omega_{\rm cw} + \widetilde{\Omega})\right] \nonumber \\ & + \gamma_X\left( d_+ d_-^* \frac{\gamma_X/2 - i\widetilde{\Omega}}{3\gamma_X - 2i \widetilde{\Omega}} + \text{c.c.}\right),
	\end{align}
		with
	\begin{equation}
	d_\pm (t) = -i\frac{F_0(t)}{2\Omega_{\rm cw}}  \pm \frac{i}{2}\left[ \frac{e^{-\gamma_X t}}{\sqrt{2}\widetilde{\Omega}} - \frac{F_0(t)}{\widetilde{\Omega}}\left(1- \frac{i\gamma_X}{2\Omega_{\rm cw}}\right) \right],
	\end{equation}
	and $F_0(t) = \frac{1}{\sqrt{2}}(\rho_X(t) - \rho_{B-X}(t))$. The $g_{\pm}^{(1)}(t,\tau)$ correlation function is
	\begin{align}
	g_{\pm}^{(1)}(t,\tau) = \frac{e^{-\frac{3}{4}\gamma_X\tau}}{\sqrt{2}}\Bigg[&d_+(t)e^{i\widetilde{\Omega} \tau/2}\left(i(\Omega_{\rm cw}\!-\!\widetilde{\Omega}) \!+\! \frac{\gamma_X}{2}\right) +\nonumber \\ &  d_-(t)e^{-i\widetilde{\Omega} \tau/2}\left(i(\Omega_{\rm cw}\!+\!\widetilde{\Omega}) \!+\! \frac{\gamma_X}{2}\right)\Bigg].
	\end{align}
	In this case, the indistinguishability $\mathcal{I}^{\pm}$ tends to 2/3 in the well-dressed limit, which is the same as what one would find for an undressed system initialized in the $\ket{B}$ state.

	\section{Interaction frame secular approximation}\label{app:secular}
	In this appendix, we discuss the secular approximation which can be made when the emission spectrum peaks are well-separated, which removes the fast oscillations from the equations of motion and makes the numerical solution of the ME vastly more efficient. This approximation also produces an ME in Lindblad form.
	
We first consider the model given by the system Hamiltonian $H_S$ of Eq.~\eqref{eq:r2} and the ME of Eq.~\eqref{eq:me1} with the additional phonon term of~\eqref{eq:pme}. We then move into an interaction frame defined by $\tilde{\rho}(t) = U^{\dagger}(t)\rho(t) U(t)$, where for our four-level system model,
	\begin{align}\label{eq:U}
	    U(t) &= \text{exp}[-i H_S t] \nonumber \\ &= \Big[\ket{G}\bra{G} + \ket{Y}\bra{Y} + \nonumber \\ & \ \  \ \ \ \ \  e^{-iE_+ t} \ket{+}\bra{+}+ e^{-iE_-t}\ket{-}\bra{-}\Big] .
	\end{align}
	The ME in this interaction frame then takes the form 
	\begin{equation}\label{eq:me_tra}
	    \dot{\tilde{\rho}}(t) = \widetilde{\mathbb{L}}_{\rm rad}(t)\tilde{\rho}(t) + \widetilde{\mathbb{L}}_{\rm PME}(t)\tilde{\rho}(t),
	\end{equation}
	where $\widetilde{\mathbb{L}}(t)\tilde{\rho}(t) = U^{\dagger}(t)\mathbb{L}\rho(t) U(t)$, in terms of the dressed state basis of Eq.~\eqref{eq:eigens} with dressed energies $E_{\pm} = \pm \eta/2$.
	
	Next, we note that applying the unitary transformation of Eq.~\eqref{eq:U} to the  radiative and phonon terms $\mathbb{L}_{\rm rad}\rho$ and $\mathbb{L}_{\rm PME}\rho$ will yield a sum over time-independent terms, as well as time-dependent terms which oscillate at frequencies given by $E_{\pm}$ and $E_{+} - E_{-}$. If $\eta$ is much larger than the characteristic rates at $\tilde{\rho}(t)$ evolves, then these time-dependent terms can be dropped, making a \emph{secular approximation} (or post-trace rotating wave approximation), as they average out to give a negligible contribution to the interaction frame density operator evolution. Intuitively, we expect this situation to occur when the peaks of the system are well-separated by much more than a linewidth or any of the phonon rates. The characteristic rates at which $\tilde{\rho}(t)$ evolves are given by the coefficients of Eq.~\eqref{eq:me_tra} which we give below.
	
	Making the secular approximation, we thus drop all of these rotating terms such that Eq.~\eqref{eq:me_tra} now has no explicit time dependence (i.e., except that coming from $\tilde{\rho}(t)$). With some work, it can be shown that the ME in the interaction frame with the secular approximation can then be written in Lindblad form as 
	\begin{equation}
\dot{\tilde{\rho}}(t) = \widetilde{\mathbb{L}}_{\rm rad}^S\tilde{\rho}(t) + \widetilde{\mathbb{L}}_{\rm PME}^S\tilde{\rho}(t) - i[\widetilde{H}_{\rm ph},\tilde{\rho}(t)],
	\end{equation}
	where the radiative contribution is
	\begin{align}
	&\widetilde{\mathbb{L}}^S_{\text{rad}}\tilde{\rho}=\frac{\gamma_X}{2}\Bigg\{\sum_{\beta = +,-}\mathcal{L}\Big[\sigma^-_X\ket{\beta}\bra{\beta}\Big] +\mathcal{L}\Big[\ket{G}\bra{Y}\Big] \Bigg\}\tilde{\rho}  \nonumber \\ &  + \frac{\gamma_B}{4}\Bigg\{\sum_{\beta = +,-}\mathcal{L}\Big[\braket{B|\beta}\ket{Y}\bra{\beta}\Big] + \left(\frac{1-\frac{\delta}{\eta}}{2}\right)^2\mathcal{L}\Big[\sigma_{+-}\Big]   \nonumber \\ & \left(\frac{1+\frac{\delta}{\eta}}{2}\right)^2\mathcal{L}\Big[\sigma^{\dagger}_{+-}\Big] +\left(\frac{\Omega_{\rm cw}}{2\eta}\right)^2\mathcal{L}\Big[[\sigma_{+-},\sigma^{\dagger}_{+-}]\Big]\Bigg\}\tilde{\rho} ,
	\end{align}
	where we have let $\sigma_{+-} = \ket{+}\bra{-}$ to simplify the notation, and the non-unitary phonon contribution is
	\begin{align}\label{eq:pme_s}
	    \widetilde{\mathbb{L}}^S_{\rm PME}\tilde{\rho}(t) &= \frac{1}{2}\text{Re}\{\tilde{\Gamma}^0_{x}\}\mathcal{L}\Big[[\sigma_{+-},\sigma^{\dagger}_{+-}]\Big] + \nonumber \\ &  \sum_{m=x,y}\! \Bigg\{\frac{\text{Re}\{\! \tilde{\Gamma}^+_{m}\}}{2}\mathcal{L}\Big[\sigma^{\dagger}_{+-}\Big] \!+\! \frac{\text{Re}\{\tilde{\Gamma}^-_{m}\}}{2}\mathcal{L}\Big[\sigma_{+-}\Big]\! \Bigg\}\tilde{\rho}.
	\end{align}
	
    We also have a unitary part of the phonon interaction, which is given by the Hamiltonian
	\begin{align}\label{eq:pme_HS}
	    \widetilde{H}_{\rm PME}&= \frac{1}{2}\text{Im}\{\tilde{\Gamma}^0_{x}\}(\sigma^+_X\sigma^-_X + \sigma^+_B \sigma^-_B) + \frac{1}{2}\sum_{m=x,y}\nonumber \\ & \Big[   \text{Im}\{\tilde{\Gamma}^+_{m}\}\sigma_{+-}\sigma^{\dagger}_{+-} + \text{Im}\{\tilde{\Gamma}^-_{m}\}\sigma^{\dagger}_{+-}\sigma_{+-} \Big].
	\end{align}
	The complex phonon scattering rates are given by 
	\begin{subequations}
	    \begin{equation}\label{eq:gypma}	    \tilde{\Gamma}^{0}_x = \frac{\Omega_{\rm cw}^4}{2\eta^2}\int_0^{\infty}\text{d}\tau G_x(\tau)
	    \end{equation}
	  \begin{equation}\label{eq:gypmb}
	  \tilde{\Gamma}^{\pm}_x = \frac{\Omega_{\rm cw}^2\delta^2}{2\eta^2}\int_0^{\infty}\text{d}\tau G_x(\tau) e^{\pm i \eta \tau} \end{equation}
	    \begin{equation}\label{eq:gypmc} \tilde{\Gamma}^{\pm}_y = \frac{\Omega_{\rm cw}^2}{2}\int_0^{\infty}\text{d}\tau G_y(\tau) e^{\pm i \eta \tau}.
	    \end{equation}
	\end{subequations}
	
	From Eq.'s~\eqref{eq:pme_s} and~\eqref{eq:pme_HS}, we see that the influence of the exciton-phonon interaction can be seen in the dressed state basis to take the form of a pure dephasing-type term with rate $\text{Re}\{\tilde{\Gamma}^0_{x}\}$, and phonon-driven transitions from $\ket{+}$ ($\ket{-}$) to $\ket{-}$ ($\ket{+}$) with rate $\text{Re}\{\! \tilde{\Gamma}^+_{m}\}$ ($\text{Re}\{\! \tilde{\Gamma}^-_{m}\}$). Additionally, the Hamiltonian term $\widetilde{H}_{\rm PME}$ gives a renormalization of the dressed state energies, which in the bare-state basis is equivalent to a small shift in the $\ket{X}$ and $\ket{B}$ state energies, as well as the drive term between them; our simulations (not shown) performed without this Hamiltonian term look very similar to the full calculations, indicating that $\widetilde{\mathbb{L}}^S_{\rm PME}\tilde{\rho}(t)$ has the dominant influence on the SPS figures of merit.
	
	It is easy to see that in the dressed-state frame, the observable figures of merit for the $\pm$ peaks are calculated the same as in the bare state frame, but with $\rho \rightarrow \tilde{\rho}$, and for the total emitted photon number, $\mathcal{N} = \mathcal{N}^+ + \mathcal{N}^-$.

	\section{Weak phonon coupling PME}\label{app:weak}
	For common values of the phonon parameters $\alpha$, $\omega_b$, and the temperature $T$, including the parameter sets we study in this work, the phonon coupling function satisfies $|\phi(\tau)| \ll 1$. In this case, we can to a good approximation expand the phonon functions $G_m(\tau)$ that appear in to leading order in $\phi$, to find $G_x(\tau) = \mathcal{O}(\phi^2)$, and $G_y(\tau) = \phi(\tau) + \mathcal{O}(\phi^3)$. 
	
	Under this \emph{weak phonon coupling approximation}, we can simplify the PME in the dressed state basis under the secular approximation of Appendix~\ref{app:secular}. The Hamiltonian part of the PME then becomes (using primes to indicate this weak phonon coupling approximation)
	\begin{equation}
	    \widetilde{H}'_{\rm PME} = \frac{1}{2}\Big[   \text{Im}\{\tilde{\Gamma}^+_{y}\}\sigma_{+-}\sigma^{\dagger}_{+-} + \text{Im}\{\tilde{\Gamma}^-_{y}\}\sigma^{\dagger}_{+-}\sigma_{+-} \Big],
	\end{equation}
	where $\tilde{\Gamma}^{\pm}_y$ are given by Eq.~\eqref{eq:gypmc}, and the incoherent part of the PME becomes
	\begin{align}
	    \widetilde{\mathbb{L}}^{'S}_{\rm PME}\tilde{\rho}(t) &=  \frac{\tilde{\Gamma}'_{0}}{2}\left[n_{\rm ph}(\eta,T) + 1\right] \mathcal{L}\Big[\sigma^{\dagger}_{+-}\Big]\tilde{\rho} \nonumber \\ & +  \frac{\tilde{\Gamma}'_{0}}{2}\left[n_{\rm ph}(\eta,T)\right] \mathcal{L}\Big[\sigma_{+-}\Big]\tilde{\rho},
	\end{align}
	where $n_{\rm ph}(\omega,T) = \left[e^{\omega/(k_B T)} -1 \right]^{-1}$ is the thermal phonon occupation number, and
	\begin{equation}\label{eq:gamma0}
	    \tilde{\Gamma}'_0 = \frac{\pi}{2} \left(\frac{\Omega_{\rm cw}}{\eta}\right)^2 J(\eta).
	\end{equation}

	While we do not use the weak phonon coupling approximation for our numerical simulations (although for the phonon parameter sets we use, it is expected to give good quantitative agreement with the full PME), it is nonetheless useful to gain analytical insight into the underlying physical processes and scaling behavior of the phonon decoherence rates.  The Hamiltonian term $\widetilde{H}'_{\rm PME}$ gives a small renormalization of the dressed state energies, such that $E_{\pm} \rightarrow E_{\pm} + \text{Im}\{\tilde{\Gamma}^{\pm}_{y}\}/2$, which gives a perturbation to the nominal splitting $\Delta_{\rm ac}$. As mentioned in the main text, we neglect this phonon drive-dependent frequency shift renormalization when we refer to the frequency shift $\Delta_{\rm ac}$, and we have checked in our simulations that its effect is small (typically $\lesssim 10 \%$ of $\Delta_{\rm ac}$).
	
\begin{figure}
		\centering
		\vspace*{2.5\baselineskip}
		\includegraphics[width=0.75\linewidth]{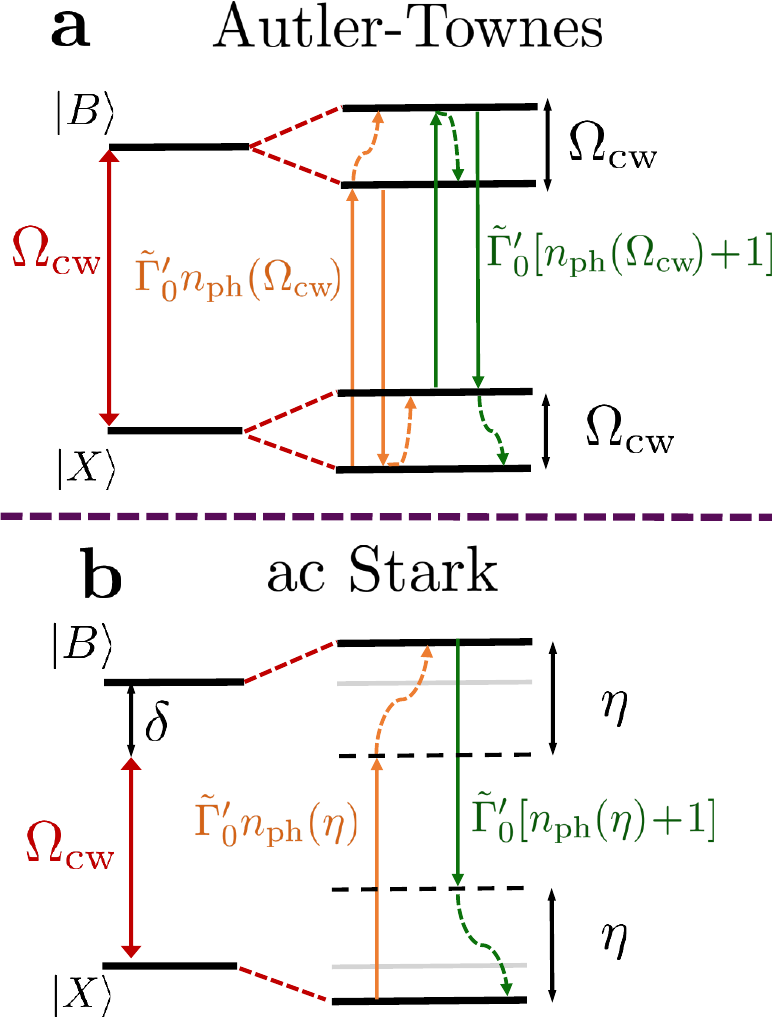}
		\caption{\label{fig:AppSchem} Schematic of phonon-assisted transitions between dressed states as given by the weak phonon coupling PME. In the AT regime (a), the phonon absorption processes with rate $\tilde{\Gamma}'_0n_{\rm ph}(\Omega_{\rm cw},T)$ and the phonon emission processes with rate $\tilde{\Gamma}'_0[n_{\rm ph}(\Omega_{\rm cw},T)+1]$ take the system to both $\ket{X}$ and $\ket{B}$ states, as the dressed states $\ket{\pm}$ are symmetric and antisymmetric combinations of these states, whereas in the ac Stark regime (b) the phonon absorption processes with rate $\tilde{\Gamma}'_0n_{\rm ph}(\eta,T)$ and the phonon emission processes with rate $\tilde{\Gamma}'_0[n_{\rm ph}(\eta,T)+1]$ take the system to higher and lower energy, respectively. In this schematic we let $n_{\rm ph}(\omega) \equiv n_{\rm ph}(\omega,T)$.
		}
	\end{figure}

	The non-unitary part of the PME (which leads to phonon-related decoherence) under the weak phonon and secular approximations is given by $\widetilde{\mathbb{L}}^{'S}_{\rm PME}\tilde{\rho}$ and is shown schematically in Fig.~\ref{fig:AppSchem}.
	
	In the AT regime, with $\delta=0$, the phonon rate is simply $\tilde{\Gamma}'_0 = \frac{\pi}{2}J(\Omega_{\rm cw})$, and we can furthermore neglect the exponential cutoff term in the phonon drive, as the regime where it becomes significant requires very high drive strengths and is difficult to realize in experiments. Then, if $\Omega_{\rm cw} \gg k_B T$ (very strong driving), the phonon dissipator simply takes the form of spontaneous emission from state $\ket{+}$ to state $\ket{-}$ with rate $\alpha \frac{\pi}{2} \Omega_{\rm cw}^3$. At lower drive strengths $\Omega_{\rm cw} \ll k_B T$, the phonon-induced transitions between $\ket{+}$ and $\ket{-}$ states occur with the same rate $\alpha \frac{\pi}{2} k_B T \Omega_{\rm cw}^2$, giving the expected $\Omega_{\rm cw}^2$ scaling.
	
In the ac Stark regime with $|\delta| \gg \Omega_{\rm cw}$, we can find the leading order behavior (i.e., in $\delta/\Delta_{\rm ac}$) of the incoherent phonon rates by approximating $\eta \approx |\delta|$, and $\Omega_{\rm cw} \approx 2\sqrt{\Delta_{\rm ac}\delta}$. Then, for $\delta \gg k_B T$, the phonon dissipator again takes the form of spontaneous emission from $\ket{+}$ to $\ket{-}$ with approximate rate $\alpha 2 \pi \Delta_{\rm ac}^3 \left(\frac{\delta}{\Delta_{\rm ac}}\right)^2 e^{-\delta^2/(2\omega_b^2)}$. At lower detunings $\delta \ll k_B T$, again the phonon-induced transitions between $\ket{+}$ and $\ket{-}$ occur at the same approximate rate $\alpha 2 \pi k_B T \Delta_{\rm ac}^2 \left(\frac{|\delta|}{\Delta_{\rm ac}}\right) e^{-\delta^2/(2\omega_b^2)}$.

	\section{Polaron master equation with pulse drive}\label{app:pulse}
	
	To extend the PME to include the excitation pulse, we add a term $\Omega_p(t)\cos{(\omega_X t)}\sigma_x^X$ to the Hamiltonian in Eq.~\eqref{eq:H1}, where $\Omega_p(t)$ is the pulse amplitude satisfying $\int_{-\infty}^{\infty} \text{d} t \Omega_p(t) = \pi$, neglecting the coupling of the pulse to the $\ket{B}$-$\ket{X}$ transition which is appropriate provided $E_B^{-1}$ is much smaller than the duration of the pulse. We also neglect the coupling of the cw drive to the $\ket{X}$-$\ket{G}$ transition. We then choose another interaction frame given by $H_0'' = \omega_X \sigma^+_X\sigma^-_X + (\omega_B-\delta)\sigma^+_B\sigma^-_B + \omega_Y\ket{Y}\bra{Y}$, giving 
	\begin{equation}
	    H_S(t) = \delta \sigma^+_B\sigma^-_B + \frac{\Omega_p(t)}{2}\sigma_x^X + \frac{\Omega_{\rm cw}}{2}\sigma_x^B.
	\end{equation}
	
	The PME superoperator is then
	\begin{align}\label{eq:pme_pulse}
& \mathbb{L}_{\rm PME}(t)\rho = \nonumber \\ & \int_0^{\infty}\!\!\!  \! \text{d}\tau \! \! \! \sum\limits_{m = x, y} \! \! G_m(\tau)   
[\widetilde{X}_m(t\! -\! \tau,t) \rho(t),X_m(t)] + {\rm H.c.},\end{align} and we have again absorbed a coherent attenuation factor $\langle B \rangle$ from the PME into our definition of $\Omega_{\rm cw}$ and $\Omega_p(t)$ for easy comparison with the no-phonon case.
The operators $\widetilde{X}_m(t-\tau,t)=U(t,t\! - \! \tau)X_m(t\! - \tau)U^{\dagger}(t,t\! - \! \tau)$ are calculated using the ``additional Markov''~\cite{Gustin2018} approximation:  $U(t,t \! - \! \tau) \approx \exp{[-i H_S(t) \tau]}$, and $X_m(t\! - \tau) \approx X_m(t)$ within the integrand, and this approximation is valid for pulse widths much greater than $\omega_b^{-1}$, and 
\begin{equation}
    X_m(t) = \frac{\Omega_{\rm cw}}{2\langle B \rangle}\sigma_m^B + \frac{\Omega_p(t)}{2\langle B \rangle}\sigma_m^X.
\end{equation}

	\section{Continuous wave excitation error rate}\label{app:cw}
		As mentioned in the main text, the presence of the $\sigma_x^X$ term in Eq.~\eqref{eq:r1} leads to weak cw excitation of the $\ket{X}$ state, by means of the far off-resonant drive. As a result, in addition to the (pulse-wise) emitted photon number $\mathcal{N}$, which is calculated in absence of this term, using instead Eq.~\eqref{eq:r2} for the system Hamiltonian, there is a small, constant in time, photon emission flux. However, this contribution produces photons which are (for $E_B+\delta >0$) blue Stark shifted by $\approx \Omega_{\rm cw}^2/4(E_B+\delta)$, which in many instances should be far-off from the desired peak of interest, and as such can be filtered out of the collected spectrum. Nonetheless, in this section, we consider the cw error rate assuming no emitted photons of the $\ket{X}$-$\ket{G}$ transition are filtered.
		
		To quantify this \emph{cw error rate}, we define a quantity $\mathcal{E}_{\rm cw}$ to be the ratio of the average number of photons emitted in the absence of any pulse excitation over a duration equal to the laser repetition rate $T_{\rm rep}$, divided by the number of photons emitted by the source with a pulse excitation. 
			
			To calculate $\mathcal{E}_{\rm cw}$, we assume as a first approximation that the pulse initializes the system in the $\ket{X}$ state. We then can simulate, using the Hamiltonian of Eq.~\eqref{eq:r1}, the photons emitted $\mathcal{N}_0$ over a duration $T_{\rm rep}$ starting from the initial condition $\rho_0$, which we choose to be the steady-state condition of the ME, and then divide this quantity by the number of photons emitted using the same Hamiltonian with instead initial condition $\ket{X}$, which we denote $\mathcal{N}_X$. Then,
			\begin{equation}\label{eq:cwerr}
			\mathcal{E}_{\rm cw} = \frac{\mathcal{N}_0}{\mathcal{N}_X}.
			\end{equation}
	
	For the case where we do not consider phonon interactions, as an additional approximation to this quantity, we can also note that so long as $\Omega_{\rm cw} \ll |E_B + \delta|$, the cw excitation of the $\ket{X}$-$\ket{G}$ transition is very weakly driven (i.e., see Fig.~\ref{fig:schematic}(d)), and as such the steady-state population of the $\ket{X}$ state, and thus the photon flux due to the cw drive will be very small. In this case, we can approximate the biexciton population (and associated coherences) to be negligible, and solve the ME in the absence of phonons analytically to find:
	\begin{align}\label{eq:cw_approx}
	    \mathcal{E}_{\rm cw} & \approx \frac{\gamma_X T_{\rm rep} \rho_X^0}{\mathcal{N}_X} + \mathcal{O}([\rho_X^0]^2)  \nonumber \\ & \approx \frac{\gamma_X T_{\rm rep} \rho_X^0}{\mathcal{N}} + \mathcal{O}([\rho_X^0]^2),
	\end{align}
	where $\rho_X^0 = \Omega_{\rm cw}^2/(4(E_B+\delta)^2)$ is the steady-state population of the $\ket{X}$ state under this approximation, and in the second line we have noted that the difference between $\mathcal{N}_X$, the emitted photon number calculated over a time duration $T_{\rm rep}$ with the Hamiltonian of Eq.~\eqref{eq:r1} and the initial condition $\rho(t=0) = \ket{X}\bra{X}$, and the total emitted photon number $\mathcal{N}$ calculated using the Hamiltonian of Eq.~\eqref{eq:r2} and the same initial condition (which does not contain any cw excitation contributions) scales with $\rho_X^0$. Specifically, the final line of Eq.~\eqref{eq:cw_approx} allows one to approximate the cw error rate in the absence of phonon couplings using only the emitted photon number $\mathcal{N}$ calculated without considering the far off-resonant driving of the $\ket{X}$-$\ket{G}$ transition. When phonons are considered at a nonzero temperature, phonon-assisted transitions as shown schematically in Fig.~\ref{fig:schematic}(d) render this approximation inappropriate.
	
\begin{figure}
		\centering
		\includegraphics[width=0.95\linewidth]{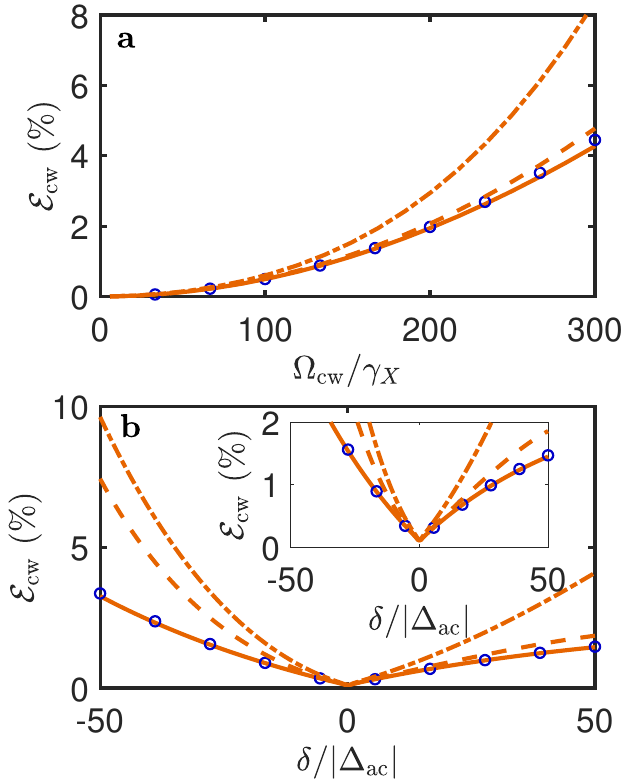}
		\caption{\label{fig:R8}
			Continuous wave error rate $\mathcal{E}_{\rm cw}$ for SPS device operating in (a) the AT regime with $\delta=0$ and (b) the ac Stark regime with $\Delta_{\rm ac} = 10\gamma_X$. The full calculations are shown in orange, and the approximate calculation in the second line of Eq.~\eqref{eq:cw_approx} is shown as blue circles.
		}
	\end{figure}

	In Fig.~\ref{fig:R8}, we plot the cw error rate $\mathcal{E}_{\rm cw}$ using both the full calculation $\mathcal{N}_0/\mathcal{N}_X$, as well as the approximate solution in the final line of Eq.~\eqref{eq:cw_approx} for the case of no phonons, for both AT and ac Stark regimes. We use biexciton binding energy $E_B = 3.24$ meV, as in our experiment in Ref.~\cite{Dusanowski2022May}. For both regimes, the approximate formula is an excellent approximation to the full solution. Also in both regimes, we find in contrast to the figures of merit studied in the main sections of this paper, the error rate is much worse for phonon parameter set \RNum{2} compared to parameter set \RNum{1} (which is close to the no-phonon case). This is due to the larger value of $\omega_{b,\RNum{2}}$ giving a more appreciable value of the phonon spectral function at the detuning from the relevant transition $J(E_B+\delta)$, meaning that the resonant process of absorption simultaneous with phonon absorption becomes more prominent, despite this process being mostly suppressed at low temperatures. We can see this clearly in Fig.~\ref{fig:R8}(b), where for negative detunings, the error rate increases drastically for \emph{both} phonon parameter sets, as the detuning $E_B + \delta$ of the $\ket{X}$-$\ket{G}$ transition becomes smaller and the spectral function becomes more and more appreciable even for the smaller phonon cutoff frequency of $\omega_{b,\RNum{1}}$. 
	This is an additional reason to prefer positive detunings (and thus red frequency shifts) when operating in the ac Stark regime, for the typical case of positive binding energy $E_B$.
	
We note, of course, that the cw error rate can be improved by using a smaller repetition rate $T_{\rm rep}$, however one has to keep in mind the relaxation rate of the SPS system, and if the repetition time were made too small the derivation and equations presented in Sec.~\ref{sec:model} on the HOM and HBT experimental procedures may need to be revised; in this manner, the introduction of a cavity provides another potential advantage through enhancement of the relaxation rate.

\clearpage
	\bibliography{thebib}

\begin{thebibliography}{93}%
\makeatletter
\providecommand \@ifxundefined [1]{%
 \@ifx{#1\undefined}
}%
\providecommand \@ifnum [1]{%
 \ifnum #1\expandafter \@firstoftwo
 \else \expandafter \@secondoftwo
 \fi
}%
\providecommand \@ifx [1]{%
 \ifx #1\expandafter \@firstoftwo
 \else \expandafter \@secondoftwo
 \fi
}%
\providecommand \natexlab [1]{#1}%
\providecommand \enquote  [1]{``#1''}%
\providecommand \bibnamefont  [1]{#1}%
\providecommand \bibfnamefont [1]{#1}%
\providecommand \citenamefont [1]{#1}%
\providecommand \href@noop [0]{\@secondoftwo}%
\providecommand \href [0]{\begingroup \@sanitize@url \@href}%
\providecommand \@href[1]{\@@startlink{#1}\@@href}%
\providecommand \@@href[1]{\endgroup#1\@@endlink}%
\providecommand \@sanitize@url [0]{\catcode `\\12\catcode `\$12\catcode
  `\&12\catcode `\#12\catcode `\^12\catcode `\_12\catcode `\%12\relax}%
\providecommand \@@startlink[1]{}%
\providecommand \@@endlink[0]{}%
\providecommand \url  [0]{\begingroup\@sanitize@url \@url }%
\providecommand \@url [1]{\endgroup\@href {#1}{\urlprefix }}%
\providecommand \urlprefix  [0]{URL }%
\providecommand \Eprint [0]{\href }%
\providecommand \doibase [0]{https://doi.org/}%
\providecommand \selectlanguage [0]{\@gobble}%
\providecommand \bibinfo  [0]{\@secondoftwo}%
\providecommand \bibfield  [0]{\@secondoftwo}%
\providecommand \translation [1]{[#1]}%
\providecommand \BibitemOpen [0]{}%
\providecommand \bibitemStop [0]{}%
\providecommand \bibitemNoStop [0]{.\EOS\space}%
\providecommand \EOS [0]{\spacefactor3000\relax}%
\providecommand \BibitemShut  [1]{\csname bibitem#1\endcsname}%
\let\auto@bib@innerbib\@empty
%</preamble>
\bibitem [{\citenamefont {Tomm}\ \emph {et~al.}(2021)\citenamefont {Tomm},
  \citenamefont {Javadi}, \citenamefont {Antoniadis}, \citenamefont {Najer},
  \citenamefont {L{\" o}bl}, \citenamefont {Korsch}, \citenamefont {Schott},
  \citenamefont {Valentin}, \citenamefont {Wieck}, \citenamefont {Ludwig},\
  and\ \citenamefont {Warburton}}]{tomm21}%
  \BibitemOpen
  \bibfield  {author} {\bibinfo {author} {\bibfnamefont {N.}~\bibnamefont
  {Tomm}}, \bibinfo {author} {\bibfnamefont {A.}~\bibnamefont {Javadi}},
  \bibinfo {author} {\bibfnamefont {N.~O.}\ \bibnamefont {Antoniadis}},
  \bibinfo {author} {\bibfnamefont {D.}~\bibnamefont {Najer}}, \bibinfo
  {author} {\bibfnamefont {M.~C.}\ \bibnamefont {L{\" o}bl}}, \bibinfo {author}
  {\bibfnamefont {A.~R.}\ \bibnamefont {Korsch}}, \bibinfo {author}
  {\bibfnamefont {R.}~\bibnamefont {Schott}}, \bibinfo {author} {\bibfnamefont
  {S.~R.}\ \bibnamefont {Valentin}}, \bibinfo {author} {\bibfnamefont {A.~D.}\
  \bibnamefont {Wieck}}, \bibinfo {author} {\bibfnamefont {A.}~\bibnamefont
  {Ludwig}},\ and\ \bibinfo {author} {\bibfnamefont {R.~J.}\ \bibnamefont
  {Warburton}},\ }\bibfield  {title} {\bibinfo {title} {{A bright and fast
  source of coherent single photons}},\ }\href
  {https://doi.org/10.1038/s41565-020-00831-x} {\bibfield  {journal} {\bibinfo
  {journal} {Nat. Nanotech.}\ }\textbf {\bibinfo {volume} {16}},\ \bibinfo
  {pages} {399} (\bibinfo {year} {2021})}\BibitemShut {NoStop}%
\bibitem [{\citenamefont {Thomas}\ \emph {et~al.}(2021)\citenamefont {Thomas},
  \citenamefont {Billard}, \citenamefont {Coste}, \citenamefont {Wein},
  \citenamefont {Priya}, \citenamefont {Ollivier}, \citenamefont {Krebs},
  \citenamefont {Taza\"{\i}rt}, \citenamefont {Harouri}, \citenamefont
  {Lemaitre}, \citenamefont {Sagnes}, \citenamefont {Anton}, \citenamefont
  {Lanco}, \citenamefont {Somaschi}, \citenamefont {Loredo},\ and\
  \citenamefont {Senellart}}]{thomas21}%
  \BibitemOpen
  \bibfield  {author} {\bibinfo {author} {\bibfnamefont {S.~E.}\ \bibnamefont
  {Thomas}}, \bibinfo {author} {\bibfnamefont {M.}~\bibnamefont {Billard}},
  \bibinfo {author} {\bibfnamefont {N.}~\bibnamefont {Coste}}, \bibinfo
  {author} {\bibfnamefont {S.~C.}\ \bibnamefont {Wein}}, \bibinfo {author}
  {\bibnamefont {Priya}}, \bibinfo {author} {\bibfnamefont {H.}~\bibnamefont
  {Ollivier}}, \bibinfo {author} {\bibfnamefont {O.}~\bibnamefont {Krebs}},
  \bibinfo {author} {\bibfnamefont {L.}~\bibnamefont {Taza\"{\i}rt}}, \bibinfo
  {author} {\bibfnamefont {A.}~\bibnamefont {Harouri}}, \bibinfo {author}
  {\bibfnamefont {A.}~\bibnamefont {Lemaitre}}, \bibinfo {author}
  {\bibfnamefont {I.}~\bibnamefont {Sagnes}}, \bibinfo {author} {\bibfnamefont
  {C.}~\bibnamefont {Anton}}, \bibinfo {author} {\bibfnamefont
  {L.}~\bibnamefont {Lanco}}, \bibinfo {author} {\bibfnamefont
  {N.}~\bibnamefont {Somaschi}}, \bibinfo {author} {\bibfnamefont {J.~C.}\
  \bibnamefont {Loredo}},\ and\ \bibinfo {author} {\bibfnamefont
  {P.}~\bibnamefont {Senellart}},\ }\bibfield  {title} {\bibinfo {title}
  {Bright polarized single-photon source based on a linear dipole},\ }\href
  {https://doi.org/10.1103/PhysRevLett.126.233601} {\bibfield  {journal}
  {\bibinfo  {journal} {Phys. Rev. Lett.}\ }\textbf {\bibinfo {volume} {126}},\
  \bibinfo {pages} {233601} (\bibinfo {year} {2021})}\BibitemShut {NoStop}%
\bibitem [{\citenamefont {Wang}\ \emph
  {et~al.}(2019{\natexlab{a}})\citenamefont {Wang}, \citenamefont {He},
  \citenamefont {Chung}, \citenamefont {Hu}, \citenamefont {Yu}, \citenamefont
  {Chen}, \citenamefont {Ding}, \citenamefont {Chen}, \citenamefont {Qin},
  \citenamefont {Yang}, \citenamefont {Liu}, \citenamefont {Duan},
  \citenamefont {Li}, \citenamefont {Gerhardt}, \citenamefont {Winkler},
  \citenamefont {Jurkat}, \citenamefont {Wang}, \citenamefont {Gregersen},
  \citenamefont {Huo}, \citenamefont {Dai}, \citenamefont {Yu}, \citenamefont
  {H{\ifmmode\ddot{o}\else\"{o}\fi}fling}, \citenamefont {Lu},\ and\
  \citenamefont {Pan}}]{Wang2019Nov}%
  \BibitemOpen
  \bibfield  {author} {\bibinfo {author} {\bibfnamefont {H.}~\bibnamefont
  {Wang}}, \bibinfo {author} {\bibfnamefont {Y.-M.}\ \bibnamefont {He}},
  \bibinfo {author} {\bibfnamefont {T.-H.}\ \bibnamefont {Chung}}, \bibinfo
  {author} {\bibfnamefont {H.}~\bibnamefont {Hu}}, \bibinfo {author}
  {\bibfnamefont {Y.}~\bibnamefont {Yu}}, \bibinfo {author} {\bibfnamefont
  {S.}~\bibnamefont {Chen}}, \bibinfo {author} {\bibfnamefont {X.}~\bibnamefont
  {Ding}}, \bibinfo {author} {\bibfnamefont {M.-C.}\ \bibnamefont {Chen}},
  \bibinfo {author} {\bibfnamefont {J.}~\bibnamefont {Qin}}, \bibinfo {author}
  {\bibfnamefont {X.}~\bibnamefont {Yang}}, \bibinfo {author} {\bibfnamefont
  {R.-Z.}\ \bibnamefont {Liu}}, \bibinfo {author} {\bibfnamefont {Z.-C.}\
  \bibnamefont {Duan}}, \bibinfo {author} {\bibfnamefont {J.-P.}\ \bibnamefont
  {Li}}, \bibinfo {author} {\bibfnamefont {S.}~\bibnamefont {Gerhardt}},
  \bibinfo {author} {\bibfnamefont {K.}~\bibnamefont {Winkler}}, \bibinfo
  {author} {\bibfnamefont {J.}~\bibnamefont {Jurkat}}, \bibinfo {author}
  {\bibfnamefont {L.-J.}\ \bibnamefont {Wang}}, \bibinfo {author}
  {\bibfnamefont {N.}~\bibnamefont {Gregersen}}, \bibinfo {author}
  {\bibfnamefont {Y.-H.}\ \bibnamefont {Huo}}, \bibinfo {author} {\bibfnamefont
  {Q.}~\bibnamefont {Dai}}, \bibinfo {author} {\bibfnamefont {S.}~\bibnamefont
  {Yu}}, \bibinfo {author} {\bibfnamefont {S.}~\bibnamefont
  {H{\ifmmode\ddot{o}\else\"{o}\fi}fling}}, \bibinfo {author} {\bibfnamefont
  {C.-Y.}\ \bibnamefont {Lu}},\ and\ \bibinfo {author} {\bibfnamefont {J.-W.}\
  \bibnamefont {Pan}},\ }\bibfield  {title} {\bibinfo {title} {{Towards optimal
  single-photon sources from polarized microcavities}},\ }\href
  {https://doi.org/10.1038/s41566-019-0494-3} {\bibfield  {journal} {\bibinfo
  {journal} {Nat. Photonics}\ }\textbf {\bibinfo {volume} {13}},\ \bibinfo
  {pages} {770} (\bibinfo {year} {2019}{\natexlab{a}})}\BibitemShut {NoStop}%
\bibitem [{\citenamefont {Somaschi}\ \emph {et~al.}(2016)\citenamefont
  {Somaschi}, \citenamefont {Giesz}, \citenamefont {De~Santis}, \citenamefont
  {Loredo}, \citenamefont {Almeida}, \citenamefont {Hornecker}, \citenamefont
  {Portalupi}, \citenamefont {Grange}, \citenamefont {Ant{\' o}n},
  \citenamefont {Demory}, \citenamefont {G{\' o}mez}, \citenamefont {Sagnes},
  \citenamefont {Lanzillotti-Kimura}, \citenamefont {Lema{\' i}tre},
  \citenamefont {Auffeves}, \citenamefont {White}, \citenamefont {Lanco},\ and\
  \citenamefont {Senellart}}]{somaschi16}%
  \BibitemOpen
  \bibfield  {author} {\bibinfo {author} {\bibfnamefont {N.}~\bibnamefont
  {Somaschi}}, \bibinfo {author} {\bibfnamefont {V.}~\bibnamefont {Giesz}},
  \bibinfo {author} {\bibfnamefont {L.}~\bibnamefont {De~Santis}}, \bibinfo
  {author} {\bibfnamefont {J.~C.}\ \bibnamefont {Loredo}}, \bibinfo {author}
  {\bibfnamefont {M.~P.}\ \bibnamefont {Almeida}}, \bibinfo {author}
  {\bibfnamefont {G.}~\bibnamefont {Hornecker}}, \bibinfo {author}
  {\bibfnamefont {S.~L.}\ \bibnamefont {Portalupi}}, \bibinfo {author}
  {\bibfnamefont {T.}~\bibnamefont {Grange}}, \bibinfo {author} {\bibfnamefont
  {C.}~\bibnamefont {Ant{\' o}n}}, \bibinfo {author} {\bibfnamefont
  {J.}~\bibnamefont {Demory}}, \bibinfo {author} {\bibfnamefont
  {C.}~\bibnamefont {G{\' o}mez}}, \bibinfo {author} {\bibfnamefont
  {I.}~\bibnamefont {Sagnes}}, \bibinfo {author} {\bibfnamefont {N.~D.}\
  \bibnamefont {Lanzillotti-Kimura}}, \bibinfo {author} {\bibfnamefont
  {A.}~\bibnamefont {Lema{\' i}tre}}, \bibinfo {author} {\bibfnamefont
  {A.}~\bibnamefont {Auffeves}}, \bibinfo {author} {\bibfnamefont {A.~G.}\
  \bibnamefont {White}}, \bibinfo {author} {\bibfnamefont {L.}~\bibnamefont
  {Lanco}},\ and\ \bibinfo {author} {\bibfnamefont {P.}~\bibnamefont
  {Senellart}},\ }\bibfield  {title} {\bibinfo {title} {{Near-optimal
  single-photon sources in the solid state}},\ }\href
  {https://doi.org/10.1038/nphoton.2016.23} {\bibfield  {journal} {\bibinfo
  {journal} {Nat. Photon.}\ }\textbf {\bibinfo {volume} {10}},\ \bibinfo
  {pages} {340} (\bibinfo {year} {2016})}\BibitemShut {NoStop}%
\bibitem [{\citenamefont {Hanschke}\ \emph {et~al.}(2018)\citenamefont
  {Hanschke}, \citenamefont {Fischer}, \citenamefont {Appel}, \citenamefont
  {Lukin}, \citenamefont {Wierzbowski}, \citenamefont {Sun}, \citenamefont
  {Trivedi}, \citenamefont {Vu{\v{c}}kovi{\'{c}}}, \citenamefont {Finley},\
  and\ \citenamefont {M{\"{u}}ller}}]{Hanschke2018}%
  \BibitemOpen
  \bibfield  {author} {\bibinfo {author} {\bibfnamefont {L.}~\bibnamefont
  {Hanschke}}, \bibinfo {author} {\bibfnamefont {K.~A.}\ \bibnamefont
  {Fischer}}, \bibinfo {author} {\bibfnamefont {S.}~\bibnamefont {Appel}},
  \bibinfo {author} {\bibfnamefont {D.}~\bibnamefont {Lukin}}, \bibinfo
  {author} {\bibfnamefont {J.}~\bibnamefont {Wierzbowski}}, \bibinfo {author}
  {\bibfnamefont {S.}~\bibnamefont {Sun}}, \bibinfo {author} {\bibfnamefont
  {R.}~\bibnamefont {Trivedi}}, \bibinfo {author} {\bibfnamefont
  {J.}~\bibnamefont {Vu{\v{c}}kovi{\'{c}}}}, \bibinfo {author} {\bibfnamefont
  {J.~J.}\ \bibnamefont {Finley}},\ and\ \bibinfo {author} {\bibfnamefont
  {K.}~\bibnamefont {M{\"{u}}ller}},\ }\bibfield  {title} {\bibinfo {title}
  {{Quantum dot single-photon sources with ultra-low multi-photon
  probability}},\ }\href {https://doi.org/10.1038/s41534-018-0092-0} {\bibfield
   {journal} {\bibinfo  {journal} {npj Quantum Information}\ }\textbf {\bibinfo
  {volume} {4}},\ \bibinfo {pages} {43} (\bibinfo {year} {2018})}\BibitemShut
  {NoStop}%
\bibitem [{\citenamefont {Sbresny}\ \emph {et~al.}(2022)\citenamefont
  {Sbresny}, \citenamefont {Hanschke}, \citenamefont
  {Sch{\ifmmode\ddot{o}\else\"{o}\fi}ll}, \citenamefont {Rauhaus},
  \citenamefont {Scaparra}, \citenamefont {Boos}, \citenamefont
  {Zubizarreta~Casalengua}, \citenamefont {Riedl}, \citenamefont {del Valle},
  \citenamefont {Finley}, \citenamefont {J{\ifmmode\ddot{o}\else\"{o}\fi}ns},\
  and\ \citenamefont {M{\ifmmode\ddot{u}\else\"{u}\fi}ller}}]{Sbresny2022Mar}%
  \BibitemOpen
  \bibfield  {author} {\bibinfo {author} {\bibfnamefont {F.}~\bibnamefont
  {Sbresny}}, \bibinfo {author} {\bibfnamefont {L.}~\bibnamefont {Hanschke}},
  \bibinfo {author} {\bibfnamefont {E.}~\bibnamefont
  {Sch{\ifmmode\ddot{o}\else\"{o}\fi}ll}}, \bibinfo {author} {\bibfnamefont
  {W.}~\bibnamefont {Rauhaus}}, \bibinfo {author} {\bibfnamefont
  {B.}~\bibnamefont {Scaparra}}, \bibinfo {author} {\bibfnamefont
  {K.}~\bibnamefont {Boos}}, \bibinfo {author} {\bibfnamefont {E.}~\bibnamefont
  {Zubizarreta~Casalengua}}, \bibinfo {author} {\bibfnamefont {H.}~\bibnamefont
  {Riedl}}, \bibinfo {author} {\bibfnamefont {E.}~\bibnamefont {del Valle}},
  \bibinfo {author} {\bibfnamefont {J.~J.}\ \bibnamefont {Finley}}, \bibinfo
  {author} {\bibfnamefont {K.~D.}\ \bibnamefont
  {J{\ifmmode\ddot{o}\else\"{o}\fi}ns}},\ and\ \bibinfo {author} {\bibfnamefont
  {K.}~\bibnamefont {M{\ifmmode\ddot{u}\else\"{u}\fi}ller}},\ }\bibfield
  {title} {\bibinfo {title} {{Stimulated Generation of Indistinguishable Single
  Photons from a Quantum Ladder System}},\ }\href
  {https://doi.org/10.1103/PhysRevLett.128.093603} {\bibfield  {journal}
  {\bibinfo  {journal} {Phys. Rev. Lett.}\ }\textbf {\bibinfo {volume} {128}},\
  \bibinfo {pages} {093603} (\bibinfo {year} {2022})}\BibitemShut {NoStop}%
\bibitem [{\citenamefont {Wang}\ \emph
  {et~al.}(2019{\natexlab{b}})\citenamefont {Wang}, \citenamefont {Qin},
  \citenamefont {Ding}, \citenamefont {Chen}, \citenamefont {Chen},
  \citenamefont {You}, \citenamefont {He}, \citenamefont {Jiang}, \citenamefont
  {You}, \citenamefont {Wang}, \citenamefont {Schneider}, \citenamefont
  {Renema}, \citenamefont {H\"ofling}, \citenamefont {Lu},\ and\ \citenamefont
  {Pan}}]{wang2019}%
  \BibitemOpen
  \bibfield  {author} {\bibinfo {author} {\bibfnamefont {H.}~\bibnamefont
  {Wang}}, \bibinfo {author} {\bibfnamefont {J.}~\bibnamefont {Qin}}, \bibinfo
  {author} {\bibfnamefont {X.}~\bibnamefont {Ding}}, \bibinfo {author}
  {\bibfnamefont {M.-C.}\ \bibnamefont {Chen}}, \bibinfo {author}
  {\bibfnamefont {S.}~\bibnamefont {Chen}}, \bibinfo {author} {\bibfnamefont
  {X.}~\bibnamefont {You}}, \bibinfo {author} {\bibfnamefont {Y.-M.}\
  \bibnamefont {He}}, \bibinfo {author} {\bibfnamefont {X.}~\bibnamefont
  {Jiang}}, \bibinfo {author} {\bibfnamefont {L.}~\bibnamefont {You}}, \bibinfo
  {author} {\bibfnamefont {Z.}~\bibnamefont {Wang}}, \bibinfo {author}
  {\bibfnamefont {C.}~\bibnamefont {Schneider}}, \bibinfo {author}
  {\bibfnamefont {J.~J.}\ \bibnamefont {Renema}}, \bibinfo {author}
  {\bibfnamefont {S.}~\bibnamefont {H\"ofling}}, \bibinfo {author}
  {\bibfnamefont {C.-Y.}\ \bibnamefont {Lu}},\ and\ \bibinfo {author}
  {\bibfnamefont {J.-W.}\ \bibnamefont {Pan}},\ }\bibfield  {title} {\bibinfo
  {title} {Boson sampling with 20 input photons and a 60-mode interferometer in
  a $1{0}^{14}$-dimensional hilbert space},\ }\href
  {https://doi.org/10.1103/PhysRevLett.123.250503} {\bibfield  {journal}
  {\bibinfo  {journal} {Phys. Rev. Lett.}\ }\textbf {\bibinfo {volume} {123}},\
  \bibinfo {pages} {250503} (\bibinfo {year} {2019}{\natexlab{b}})}\BibitemShut
  {NoStop}%
\bibitem [{\citenamefont {Basso~Basset}\ \emph {et~al.}(2021)\citenamefont
  {Basso~Basset}, \citenamefont {Valeri}, \citenamefont {Roccia}, \citenamefont
  {Muredda}, \citenamefont {Poderini}, \citenamefont {Neuwirth}, \citenamefont
  {Spagnolo}, \citenamefont {Rota}, \citenamefont {Carvacho}, \citenamefont
  {Sciarrino},\ and\ \citenamefont {Trotta}}]{francesco2021}%
  \BibitemOpen
  \bibfield  {author} {\bibinfo {author} {\bibfnamefont {F.}~\bibnamefont
  {Basso~Basset}}, \bibinfo {author} {\bibfnamefont {M.}~\bibnamefont
  {Valeri}}, \bibinfo {author} {\bibfnamefont {E.}~\bibnamefont {Roccia}},
  \bibinfo {author} {\bibfnamefont {V.}~\bibnamefont {Muredda}}, \bibinfo
  {author} {\bibfnamefont {D.}~\bibnamefont {Poderini}}, \bibinfo {author}
  {\bibfnamefont {J.}~\bibnamefont {Neuwirth}}, \bibinfo {author}
  {\bibfnamefont {N.}~\bibnamefont {Spagnolo}}, \bibinfo {author}
  {\bibfnamefont {M.~B.}\ \bibnamefont {Rota}}, \bibinfo {author}
  {\bibfnamefont {G.}~\bibnamefont {Carvacho}}, \bibinfo {author}
  {\bibfnamefont {F.}~\bibnamefont {Sciarrino}},\ and\ \bibinfo {author}
  {\bibfnamefont {R.}~\bibnamefont {Trotta}},\ }\bibfield  {title} {\bibinfo
  {title} {{Quantum key distribution with entangled photons generated on demand
  by a quantum dot}},\ }\href {https://doi.org/10.1126/sciadv.abe6379}
  {\bibfield  {journal} {\bibinfo  {journal} {Science Advances}\ }\textbf
  {\bibinfo {volume} {7}},\ \bibinfo {pages} {eabe6379} (\bibinfo {year}
  {2021})}\BibitemShut {NoStop}%
\bibitem [{\citenamefont {Kupko}\ \emph {et~al.}(2020)\citenamefont {Kupko},
  \citenamefont {Helverson}, \citenamefont {Rickert}, \citenamefont {Schulze},
  \citenamefont {Strittmatter}, \citenamefont {Gschrey}, \citenamefont {Rodt},
  \citenamefont {Reitzenstein},\ and\ \citenamefont {Heindel}}]{kupko2020}%
  \BibitemOpen
  \bibfield  {author} {\bibinfo {author} {\bibfnamefont {T.}~\bibnamefont
  {Kupko}}, \bibinfo {author} {\bibfnamefont {M.~v.}\ \bibnamefont
  {Helverson}}, \bibinfo {author} {\bibfnamefont {L.}~\bibnamefont {Rickert}},
  \bibinfo {author} {\bibfnamefont {J.-H.}\ \bibnamefont {Schulze}}, \bibinfo
  {author} {\bibfnamefont {A.}~\bibnamefont {Strittmatter}}, \bibinfo {author}
  {\bibfnamefont {M.}~\bibnamefont {Gschrey}}, \bibinfo {author} {\bibfnamefont
  {S.}~\bibnamefont {Rodt}}, \bibinfo {author} {\bibfnamefont {S.}~\bibnamefont
  {Reitzenstein}},\ and\ \bibinfo {author} {\bibfnamefont {T.}~\bibnamefont
  {Heindel}},\ }\bibfield  {title} {\bibinfo {title} {{Tools for the
  performance optimization of single-photon quantum key distribution}},\ }\href
  {https://doi.org/10.1038/s41534-020-0262-8} {\bibfield  {journal} {\bibinfo
  {journal} {npj Quantum Inf.}\ }\textbf {\bibinfo {volume} {6}},\ \bibinfo
  {pages} {29} (\bibinfo {year} {2020})}\BibitemShut {NoStop}%
\bibitem [{\citenamefont {Ko\l{}ody{\'n}ski}\ \emph {et~al.}(2020)\citenamefont
  {Ko\l{}ody{\'n}ski}, \citenamefont {M{\'a}ttar}, \citenamefont {Skrzypczyk},
  \citenamefont {Woodhead}, \citenamefont {Cavalcanti}, \citenamefont
  {Banaszek},\ and\ \citenamefont {Ac{\' i}n}}]{kolodynski2020}%
  \BibitemOpen
  \bibfield  {author} {\bibinfo {author} {\bibfnamefont {J.}~\bibnamefont
  {Ko\l{}ody{\'n}ski}}, \bibinfo {author} {\bibfnamefont {A.}~\bibnamefont
  {M{\'a}ttar}}, \bibinfo {author} {\bibfnamefont {P.}~\bibnamefont
  {Skrzypczyk}}, \bibinfo {author} {\bibfnamefont {E.}~\bibnamefont
  {Woodhead}}, \bibinfo {author} {\bibfnamefont {D.}~\bibnamefont
  {Cavalcanti}}, \bibinfo {author} {\bibfnamefont {K.}~\bibnamefont
  {Banaszek}},\ and\ \bibinfo {author} {\bibfnamefont {A.}~\bibnamefont {Ac{\'
  i}n}},\ }\bibfield  {title} {\bibinfo {title} {{Device-independent quantum
  key distribution with single-photon sources}},\ }\href
  {https://doi.org/10.22331/q-2020-04-30-260} {\bibfield  {journal} {\bibinfo
  {journal} {Quantum}\ }\textbf {\bibinfo {volume} {4}},\ \bibinfo {pages}
  {260} (\bibinfo {year} {2020})}\BibitemShut {NoStop}%
\bibitem [{\citenamefont {Jennewein}\ \emph {et~al.}(2011)\citenamefont
  {Jennewein}, \citenamefont {Barbieri},\ and\ \citenamefont
  {White}}]{Jennewein2011Feb}%
  \BibitemOpen
  \bibfield  {author} {\bibinfo {author} {\bibfnamefont {T.}~\bibnamefont
  {Jennewein}}, \bibinfo {author} {\bibfnamefont {M.}~\bibnamefont
  {Barbieri}},\ and\ \bibinfo {author} {\bibfnamefont {A.~G.}\ \bibnamefont
  {White}},\ }\bibfield  {title} {\bibinfo {title} {{Single-photon device
  requirements for operating linear optics quantum computing outside the
  post-selection basis}},\ }\href
  {https://doi.org/10.1080/09500340.2010.546894} {\bibfield  {journal}
  {\bibinfo  {journal} {J. Mod. Opt.}\ }\textbf {\bibinfo {volume} {58}},\
  \bibinfo {pages} {276} (\bibinfo {year} {2011})}\BibitemShut {NoStop}%
\bibitem [{\citenamefont {Varnava}\ \emph {et~al.}(2008)\citenamefont
  {Varnava}, \citenamefont {Browne},\ and\ \citenamefont
  {Rudolph}}]{Varnava2008}%
  \BibitemOpen
  \bibfield  {author} {\bibinfo {author} {\bibfnamefont {M.}~\bibnamefont
  {Varnava}}, \bibinfo {author} {\bibfnamefont {D.~E.}\ \bibnamefont
  {Browne}},\ and\ \bibinfo {author} {\bibfnamefont {T.}~\bibnamefont
  {Rudolph}},\ }\bibfield  {title} {\bibinfo {title} {{How Good Must Single
  Photon Sources and Detectors Be for Efficient Linear Optical Quantum
  Computation?}},\ }\href {https://doi.org/10.1103/PhysRevLett.100.060502}
  {\bibfield  {journal} {\bibinfo  {journal} {Phys. Rev. Lett.}\ }\textbf
  {\bibinfo {volume} {100}},\ \bibinfo {pages} {060502} (\bibinfo {year}
  {2008})}\BibitemShut {NoStop}%
\bibitem [{\citenamefont {Uppu}\ \emph
  {et~al.}(2020{\natexlab{a}})\citenamefont {Uppu}, \citenamefont {Eriksen},
  \citenamefont {Thyrrestrup}, \citenamefont {U\u{g}urlu}, \citenamefont
  {Wang}, \citenamefont {Scholz}, \citenamefont {Wieck}, \citenamefont {Arne},
  \citenamefont {L{\" o}bl}, \citenamefont {Warburton}, \citenamefont
  {Lodahl},\ and\ \citenamefont {Midolo}}]{uppu2020_2}%
  \BibitemOpen
  \bibfield  {author} {\bibinfo {author} {\bibfnamefont {R.}~\bibnamefont
  {Uppu}}, \bibinfo {author} {\bibfnamefont {H.~T.}\ \bibnamefont {Eriksen}},
  \bibinfo {author} {\bibfnamefont {H.}~\bibnamefont {Thyrrestrup}}, \bibinfo
  {author} {\bibfnamefont {A.~D.}\ \bibnamefont {U\u{g}urlu}}, \bibinfo
  {author} {\bibfnamefont {Y.}~\bibnamefont {Wang}}, \bibinfo {author}
  {\bibfnamefont {S.}~\bibnamefont {Scholz}}, \bibinfo {author} {\bibfnamefont
  {A.~D.}\ \bibnamefont {Wieck}}, \bibinfo {author} {\bibfnamefont
  {L.}~\bibnamefont {Arne}}, \bibinfo {author} {\bibfnamefont {M.~C.}\
  \bibnamefont {L{\" o}bl}}, \bibinfo {author} {\bibfnamefont {R.~J.}\
  \bibnamefont {Warburton}}, \bibinfo {author} {\bibfnamefont {P.}~\bibnamefont
  {Lodahl}},\ and\ \bibinfo {author} {\bibfnamefont {L.}~\bibnamefont
  {Midolo}},\ }\bibfield  {title} {\bibinfo {title} {{On-chip deterministic
  operation of quantum dots in dual-mode waveguides for a plug-and-play
  single-photon source}},\ }\href {https://doi.org/10.1038/s41467-020-17603-9}
  {\bibfield  {journal} {\bibinfo  {journal} {Nat. Commun.}\ }\textbf {\bibinfo
  {volume} {11}},\ \bibinfo {pages} {3782} (\bibinfo {year}
  {2020}{\natexlab{a}})}\BibitemShut {NoStop}%
\bibitem [{\citenamefont {Uppu}\ \emph
  {et~al.}(2020{\natexlab{b}})\citenamefont {Uppu}, \citenamefont {Pedersen},
  \citenamefont {Wang}, \citenamefont {Olesen}, \citenamefont {Papon},
  \citenamefont {Zhou}, \citenamefont {Midolo}, \citenamefont {Scholz},
  \citenamefont {Wieck}, \citenamefont {Ludwig},\ and\ \citenamefont
  {Lodahl}}]{uppu2020}%
  \BibitemOpen
  \bibfield  {author} {\bibinfo {author} {\bibfnamefont {R.}~\bibnamefont
  {Uppu}}, \bibinfo {author} {\bibfnamefont {F.~T.}\ \bibnamefont {Pedersen}},
  \bibinfo {author} {\bibfnamefont {Y.}~\bibnamefont {Wang}}, \bibinfo {author}
  {\bibfnamefont {C.~T.}\ \bibnamefont {Olesen}}, \bibinfo {author}
  {\bibfnamefont {C.}~\bibnamefont {Papon}}, \bibinfo {author} {\bibfnamefont
  {X.}~\bibnamefont {Zhou}}, \bibinfo {author} {\bibfnamefont {L.}~\bibnamefont
  {Midolo}}, \bibinfo {author} {\bibfnamefont {S.}~\bibnamefont {Scholz}},
  \bibinfo {author} {\bibfnamefont {A.~D.}\ \bibnamefont {Wieck}}, \bibinfo
  {author} {\bibfnamefont {A.}~\bibnamefont {Ludwig}},\ and\ \bibinfo {author}
  {\bibfnamefont {P.}~\bibnamefont {Lodahl}},\ }\bibfield  {title} {\bibinfo
  {title} {{Scalable integrated single-photon source}},\ }\href
  {https://doi.org/10.1126/sciadv.abc8268} {\bibfield  {journal} {\bibinfo
  {journal} {Science Advances}\ }\textbf {\bibinfo {volume} {6}},\ \bibinfo
  {pages} {eabc8268} (\bibinfo {year} {2020}{\natexlab{b}})}\BibitemShut
  {NoStop}%
\bibitem [{\citenamefont {Dusanowski}\ \emph {et~al.}(2019)\citenamefont
  {Dusanowski}, \citenamefont {Kwon}, \citenamefont {Schneider},\ and\
  \citenamefont {H{\"{o}}fling}}]{Dusanowski2019}%
  \BibitemOpen
  \bibfield  {author} {\bibinfo {author} {\bibfnamefont {{\L}.}~\bibnamefont
  {Dusanowski}}, \bibinfo {author} {\bibfnamefont {S.-H.}\ \bibnamefont
  {Kwon}}, \bibinfo {author} {\bibfnamefont {C.}~\bibnamefont {Schneider}},\
  and\ \bibinfo {author} {\bibfnamefont {S.}~\bibnamefont {H{\"{o}}fling}},\
  }\bibfield  {title} {\bibinfo {title} {{Near-Unity Indistinguishability
  Single Photon Source for Large-Scale Integrated Quantum Optics}},\ }\href
  {https://doi.org/10.1103/PhysRevLett.122.173602} {\bibfield  {journal}
  {\bibinfo  {journal} {Physical Review Letters}\ }\textbf {\bibinfo {volume}
  {122}},\ \bibinfo {pages} {173602} (\bibinfo {year} {2019})}\BibitemShut
  {NoStop}%
\bibitem [{\citenamefont {Ollivier}\ \emph {et~al.}(2020)\citenamefont
  {Ollivier}, \citenamefont {de~Buy~Wenniger}, \citenamefont {Thomas},
  \citenamefont {Wein}, \citenamefont {Harouri}, \citenamefont {Coppola},
  \citenamefont {Hilaire}, \citenamefont {Millet}, \citenamefont
  {Lema{\ifmmode\hat{\imath}\else\^{\i}\fi}tre}, \citenamefont {Sagnes},
  \citenamefont {Krebs}, \citenamefont {Lanco}, \citenamefont {Loredo},
  \citenamefont {Ant{\ifmmode\acute{o}\else\'{o}\fi}n}, \citenamefont
  {Somaschi},\ and\ \citenamefont {Senellart}}]{Ollivier2020Mar}%
  \BibitemOpen
  \bibfield  {author} {\bibinfo {author} {\bibfnamefont {H.}~\bibnamefont
  {Ollivier}}, \bibinfo {author} {\bibfnamefont {I.~M.}\ \bibnamefont
  {de~Buy~Wenniger}}, \bibinfo {author} {\bibfnamefont {S.}~\bibnamefont
  {Thomas}}, \bibinfo {author} {\bibfnamefont {S.~C.}\ \bibnamefont {Wein}},
  \bibinfo {author} {\bibfnamefont {A.}~\bibnamefont {Harouri}}, \bibinfo
  {author} {\bibfnamefont {G.}~\bibnamefont {Coppola}}, \bibinfo {author}
  {\bibfnamefont {P.}~\bibnamefont {Hilaire}}, \bibinfo {author} {\bibfnamefont
  {C.}~\bibnamefont {Millet}}, \bibinfo {author} {\bibfnamefont
  {A.}~\bibnamefont {Lema{\ifmmode\hat{\imath}\else\^{\i}\fi}tre}}, \bibinfo
  {author} {\bibfnamefont {I.}~\bibnamefont {Sagnes}}, \bibinfo {author}
  {\bibfnamefont {O.}~\bibnamefont {Krebs}}, \bibinfo {author} {\bibfnamefont
  {L.}~\bibnamefont {Lanco}}, \bibinfo {author} {\bibfnamefont {J.~C.}\
  \bibnamefont {Loredo}}, \bibinfo {author} {\bibfnamefont {C.}~\bibnamefont
  {Ant{\ifmmode\acute{o}\else\'{o}\fi}n}}, \bibinfo {author} {\bibfnamefont
  {N.}~\bibnamefont {Somaschi}},\ and\ \bibinfo {author} {\bibfnamefont
  {P.}~\bibnamefont {Senellart}},\ }\bibfield  {title} {\bibinfo {title}
  {{Reproducibility of High-Performance Quantum Dot Single-Photon Sources}},\
  }\bibfield  {journal} {\bibinfo  {journal} {ACS Photonics}\ }\href
  {https://doi.org/10.1021/acsphotonics.9b01805} {10.1021/acsphotonics.9b01805}
  (\bibinfo {year} {2020})\BibitemShut {NoStop}%
\bibitem [{\citenamefont {Lee}\ \emph {et~al.}(2020)\citenamefont {Lee},
  \citenamefont {Leong}, \citenamefont {Kalashnikov}, \citenamefont {Dai},
  \citenamefont {Gandhi},\ and\ \citenamefont {Krivitsky}}]{Lee2020}%
  \BibitemOpen
  \bibfield  {author} {\bibinfo {author} {\bibfnamefont {J.}~\bibnamefont
  {Lee}}, \bibinfo {author} {\bibfnamefont {V.}~\bibnamefont {Leong}}, \bibinfo
  {author} {\bibfnamefont {D.}~\bibnamefont {Kalashnikov}}, \bibinfo {author}
  {\bibfnamefont {J.}~\bibnamefont {Dai}}, \bibinfo {author} {\bibfnamefont
  {A.}~\bibnamefont {Gandhi}},\ and\ \bibinfo {author} {\bibfnamefont {L.~A.}\
  \bibnamefont {Krivitsky}},\ }\bibfield  {title} {\bibinfo {title}
  {{Integrated single photon emitters}},\ }\href
  {https://doi.org/10.1116/5.0011316} {\bibfield  {journal} {\bibinfo
  {journal} {AVS Quantum Sci.}\ }\textbf {\bibinfo {volume} {2}},\ \bibinfo
  {pages} {031701} (\bibinfo {year} {2020})}\BibitemShut {NoStop}%
\bibitem [{\citenamefont {Schnauber}\ \emph {et~al.}(2021)\citenamefont
  {Schnauber}, \citenamefont {Gro{\ss}e}, \citenamefont {Kaganskiy},
  \citenamefont {Ott}, \citenamefont {Anikin}, \citenamefont {Schmidt},
  \citenamefont {Rodt},\ and\ \citenamefont {Reitzenstein}}]{Schnauber2021}%
  \BibitemOpen
  \bibfield  {author} {\bibinfo {author} {\bibfnamefont {P.}~\bibnamefont
  {Schnauber}}, \bibinfo {author} {\bibfnamefont {J.}~\bibnamefont
  {Gro{\ss}e}}, \bibinfo {author} {\bibfnamefont {A.}~\bibnamefont
  {Kaganskiy}}, \bibinfo {author} {\bibfnamefont {M.}~\bibnamefont {Ott}},
  \bibinfo {author} {\bibfnamefont {P.}~\bibnamefont {Anikin}}, \bibinfo
  {author} {\bibfnamefont {R.}~\bibnamefont {Schmidt}}, \bibinfo {author}
  {\bibfnamefont {S.}~\bibnamefont {Rodt}},\ and\ \bibinfo {author}
  {\bibfnamefont {S.}~\bibnamefont {Reitzenstein}},\ }\bibfield  {title}
  {\bibinfo {title} {{Spectral control of deterministically fabricated quantum
  dot waveguide systems using the quantum confined Stark effect}},\ }\href
  {https://doi.org/10.1063/5.0050152} {\bibfield  {journal} {\bibinfo
  {journal} {APL Photonics}\ }\textbf {\bibinfo {volume} {6}},\ \bibinfo
  {pages} {050801} (\bibinfo {year} {2021})}\BibitemShut {NoStop}%
\bibitem [{\citenamefont {Nowak}\ \emph {et~al.}(2014)\citenamefont {Nowak},
  \citenamefont {Portalupi}, \citenamefont {Giesz}, \citenamefont {Gazzano},
  \citenamefont {Dal~Savio}, \citenamefont {Braun}, \citenamefont {Karrai},
  \citenamefont {Arnold}, \citenamefont {Lanco}, \citenamefont {Sagnes},
  \citenamefont {Lema{\ifmmode\hat{\imath}\else\^{\i}\fi}tre},\ and\
  \citenamefont {Senellart}}]{Nowak2014Feb}%
  \BibitemOpen
  \bibfield  {author} {\bibinfo {author} {\bibfnamefont {A.~K.}\ \bibnamefont
  {Nowak}}, \bibinfo {author} {\bibfnamefont {S.~L.}\ \bibnamefont
  {Portalupi}}, \bibinfo {author} {\bibfnamefont {V.}~\bibnamefont {Giesz}},
  \bibinfo {author} {\bibfnamefont {O.}~\bibnamefont {Gazzano}}, \bibinfo
  {author} {\bibfnamefont {C.}~\bibnamefont {Dal~Savio}}, \bibinfo {author}
  {\bibfnamefont {P.-F.}\ \bibnamefont {Braun}}, \bibinfo {author}
  {\bibfnamefont {K.}~\bibnamefont {Karrai}}, \bibinfo {author} {\bibfnamefont
  {C.}~\bibnamefont {Arnold}}, \bibinfo {author} {\bibfnamefont
  {L.}~\bibnamefont {Lanco}}, \bibinfo {author} {\bibfnamefont
  {I.}~\bibnamefont {Sagnes}}, \bibinfo {author} {\bibfnamefont
  {A.}~\bibnamefont {Lema{\ifmmode\hat{\imath}\else\^{\i}\fi}tre}},\ and\
  \bibinfo {author} {\bibfnamefont {P.}~\bibnamefont {Senellart}},\ }\bibfield
  {title} {\bibinfo {title} {{Deterministic and electrically tunable bright
  single-photon source - Nature Communications}},\ }\href
  {https://doi.org/10.1038/ncomms4240} {\bibfield  {journal} {\bibinfo
  {journal} {Nat. Commun.}\ }\textbf {\bibinfo {volume} {5}},\ \bibinfo {pages}
  {1} (\bibinfo {year} {2014})}\BibitemShut {NoStop}%
\bibitem [{\citenamefont {Mocza{\l}a-Dusanowska}\ \emph
  {et~al.}(2020)\citenamefont {Mocza{\l}a-Dusanowska}, \citenamefont
  {Dusanowski}, \citenamefont {Iff}, \citenamefont {Huber}, \citenamefont
  {Kuhn}, \citenamefont {Czyszanowski}, \citenamefont {Schneider},\ and\
  \citenamefont {H{\"o}fling}}]{dusanowska2020}%
  \BibitemOpen
  \bibfield  {author} {\bibinfo {author} {\bibfnamefont {M.}~\bibnamefont
  {Mocza{\l}a-Dusanowska}}, \bibinfo {author} {\bibfnamefont
  {{\L}.}~\bibnamefont {Dusanowski}}, \bibinfo {author} {\bibfnamefont
  {O.}~\bibnamefont {Iff}}, \bibinfo {author} {\bibfnamefont {T.}~\bibnamefont
  {Huber}}, \bibinfo {author} {\bibfnamefont {S.}~\bibnamefont {Kuhn}},
  \bibinfo {author} {\bibfnamefont {T.}~\bibnamefont {Czyszanowski}}, \bibinfo
  {author} {\bibfnamefont {C.}~\bibnamefont {Schneider}},\ and\ \bibinfo
  {author} {\bibfnamefont {S.}~\bibnamefont {H{\"o}fling}},\ }\bibfield
  {title} {\bibinfo {title} {{Strain-Tunable Single-Photon Source Based on a
  Circular Bragg Grating Cavity with Embedded Quantum Dots}},\ }\bibfield
  {journal} {\bibinfo  {journal} {ACS Photonics}\ }\href
  {https://doi.org/10.1021/acsphotonics.0c01465} {10.1021/acsphotonics.0c01465}
  (\bibinfo {year} {2020})\BibitemShut {NoStop}%
\bibitem [{\citenamefont {Grim}\ \emph {et~al.}(2019)\citenamefont {Grim},
  \citenamefont {Bracker}, \citenamefont {Zalalutdinov}, \citenamefont
  {Carter}, \citenamefont {Kozen}, \citenamefont {Kim}, \citenamefont {Kim},
  \citenamefont {Mlack}, \citenamefont {Yakes}, \citenamefont {Lee},\ and\
  \citenamefont {Gammon}}]{Grim2019}%
  \BibitemOpen
  \bibfield  {author} {\bibinfo {author} {\bibfnamefont {J.~Q.}\ \bibnamefont
  {Grim}}, \bibinfo {author} {\bibfnamefont {A.~S.}\ \bibnamefont {Bracker}},
  \bibinfo {author} {\bibfnamefont {M.}~\bibnamefont {Zalalutdinov}}, \bibinfo
  {author} {\bibfnamefont {S.~G.}\ \bibnamefont {Carter}}, \bibinfo {author}
  {\bibfnamefont {A.~C.}\ \bibnamefont {Kozen}}, \bibinfo {author}
  {\bibfnamefont {M.}~\bibnamefont {Kim}}, \bibinfo {author} {\bibfnamefont
  {C.~S.}\ \bibnamefont {Kim}}, \bibinfo {author} {\bibfnamefont {J.~T.}\
  \bibnamefont {Mlack}}, \bibinfo {author} {\bibfnamefont {M.}~\bibnamefont
  {Yakes}}, \bibinfo {author} {\bibfnamefont {B.}~\bibnamefont {Lee}},\ and\
  \bibinfo {author} {\bibfnamefont {D.}~\bibnamefont {Gammon}},\ }\bibfield
  {title} {\bibinfo {title} {{Scalable in operando strain tuning in
  nanophotonic waveguides enabling three-quantum-dot superradiance}},\ }\href
  {https://doi.org/10.1038/s41563-019-0418-0} {\bibfield  {journal} {\bibinfo
  {journal} {Nat. Mater.}\ }\textbf {\bibinfo {volume} {18}},\ \bibinfo {pages}
  {963} (\bibinfo {year} {2019})}\BibitemShut {NoStop}%
\bibitem [{\citenamefont {Elshaari}\ \emph {et~al.}(2018)\citenamefont
  {Elshaari}, \citenamefont
  {B{\ifmmode\ddot{u}\else\"{u}\fi}y{\ifmmode\ddot{u}\else\"{u}\fi}k{\ifmmode\ddot{o}\else\"{o}\fi}zer},
  \citenamefont {Zadeh}, \citenamefont {Lettner}, \citenamefont {Zhao},
  \citenamefont {Sch{\ifmmode\ddot{o}\else\"{o}\fi}ll}, \citenamefont {Gyger},
  \citenamefont {Reimer}, \citenamefont {Dalacu}, \citenamefont {Poole},
  \citenamefont {J{\ifmmode\ddot{o}\else\"{o}\fi}ns},\ and\ \citenamefont
  {Zwiller}}]{ElShaari2018}%
  \BibitemOpen
  \bibfield  {author} {\bibinfo {author} {\bibfnamefont {A.~W.}\ \bibnamefont
  {Elshaari}}, \bibinfo {author} {\bibfnamefont {E.}~\bibnamefont
  {B{\ifmmode\ddot{u}\else\"{u}\fi}y{\ifmmode\ddot{u}\else\"{u}\fi}k{\ifmmode\ddot{o}\else\"{o}\fi}zer}},
  \bibinfo {author} {\bibfnamefont {I.~E.}\ \bibnamefont {Zadeh}}, \bibinfo
  {author} {\bibfnamefont {T.}~\bibnamefont {Lettner}}, \bibinfo {author}
  {\bibfnamefont {P.}~\bibnamefont {Zhao}}, \bibinfo {author} {\bibfnamefont
  {E.}~\bibnamefont {Sch{\ifmmode\ddot{o}\else\"{o}\fi}ll}}, \bibinfo {author}
  {\bibfnamefont {S.}~\bibnamefont {Gyger}}, \bibinfo {author} {\bibfnamefont
  {M.~E.}\ \bibnamefont {Reimer}}, \bibinfo {author} {\bibfnamefont
  {D.}~\bibnamefont {Dalacu}}, \bibinfo {author} {\bibfnamefont {P.~J.}\
  \bibnamefont {Poole}}, \bibinfo {author} {\bibfnamefont {K.~D.}\ \bibnamefont
  {J{\ifmmode\ddot{o}\else\"{o}\fi}ns}},\ and\ \bibinfo {author} {\bibfnamefont
  {V.}~\bibnamefont {Zwiller}},\ }\bibfield  {title} {\bibinfo {title}
  {{Strain-Tunable Quantum Integrated Photonics}},\ }\bibfield  {journal}
  {\bibinfo  {journal} {Nano Lett.}\ }\href
  {https://doi.org/10.1021/acs.nanolett.8b03937} {10.1021/acs.nanolett.8b03937}
  (\bibinfo {year} {2018})\BibitemShut {NoStop}%
\bibitem [{\citenamefont {Singh}\ \emph {et~al.}(2019)\citenamefont {Singh},
  \citenamefont {Li}, \citenamefont {Liu}, \citenamefont {Yu}, \citenamefont
  {Lu}, \citenamefont {Schneider}, \citenamefont
  {H{\ifmmode\ddot{o}\else\"{o}\fi}fling}, \citenamefont {Lawall},
  \citenamefont {Verma}, \citenamefont {Mirin}, \citenamefont {Nam},
  \citenamefont {Liu},\ and\ \citenamefont {Srinivasan}}]{Singh2019}%
  \BibitemOpen
  \bibfield  {author} {\bibinfo {author} {\bibfnamefont {A.}~\bibnamefont
  {Singh}}, \bibinfo {author} {\bibfnamefont {Q.}~\bibnamefont {Li}}, \bibinfo
  {author} {\bibfnamefont {S.}~\bibnamefont {Liu}}, \bibinfo {author}
  {\bibfnamefont {Y.}~\bibnamefont {Yu}}, \bibinfo {author} {\bibfnamefont
  {X.}~\bibnamefont {Lu}}, \bibinfo {author} {\bibfnamefont {C.}~\bibnamefont
  {Schneider}}, \bibinfo {author} {\bibfnamefont {S.}~\bibnamefont
  {H{\ifmmode\ddot{o}\else\"{o}\fi}fling}}, \bibinfo {author} {\bibfnamefont
  {J.}~\bibnamefont {Lawall}}, \bibinfo {author} {\bibfnamefont
  {V.}~\bibnamefont {Verma}}, \bibinfo {author} {\bibfnamefont
  {R.}~\bibnamefont {Mirin}}, \bibinfo {author} {\bibfnamefont {S.~W.}\
  \bibnamefont {Nam}}, \bibinfo {author} {\bibfnamefont {J.}~\bibnamefont
  {Liu}},\ and\ \bibinfo {author} {\bibfnamefont {K.}~\bibnamefont
  {Srinivasan}},\ }\bibfield  {title} {\bibinfo {title} {{Quantum frequency
  conversion of a quantum dot single-photon source on a nanophotonic chip}},\
  }\href {https://doi.org/10.1364/OPTICA.6.000563} {\bibfield  {journal}
  {\bibinfo  {journal} {Optica}\ }\textbf {\bibinfo {volume} {6}},\ \bibinfo
  {pages} {563} (\bibinfo {year} {2019})}\BibitemShut {NoStop}%
\bibitem [{\citenamefont {Breddermann}\ \emph {et~al.}(2016)\citenamefont
  {Breddermann}, \citenamefont {Heinze}, \citenamefont {Binder}, \citenamefont
  {Zrenner},\ and\ \citenamefont {Schumacher}}]{Breddermann2016}%
  \BibitemOpen
  \bibfield  {author} {\bibinfo {author} {\bibfnamefont {D.}~\bibnamefont
  {Breddermann}}, \bibinfo {author} {\bibfnamefont {D.}~\bibnamefont {Heinze}},
  \bibinfo {author} {\bibfnamefont {R.}~\bibnamefont {Binder}}, \bibinfo
  {author} {\bibfnamefont {A.}~\bibnamefont {Zrenner}},\ and\ \bibinfo {author}
  {\bibfnamefont {S.}~\bibnamefont {Schumacher}},\ }\bibfield  {title}
  {\bibinfo {title} {{All-optical tailoring of single-photon spectra in a
  quantum-dot microcavity system}},\ }\href
  {https://doi.org/10.1103/PhysRevB.94.165310} {\bibfield  {journal} {\bibinfo
  {journal} {Phys. Rev. B}\ }\textbf {\bibinfo {volume} {94}},\ \bibinfo
  {pages} {165310} (\bibinfo {year} {2016})}\BibitemShut {NoStop}%
\bibitem [{\citenamefont {Gustin}\ and\ \citenamefont
  {Hughes}(2017)}]{Gustin2017}%
  \BibitemOpen
  \bibfield  {author} {\bibinfo {author} {\bibfnamefont {C.}~\bibnamefont
  {Gustin}}\ and\ \bibinfo {author} {\bibfnamefont {S.}~\bibnamefont
  {Hughes}},\ }\bibfield  {title} {\bibinfo {title} {{Influence of
  electron-phonon scattering for an on-demand quantum dot single-photon source
  using cavity-assisted adiabatic passage}},\ }\href
  {https://doi.org/10.1103/PhysRevB.96.085305} {\bibfield  {journal} {\bibinfo
  {journal} {Phys. Rev. B}\ }\textbf {\bibinfo {volume} {96}},\ \bibinfo
  {pages} {085305} (\bibinfo {year} {2017})}\BibitemShut {NoStop}%
\bibitem [{\citenamefont {Jonas}\ \emph {et~al.}(2022)\citenamefont {Jonas},
  \citenamefont {Heinze}, \citenamefont {Sch{\ifmmode\ddot{o}\else\"{o}\fi}ll},
  \citenamefont {Kallert}, \citenamefont {Langer}, \citenamefont {Krehs},
  \citenamefont {Widhalm}, \citenamefont {J{\ifmmode\ddot{o}\else\"{o}\fi}ns},
  \citenamefont {Reuter}, \citenamefont {Schumacher},\ and\ \citenamefont
  {Zrenner}}]{Jonas2022Mar}%
  \BibitemOpen
  \bibfield  {author} {\bibinfo {author} {\bibfnamefont {B.}~\bibnamefont
  {Jonas}}, \bibinfo {author} {\bibfnamefont {D.}~\bibnamefont {Heinze}},
  \bibinfo {author} {\bibfnamefont {E.}~\bibnamefont
  {Sch{\ifmmode\ddot{o}\else\"{o}\fi}ll}}, \bibinfo {author} {\bibfnamefont
  {P.}~\bibnamefont {Kallert}}, \bibinfo {author} {\bibfnamefont
  {T.}~\bibnamefont {Langer}}, \bibinfo {author} {\bibfnamefont
  {S.}~\bibnamefont {Krehs}}, \bibinfo {author} {\bibfnamefont
  {A.}~\bibnamefont {Widhalm}}, \bibinfo {author} {\bibfnamefont {K.~D.}\
  \bibnamefont {J{\ifmmode\ddot{o}\else\"{o}\fi}ns}}, \bibinfo {author}
  {\bibfnamefont {D.}~\bibnamefont {Reuter}}, \bibinfo {author} {\bibfnamefont
  {S.}~\bibnamefont {Schumacher}},\ and\ \bibinfo {author} {\bibfnamefont
  {A.}~\bibnamefont {Zrenner}},\ }\bibfield  {title} {\bibinfo {title}
  {{Nonlinear down-conversion in a single quantum dot}},\ }\href
  {https://doi.org/10.1038/s41467-022-28993-3} {\bibfield  {journal} {\bibinfo
  {journal} {Nat. Commun.}\ }\textbf {\bibinfo {volume} {13}},\ \bibinfo
  {pages} {1} (\bibinfo {year} {2022})}\BibitemShut {NoStop}%
\bibitem [{\citenamefont {Dusanowski}\ \emph {et~al.}(2022)\citenamefont
  {Dusanowski}, \citenamefont {Gustin}, \citenamefont {Hughes}, \citenamefont
  {Schneider},\ and\ \citenamefont
  {H{\ifmmode\ddot{o}\else\"{o}\fi}fling}}]{Dusanowski2022May}%
  \BibitemOpen
  \bibfield  {author} {\bibinfo {author} {\bibfnamefont {{\L}.}~\bibnamefont
  {Dusanowski}}, \bibinfo {author} {\bibfnamefont {C.}~\bibnamefont {Gustin}},
  \bibinfo {author} {\bibfnamefont {S.}~\bibnamefont {Hughes}}, \bibinfo
  {author} {\bibfnamefont {C.}~\bibnamefont {Schneider}},\ and\ \bibinfo
  {author} {\bibfnamefont {S.}~\bibnamefont
  {H{\ifmmode\ddot{o}\else\"{o}\fi}fling}},\ }\bibfield  {title} {\bibinfo
  {title} {{All-Optical Tuning of Indistinguishable Single Photons Generated in
  Three-Level Quantum Systems}},\ }\href
  {https://doi.org/10.1021/acs.nanolett.1c04700} {\bibfield  {journal}
  {\bibinfo  {journal} {Nano Lett.}\ }\textbf {\bibinfo {volume} {22}},\
  \bibinfo {pages} {3562} (\bibinfo {year} {2022})}\BibitemShut {NoStop}%
\bibitem [{\citenamefont {Lukin}\ \emph {et~al.}(2020)\citenamefont {Lukin},
  \citenamefont {White}, \citenamefont {Trivedi}, \citenamefont {Guidry},
  \citenamefont {Morioka}, \citenamefont {Babin}, \citenamefont {Soykal},
  \citenamefont {Ul-Hassan}, \citenamefont {Son}, \citenamefont {Ohshima},
  \citenamefont {Vasireddy}, \citenamefont {Nasr}, \citenamefont {Sun},
  \citenamefont {MacLean}, \citenamefont {Dory}, \citenamefont {Nanni},
  \citenamefont {Wrachtrup}, \citenamefont {Kaiser},\ and\ \citenamefont
  {Vu{\ifmmode\check{c}\else\v{c}\fi}kovi{\ifmmode\acute{c}\else\'{c}\fi}}}]{Lukin2020}%
  \BibitemOpen
  \bibfield  {author} {\bibinfo {author} {\bibfnamefont {D.~M.}\ \bibnamefont
  {Lukin}}, \bibinfo {author} {\bibfnamefont {A.~D.}\ \bibnamefont {White}},
  \bibinfo {author} {\bibfnamefont {R.}~\bibnamefont {Trivedi}}, \bibinfo
  {author} {\bibfnamefont {M.~A.}\ \bibnamefont {Guidry}}, \bibinfo {author}
  {\bibfnamefont {N.}~\bibnamefont {Morioka}}, \bibinfo {author} {\bibfnamefont
  {C.}~\bibnamefont {Babin}}, \bibinfo {author} {\bibfnamefont
  {{\ifmmode\ddot{O}\else\"{O}\fi}.~O.}\ \bibnamefont {Soykal}}, \bibinfo
  {author} {\bibfnamefont {J.}~\bibnamefont {Ul-Hassan}}, \bibinfo {author}
  {\bibfnamefont {N.~T.}\ \bibnamefont {Son}}, \bibinfo {author} {\bibfnamefont
  {T.}~\bibnamefont {Ohshima}}, \bibinfo {author} {\bibfnamefont {P.~K.}\
  \bibnamefont {Vasireddy}}, \bibinfo {author} {\bibfnamefont {M.~H.}\
  \bibnamefont {Nasr}}, \bibinfo {author} {\bibfnamefont {S.}~\bibnamefont
  {Sun}}, \bibinfo {author} {\bibfnamefont {J.-P.~W.}\ \bibnamefont {MacLean}},
  \bibinfo {author} {\bibfnamefont {C.}~\bibnamefont {Dory}}, \bibinfo {author}
  {\bibfnamefont {E.~A.}\ \bibnamefont {Nanni}}, \bibinfo {author}
  {\bibfnamefont {J.}~\bibnamefont {Wrachtrup}}, \bibinfo {author}
  {\bibfnamefont {F.}~\bibnamefont {Kaiser}},\ and\ \bibinfo {author}
  {\bibfnamefont {J.}~\bibnamefont
  {Vu{\ifmmode\check{c}\else\v{c}\fi}kovi{\ifmmode\acute{c}\else\'{c}\fi}}},\
  }\bibfield  {title} {\bibinfo {title} {{Spectrally reconfigurable quantum
  emitters enabled by optimized fast modulation}},\ }\href
  {https://doi.org/10.1038/s41534-020-00310-0} {\bibfield  {journal} {\bibinfo
  {journal} {npj Quantum Inf.}\ }\textbf {\bibinfo {volume} {6}},\ \bibinfo
  {pages} {1} (\bibinfo {year} {2020})}\BibitemShut {NoStop}%
\bibitem [{\citenamefont {Kues}\ \emph {et~al.}(2017)\citenamefont {Kues},
  \citenamefont {Reimer}, \citenamefont {Roztocki}, \citenamefont
  {Cort{\ifmmode\acute{e}\else\'{e}\fi}s}, \citenamefont {Sciara},
  \citenamefont {Wetzel}, \citenamefont {Zhang}, \citenamefont {Cino},
  \citenamefont {Chu}, \citenamefont {Little}, \citenamefont {Moss},
  \citenamefont {Caspani}, \citenamefont
  {Aza{\ifmmode\tilde{n}\else\~{n}\fi}a},\ and\ \citenamefont
  {Morandotti}}]{Kues2017}%
  \BibitemOpen
  \bibfield  {author} {\bibinfo {author} {\bibfnamefont {M.}~\bibnamefont
  {Kues}}, \bibinfo {author} {\bibfnamefont {C.}~\bibnamefont {Reimer}},
  \bibinfo {author} {\bibfnamefont {P.}~\bibnamefont {Roztocki}}, \bibinfo
  {author} {\bibfnamefont {L.~R.}\ \bibnamefont
  {Cort{\ifmmode\acute{e}\else\'{e}\fi}s}}, \bibinfo {author} {\bibfnamefont
  {S.}~\bibnamefont {Sciara}}, \bibinfo {author} {\bibfnamefont
  {B.}~\bibnamefont {Wetzel}}, \bibinfo {author} {\bibfnamefont
  {Y.}~\bibnamefont {Zhang}}, \bibinfo {author} {\bibfnamefont
  {A.}~\bibnamefont {Cino}}, \bibinfo {author} {\bibfnamefont {S.~T.}\
  \bibnamefont {Chu}}, \bibinfo {author} {\bibfnamefont {B.~E.}\ \bibnamefont
  {Little}}, \bibinfo {author} {\bibfnamefont {D.~J.}\ \bibnamefont {Moss}},
  \bibinfo {author} {\bibfnamefont {L.}~\bibnamefont {Caspani}}, \bibinfo
  {author} {\bibfnamefont {J.}~\bibnamefont
  {Aza{\ifmmode\tilde{n}\else\~{n}\fi}a}},\ and\ \bibinfo {author}
  {\bibfnamefont {R.}~\bibnamefont {Morandotti}},\ }\bibfield  {title}
  {\bibinfo {title} {{On-chip generation of high-dimensional entangled quantum
  states and their coherent control}},\ }\href
  {https://doi.org/10.1038/nature22986} {\bibfield  {journal} {\bibinfo
  {journal} {Nature}\ }\textbf {\bibinfo {volume} {546}},\ \bibinfo {pages}
  {622} (\bibinfo {year} {2017})}\BibitemShut {NoStop}%
\bibitem [{\citenamefont {Lukens}\ and\ \citenamefont
  {Lougovski}(2017)}]{Lukens2017}%
  \BibitemOpen
  \bibfield  {author} {\bibinfo {author} {\bibfnamefont {J.~M.}\ \bibnamefont
  {Lukens}}\ and\ \bibinfo {author} {\bibfnamefont {P.}~\bibnamefont
  {Lougovski}},\ }\bibfield  {title} {\bibinfo {title} {{Frequency-encoded
  photonic qubits for scalable quantum information processing}},\ }\href
  {https://doi.org/10.1364/OPTICA.4.000008} {\bibfield  {journal} {\bibinfo
  {journal} {Optica}\ }\textbf {\bibinfo {volume} {4}},\ \bibinfo {pages} {8}
  (\bibinfo {year} {2017})}\BibitemShut {NoStop}%
\bibitem [{\citenamefont {Silveri}\ \emph {et~al.}(2017)\citenamefont
  {Silveri}, \citenamefont {Tuorila}, \citenamefont {Thuneberg},\ and\
  \citenamefont {Paraoanu}}]{Silveri2017}%
  \BibitemOpen
  \bibfield  {author} {\bibinfo {author} {\bibfnamefont {M.~P.}\ \bibnamefont
  {Silveri}}, \bibinfo {author} {\bibfnamefont {J.~A.}\ \bibnamefont
  {Tuorila}}, \bibinfo {author} {\bibfnamefont {E.~V.}\ \bibnamefont
  {Thuneberg}},\ and\ \bibinfo {author} {\bibfnamefont {G.~S.}\ \bibnamefont
  {Paraoanu}},\ }\bibfield  {title} {\bibinfo {title} {{Quantum systems under
  frequency modulation}},\ }\href {https://doi.org/10.1088/1361-6633/aa5170}
  {\bibfield  {journal} {\bibinfo  {journal} {Rep. Prog. Phys.}\ }\textbf
  {\bibinfo {volume} {80}},\ \bibinfo {pages} {056002} (\bibinfo {year}
  {2017})}\BibitemShut {NoStop}%
\bibitem [{\citenamefont {Schimpf}\ \emph {et~al.}(2021)\citenamefont
  {Schimpf}, \citenamefont {Reindl}, \citenamefont {Basset}, \citenamefont
  {J{\ifmmode\ddot{o}\else\"{o}\fi}ns}, \citenamefont {Trotta},\ and\
  \citenamefont {Rastelli}}]{Schimpf2021Mar}%
  \BibitemOpen
  \bibfield  {author} {\bibinfo {author} {\bibfnamefont {C.}~\bibnamefont
  {Schimpf}}, \bibinfo {author} {\bibfnamefont {M.}~\bibnamefont {Reindl}},
  \bibinfo {author} {\bibfnamefont {F.~B.}\ \bibnamefont {Basset}}, \bibinfo
  {author} {\bibfnamefont {K.~D.}\ \bibnamefont
  {J{\ifmmode\ddot{o}\else\"{o}\fi}ns}}, \bibinfo {author} {\bibfnamefont
  {R.}~\bibnamefont {Trotta}},\ and\ \bibinfo {author} {\bibfnamefont
  {A.}~\bibnamefont {Rastelli}},\ }\bibfield  {title} {\bibinfo {title}
  {{Quantum dots as potential sources of strongly entangled photons:
  Perspectives and challenges for applications in quantum networks}},\ }\href
  {https://doi.org/10.1063/5.0038729} {\bibfield  {journal} {\bibinfo
  {journal} {Appl. Phys. Lett.}\ }\textbf {\bibinfo {volume} {118}},\ \bibinfo
  {pages} {100502} (\bibinfo {year} {2021})}\BibitemShut {NoStop}%
\bibitem [{\citenamefont {Liu}\ \emph {et~al.}(2019)\citenamefont {Liu},
  \citenamefont {Su}, \citenamefont {Wei}, \citenamefont {Yao}, \citenamefont
  {Silva}, \citenamefont {Yu}, \citenamefont {Iles-Smith}, \citenamefont
  {Srinivasan}, \citenamefont {Rastelli}, \citenamefont {Li},\ and\
  \citenamefont {Wang}}]{Liu2019Jun}%
  \BibitemOpen
  \bibfield  {author} {\bibinfo {author} {\bibfnamefont {J.}~\bibnamefont
  {Liu}}, \bibinfo {author} {\bibfnamefont {R.}~\bibnamefont {Su}}, \bibinfo
  {author} {\bibfnamefont {Y.}~\bibnamefont {Wei}}, \bibinfo {author}
  {\bibfnamefont {B.}~\bibnamefont {Yao}}, \bibinfo {author} {\bibfnamefont
  {S.~F. C.~d.}\ \bibnamefont {Silva}}, \bibinfo {author} {\bibfnamefont
  {Y.}~\bibnamefont {Yu}}, \bibinfo {author} {\bibfnamefont {J.}~\bibnamefont
  {Iles-Smith}}, \bibinfo {author} {\bibfnamefont {K.}~\bibnamefont
  {Srinivasan}}, \bibinfo {author} {\bibfnamefont {A.}~\bibnamefont
  {Rastelli}}, \bibinfo {author} {\bibfnamefont {J.}~\bibnamefont {Li}},\ and\
  \bibinfo {author} {\bibfnamefont {X.}~\bibnamefont {Wang}},\ }\bibfield
  {title} {\bibinfo {title} {{A solid-state source of strongly entangled photon
  pairs with high brightness and indistinguishability}},\ }\href
  {https://doi.org/10.1038/s41565-019-0435-9} {\bibfield  {journal} {\bibinfo
  {journal} {Nat. Nanotechnol.}\ }\textbf {\bibinfo {volume} {14}},\ \bibinfo
  {pages} {586} (\bibinfo {year} {2019})}\BibitemShut {NoStop}%
\bibitem [{\citenamefont {Olbrich}\ \emph {et~al.}(2017)\citenamefont
  {Olbrich}, \citenamefont {H{\ifmmode\ddot{o}\else\"{o}\fi}schele},
  \citenamefont {M{\ifmmode\ddot{u}\else\"{u}\fi}ller}, \citenamefont
  {Kettler}, \citenamefont {Portalupi}, \citenamefont {Paul}, \citenamefont
  {Jetter},\ and\ \citenamefont {Michler}}]{Olbrich2017Sep}%
  \BibitemOpen
  \bibfield  {author} {\bibinfo {author} {\bibfnamefont {F.}~\bibnamefont
  {Olbrich}}, \bibinfo {author} {\bibfnamefont {J.}~\bibnamefont
  {H{\ifmmode\ddot{o}\else\"{o}\fi}schele}}, \bibinfo {author} {\bibfnamefont
  {M.}~\bibnamefont {M{\ifmmode\ddot{u}\else\"{u}\fi}ller}}, \bibinfo {author}
  {\bibfnamefont {J.}~\bibnamefont {Kettler}}, \bibinfo {author} {\bibfnamefont
  {S.~L.}\ \bibnamefont {Portalupi}}, \bibinfo {author} {\bibfnamefont
  {M.}~\bibnamefont {Paul}}, \bibinfo {author} {\bibfnamefont {M.}~\bibnamefont
  {Jetter}},\ and\ \bibinfo {author} {\bibfnamefont {P.}~\bibnamefont
  {Michler}},\ }\bibfield  {title} {\bibinfo {title} {{Polarization-entangled
  photons from an InGaAs-based quantum dot emitting in the telecom C-band}},\
  }\href {https://doi.org/10.1063/1.4994145} {\bibfield  {journal} {\bibinfo
  {journal} {Appl. Phys. Lett.}\ }\textbf {\bibinfo {volume} {111}},\ \bibinfo
  {pages} {133106} (\bibinfo {year} {2017})}\BibitemShut {NoStop}%
\bibitem [{\citenamefont {Huber}\ \emph {et~al.}(2014)\citenamefont {Huber},
  \citenamefont {Predojevi{\ifmmode\acute{c}\else\'{c}\fi}}, \citenamefont
  {Khoshnegar}, \citenamefont {Dalacu}, \citenamefont {Poole}, \citenamefont
  {Majedi},\ and\ \citenamefont {Weihs}}]{Huber2014Nov}%
  \BibitemOpen
  \bibfield  {author} {\bibinfo {author} {\bibfnamefont {T.}~\bibnamefont
  {Huber}}, \bibinfo {author} {\bibfnamefont {A.}~\bibnamefont
  {Predojevi{\ifmmode\acute{c}\else\'{c}\fi}}}, \bibinfo {author}
  {\bibfnamefont {M.}~\bibnamefont {Khoshnegar}}, \bibinfo {author}
  {\bibfnamefont {D.}~\bibnamefont {Dalacu}}, \bibinfo {author} {\bibfnamefont
  {P.~J.}\ \bibnamefont {Poole}}, \bibinfo {author} {\bibfnamefont
  {H.}~\bibnamefont {Majedi}},\ and\ \bibinfo {author} {\bibfnamefont
  {G.}~\bibnamefont {Weihs}},\ }\bibfield  {title} {\bibinfo {title}
  {{Polarization Entangled Photons from Quantum Dots Embedded in Nanowires}},\
  }\bibfield  {journal} {\bibinfo  {journal} {ACS Publications}\ }\href
  {https://doi.org/10.1021/nl503581d} {10.1021/nl503581d} (\bibinfo {year}
  {2014})\BibitemShut {NoStop}%
\bibitem [{\citenamefont {Zeuner}\ \emph {et~al.}(2021)\citenamefont {Zeuner},
  \citenamefont {J{\ifmmode\ddot{o}\else\"{o}\fi}ns}, \citenamefont
  {Schweickert}, \citenamefont {Hedlund}, \citenamefont {Lobato}, \citenamefont
  {Lettner}, \citenamefont {Wang}, \citenamefont {Gyger}, \citenamefont
  {Sch{\ifmmode\ddot{o}\else\"{o}\fi}ll}, \citenamefont {Steinhauer},
  \citenamefont {Hammar},\ and\ \citenamefont {Zwiller}}]{Zeuner2021Jul}%
  \BibitemOpen
  \bibfield  {author} {\bibinfo {author} {\bibfnamefont {K.~D.}\ \bibnamefont
  {Zeuner}}, \bibinfo {author} {\bibfnamefont {K.~D.}\ \bibnamefont
  {J{\ifmmode\ddot{o}\else\"{o}\fi}ns}}, \bibinfo {author} {\bibfnamefont
  {L.}~\bibnamefont {Schweickert}}, \bibinfo {author} {\bibfnamefont {C.~R.}\
  \bibnamefont {Hedlund}}, \bibinfo {author} {\bibfnamefont {C.~N.}\
  \bibnamefont {Lobato}}, \bibinfo {author} {\bibfnamefont {T.}~\bibnamefont
  {Lettner}}, \bibinfo {author} {\bibfnamefont {K.}~\bibnamefont {Wang}},
  \bibinfo {author} {\bibfnamefont {S.}~\bibnamefont {Gyger}}, \bibinfo
  {author} {\bibfnamefont {E.}~\bibnamefont
  {Sch{\ifmmode\ddot{o}\else\"{o}\fi}ll}}, \bibinfo {author} {\bibfnamefont
  {S.}~\bibnamefont {Steinhauer}}, \bibinfo {author} {\bibfnamefont
  {M.}~\bibnamefont {Hammar}},\ and\ \bibinfo {author} {\bibfnamefont
  {V.}~\bibnamefont {Zwiller}},\ }\bibfield  {title} {\bibinfo {title}
  {{On-Demand Generation of Entangled Photon Pairs in the Telecom C-Band with
  InAs Quantum Dots}},\ }\bibfield  {journal} {\bibinfo  {journal} {ACS
  Photonics}\ }\href {https://doi.org/10.1021/acsphotonics.1c00504}
  {10.1021/acsphotonics.1c00504} (\bibinfo {year} {2021})\BibitemShut {NoStop}%
\bibitem [{\citenamefont {Gustin}\ and\ \citenamefont
  {Hughes}(2018)}]{Gustin2018}%
  \BibitemOpen
  \bibfield  {author} {\bibinfo {author} {\bibfnamefont {C.}~\bibnamefont
  {Gustin}}\ and\ \bibinfo {author} {\bibfnamefont {S.}~\bibnamefont
  {Hughes}},\ }\bibfield  {title} {\bibinfo {title} {{Pulsed excitation
  dynamics in quantum-dot–cavity systems: Limits to optimizing the fidelity
  of on-demand single-photon sources}},\ }\href
  {https://doi.org/10.1103/PhysRevB.98.045309} {\bibfield  {journal} {\bibinfo
  {journal} {Physical Review B}\ }\textbf {\bibinfo {volume} {98}},\ \bibinfo
  {pages} {045309} (\bibinfo {year} {2018})}\BibitemShut {NoStop}%
\bibitem [{\citenamefont {Gustin}\ and\ \citenamefont
  {Hughes}(2020)}]{Gustin2019}%
  \BibitemOpen
  \bibfield  {author} {\bibinfo {author} {\bibfnamefont {C.}~\bibnamefont
  {Gustin}}\ and\ \bibinfo {author} {\bibfnamefont {S.}~\bibnamefont
  {Hughes}},\ }\bibfield  {title} {\bibinfo {title} {{Efficient
  Pulse‐Excitation Techniques for Single Photon Sources from Quantum Dots in
  Optical Cavities}},\ }\href {https://doi.org/10.1002/qute.201900073}
  {\bibfield  {journal} {\bibinfo  {journal} {Advanced Quantum Technologies}\
  }\textbf {\bibinfo {volume} {3}},\ \bibinfo {pages} {1900073} (\bibinfo
  {year} {2020})}\BibitemShut {NoStop}%
\bibitem [{\citenamefont {{J. Iles-Smith, D. P. S. McCutcheon A. Nazir, and J.
  M\o{}rk}}(2017)}]{ilessmith17}%
  \BibitemOpen
  \bibfield  {author} {\bibinfo {author} {\bibnamefont {{J. Iles-Smith, D. P.
  S. McCutcheon A. Nazir, and J. M\o{}rk}}},\ }\bibfield  {title} {\bibinfo
  {title} {Phonon scattering inhibits simultaneous near-unity efficiency and
  indistinguishability in semiconductor single-photon sources},\ }\href
  {https://doi.org/nphoton.2017.101} {\bibfield  {journal} {\bibinfo  {journal}
  {Nat. Photonics}\ }\textbf {\bibinfo {volume} {11}},\ \bibinfo {pages} {521}
  (\bibinfo {year} {2017})}\BibitemShut {NoStop}%
\bibitem [{\citenamefont {{M. Cosacchi, F. Ungar, M. Cygorek, A. Vagov, and V.
  M. Axt}}(2019)}]{cosacchi19}%
  \BibitemOpen
  \bibfield  {author} {\bibinfo {author} {\bibnamefont {{M. Cosacchi, F. Ungar,
  M. Cygorek, A. Vagov, and V. M. Axt}}},\ }\bibfield  {title} {\bibinfo
  {title} {{Emission-Frequency Separated High Quality Single-Photon Sources
  Enabled by Phonons}},\ }\href
  {https://doi.org/10.1103/PhysRevLett.123.017403} {\bibfield  {journal}
  {\bibinfo  {journal} {Physical Review Letters}\ }\textbf {\bibinfo {volume}
  {123}},\ \bibinfo {pages} {017403} (\bibinfo {year} {2019})}\BibitemShut
  {NoStop}%
\bibitem [{\citenamefont {Santori}\ \emph {et~al.}(2002)\citenamefont
  {Santori}, \citenamefont {Fattal}, \citenamefont {Vuckovi{\'{c}}},
  \citenamefont {Solomon},\ and\ \citenamefont {Yamamoto}}]{santori2002}%
  \BibitemOpen
  \bibfield  {author} {\bibinfo {author} {\bibfnamefont {C.}~\bibnamefont
  {Santori}}, \bibinfo {author} {\bibfnamefont {D.}~\bibnamefont {Fattal}},
  \bibinfo {author} {\bibfnamefont {J.}~\bibnamefont {Vuckovi{\'{c}}}},
  \bibinfo {author} {\bibfnamefont {G.~S.}\ \bibnamefont {Solomon}},\ and\
  \bibinfo {author} {\bibfnamefont {Y.}~\bibnamefont {Yamamoto}},\ }\bibfield
  {title} {\bibinfo {title} {{Indistinguishable photons from a single-photon
  device.}},\ }\href {https://doi.org/10.1038/419568a} {\bibfield  {journal}
  {\bibinfo  {journal} {Nature}\ }\textbf {\bibinfo {volume} {419}},\ \bibinfo
  {pages} {594} (\bibinfo {year} {2002})}\BibitemShut {NoStop}%
\bibitem [{\citenamefont {Ollivier}\ \emph {et~al.}(2021)\citenamefont
  {Ollivier}, \citenamefont {Thomas}, \citenamefont {Wein}, \citenamefont
  {de~Buy~Wenniger}, \citenamefont {Coste}, \citenamefont {Loredo},
  \citenamefont {Somaschi}, \citenamefont {Harouri}, \citenamefont {Lemaitre},
  \citenamefont {Sagnes}, \citenamefont {Lanco}, \citenamefont {Simon},
  \citenamefont {Anton}, \citenamefont {Krebs},\ and\ \citenamefont
  {Senellart}}]{Ollivier2021}%
  \BibitemOpen
  \bibfield  {author} {\bibinfo {author} {\bibfnamefont {H.}~\bibnamefont
  {Ollivier}}, \bibinfo {author} {\bibfnamefont {S.~E.}\ \bibnamefont
  {Thomas}}, \bibinfo {author} {\bibfnamefont {S.~C.}\ \bibnamefont {Wein}},
  \bibinfo {author} {\bibfnamefont {I.~M.}\ \bibnamefont {de~Buy~Wenniger}},
  \bibinfo {author} {\bibfnamefont {N.}~\bibnamefont {Coste}}, \bibinfo
  {author} {\bibfnamefont {J.~C.}\ \bibnamefont {Loredo}}, \bibinfo {author}
  {\bibfnamefont {N.}~\bibnamefont {Somaschi}}, \bibinfo {author}
  {\bibfnamefont {A.}~\bibnamefont {Harouri}}, \bibinfo {author} {\bibfnamefont
  {A.}~\bibnamefont {Lemaitre}}, \bibinfo {author} {\bibfnamefont
  {I.}~\bibnamefont {Sagnes}}, \bibinfo {author} {\bibfnamefont
  {L.}~\bibnamefont {Lanco}}, \bibinfo {author} {\bibfnamefont
  {C.}~\bibnamefont {Simon}}, \bibinfo {author} {\bibfnamefont
  {C.}~\bibnamefont {Anton}}, \bibinfo {author} {\bibfnamefont
  {O.}~\bibnamefont {Krebs}},\ and\ \bibinfo {author} {\bibfnamefont
  {P.}~\bibnamefont {Senellart}},\ }\bibfield  {title} {\bibinfo {title}
  {{{Hong-Ou-Mandel} Interference with Imperfect Single Photon Sources}},\
  }\href {https://doi.org/10.1103/PhysRevLett.126.063602} {\bibfield  {journal}
  {\bibinfo  {journal} {Phys. Rev. Lett.}\ }\textbf {\bibinfo {volume} {126}},\
  \bibinfo {pages} {063602} (\bibinfo {year} {2021})}\BibitemShut {NoStop}%
\bibitem [{\citenamefont {Besombes}\ \emph {et~al.}(2001)\citenamefont
  {Besombes}, \citenamefont {Kheng}, \citenamefont {Marsal},\ and\
  \citenamefont {Mariette}}]{Besombes2001Mar}%
  \BibitemOpen
  \bibfield  {author} {\bibinfo {author} {\bibfnamefont {L.}~\bibnamefont
  {Besombes}}, \bibinfo {author} {\bibfnamefont {K.}~\bibnamefont {Kheng}},
  \bibinfo {author} {\bibfnamefont {L.}~\bibnamefont {Marsal}},\ and\ \bibinfo
  {author} {\bibfnamefont {H.}~\bibnamefont {Mariette}},\ }\bibfield  {title}
  {\bibinfo {title} {{Acoustic phonon broadening mechanism in single quantum
  dot emission}},\ }\href {https://doi.org/10.1103/PhysRevB.63.155307}
  {\bibfield  {journal} {\bibinfo  {journal} {Phys. Rev. B}\ }\textbf {\bibinfo
  {volume} {63}},\ \bibinfo {pages} {155307} (\bibinfo {year}
  {2001})}\BibitemShut {NoStop}%
\bibitem [{\citenamefont {Krummheuer}\ \emph {et~al.}(2002)\citenamefont
  {Krummheuer}, \citenamefont {Axt},\ and\ \citenamefont
  {Kuhn}}]{Krummheuer2002May}%
  \BibitemOpen
  \bibfield  {author} {\bibinfo {author} {\bibfnamefont {B.}~\bibnamefont
  {Krummheuer}}, \bibinfo {author} {\bibfnamefont {V.~M.}\ \bibnamefont
  {Axt}},\ and\ \bibinfo {author} {\bibfnamefont {T.}~\bibnamefont {Kuhn}},\
  }\bibfield  {title} {\bibinfo {title} {{Theory of pure dephasing and the
  resulting absorption line shape in semiconductor quantum dots}},\ }\href
  {https://doi.org/10.1103/PhysRevB.65.195313} {\bibfield  {journal} {\bibinfo
  {journal} {Phys. Rev. B}\ }\textbf {\bibinfo {volume} {65}},\ \bibinfo
  {pages} {195313} (\bibinfo {year} {2002})}\BibitemShut {NoStop}%
\bibitem [{\citenamefont {F{\"{o}}rstner}\ \emph {et~al.}(2003)\citenamefont
  {F{\"{o}}rstner}, \citenamefont {Weber}, \citenamefont {Danckwerts},\ and\
  \citenamefont {Knorr}}]{Forstner2003}%
  \BibitemOpen
  \bibfield  {author} {\bibinfo {author} {\bibfnamefont {J.}~\bibnamefont
  {F{\"{o}}rstner}}, \bibinfo {author} {\bibfnamefont {C.}~\bibnamefont
  {Weber}}, \bibinfo {author} {\bibfnamefont {J.}~\bibnamefont {Danckwerts}},\
  and\ \bibinfo {author} {\bibfnamefont {A.}~\bibnamefont {Knorr}},\ }\bibfield
   {title} {\bibinfo {title} {{Phonon-Assisted Damping of Rabi Oscillations in
  Semiconductor Quantum Dots}},\ }\href
  {https://doi.org/10.1103/PhysRevLett.91.127401} {\bibfield  {journal}
  {\bibinfo  {journal} {Physical Review Letters}\ }\textbf {\bibinfo {volume}
  {91}},\ \bibinfo {pages} {127401} (\bibinfo {year} {2003})}\BibitemShut
  {NoStop}%
\bibitem [{\citenamefont {{A. J. Ramsay, A. V. Gopal, E. M. Gauger, A. Nazir,
  B. W. Lovett, A. M. Fox, and M. S. Skolnick}}(2010)}]{Ramsay10}%
  \BibitemOpen
  \bibfield  {author} {\bibinfo {author} {\bibnamefont {{A. J. Ramsay, A. V.
  Gopal, E. M. Gauger, A. Nazir, B. W. Lovett, A. M. Fox, and M. S.
  Skolnick}}},\ }\bibfield  {title} {\bibinfo {title} {{Damping of exciton Rabi
  rotations by acoustic phonons in optically excited InGaAs/GaAs quantum
  dots}},\ }\href {https://doi.org/PhysRevLett.104.017402} {\bibfield
  {journal} {\bibinfo  {journal} {Physical Review Letters}\ }\textbf {\bibinfo
  {volume} {104}},\ \bibinfo {pages} {017402} (\bibinfo {year}
  {2010})}\BibitemShut {NoStop}%
\bibitem [{\citenamefont {{D. P. S. McCutcheon, and A.
  Nazir}}(2010)}]{mccutcheon10}%
  \BibitemOpen
  \bibfield  {author} {\bibinfo {author} {\bibnamefont {{D. P. S. McCutcheon,
  and A. Nazir}}},\ }\bibfield  {title} {\bibinfo {title} {{Quantum dot Rabi
  rotations beyond the weak exciton-phonon coupling regime}},\ }\href
  {https://doi.org/10.1088/1367-2630/12/11/113042} {\bibfield  {journal}
  {\bibinfo  {journal} {New Journal of Physics}\ }\textbf {\bibinfo {volume}
  {12}},\ \bibinfo {pages} {113042} (\bibinfo {year} {2010})}\BibitemShut
  {NoStop}%
\bibitem [{\citenamefont {{C. Roy and S. Hughes}}(2011)}]{roy11}%
  \BibitemOpen
  \bibfield  {author} {\bibinfo {author} {\bibnamefont {{C. Roy and S.
  Hughes}}},\ }\bibfield  {title} {\bibinfo {title} {{Phonon-Dressed Mollow
  Triplet in the Regime of Cavity Quantum Electrodynamics: Excitation-Induced
  Dephasing and Nonperturbative Cavity Feeding Effects}},\ }\href
  {https://doi.org/10.1103/PhysRevLett.106.247403} {\bibfield  {journal}
  {\bibinfo  {journal} {Physical Review Letters}\ }\textbf {\bibinfo {volume}
  {106}},\ \bibinfo {pages} {247403} (\bibinfo {year} {2011})}\BibitemShut
  {NoStop}%
\bibitem [{\citenamefont {{S. Weiler, A. Ulhaq, S. M. Ulrich, D. Richter, M.
  Jetter, P. Michler, C. Roy, and S. Hughes}}(2012)}]{Weiler2012}%
  \BibitemOpen
  \bibfield  {author} {\bibinfo {author} {\bibnamefont {{S. Weiler, A. Ulhaq,
  S. M. Ulrich, D. Richter, M. Jetter, P. Michler, C. Roy, and S. Hughes}}},\
  }\bibfield  {title} {\bibinfo {title} {{Phonon-assisted incoherent excitation
  of a quantum dot and its emission properties}},\ }\href
  {https://doi.org/10.1103/PhysRevB.86.241304} {\bibfield  {journal} {\bibinfo
  {journal} {Physical Review B}\ }\textbf {\bibinfo {volume} {86}},\ \bibinfo
  {pages} {241304(R)} (\bibinfo {year} {2012})}\BibitemShut {NoStop}%
\bibitem [{\citenamefont {Hughes}\ \emph {et~al.}(2011)\citenamefont {Hughes},
  \citenamefont {Yao}, \citenamefont {Milde}, \citenamefont {Knorr},
  \citenamefont {Dalacu}, \citenamefont {Mnaymneh}, \citenamefont {Sazonova},
  \citenamefont {Poole}, \citenamefont {Aers}, \citenamefont {Lapointe},
  \citenamefont {Cheriton},\ and\ \citenamefont {Williams}}]{Hughes2011Apr}%
  \BibitemOpen
  \bibfield  {author} {\bibinfo {author} {\bibfnamefont {S.}~\bibnamefont
  {Hughes}}, \bibinfo {author} {\bibfnamefont {P.}~\bibnamefont {Yao}},
  \bibinfo {author} {\bibfnamefont {F.}~\bibnamefont {Milde}}, \bibinfo
  {author} {\bibfnamefont {A.}~\bibnamefont {Knorr}}, \bibinfo {author}
  {\bibfnamefont {D.}~\bibnamefont {Dalacu}}, \bibinfo {author} {\bibfnamefont
  {K.}~\bibnamefont {Mnaymneh}}, \bibinfo {author} {\bibfnamefont
  {V.}~\bibnamefont {Sazonova}}, \bibinfo {author} {\bibfnamefont {P.~J.}\
  \bibnamefont {Poole}}, \bibinfo {author} {\bibfnamefont {G.~C.}\ \bibnamefont
  {Aers}}, \bibinfo {author} {\bibfnamefont {J.}~\bibnamefont {Lapointe}},
  \bibinfo {author} {\bibfnamefont {R.}~\bibnamefont {Cheriton}},\ and\
  \bibinfo {author} {\bibfnamefont {R.~L.}\ \bibnamefont {Williams}},\
  }\bibfield  {title} {\bibinfo {title} {{Influence of electron-acoustic phonon
  scattering on off-resonant cavity feeding within a strongly coupled
  quantum-dot cavity system}},\ }\href
  {https://doi.org/10.1103/PhysRevB.83.165313} {\bibfield  {journal} {\bibinfo
  {journal} {Phys. Rev. B}\ }\textbf {\bibinfo {volume} {83}},\ \bibinfo
  {pages} {165313} (\bibinfo {year} {2011})}\BibitemShut {NoStop}%
\bibitem [{\citenamefont {{J. H. Quilter, A. J. Brash, F. Liu, M Gl{\"a}ssl, A.
  M. Barth, V. M. Axt, A. J. Ramsay, M. S. Skolnick, and A. M.
  Fox}}(2015)}]{Quilter2015}%
  \BibitemOpen
  \bibfield  {author} {\bibinfo {author} {\bibnamefont {{J. H. Quilter, A. J.
  Brash, F. Liu, M Gl{\"a}ssl, A. M. Barth, V. M. Axt, A. J. Ramsay, M. S.
  Skolnick, and A. M. Fox}}},\ }\bibfield  {title} {\bibinfo {title}
  {{Phonon-Assisted Population Inversion of a Single InGaAs/GaAs Quantum Dot by
  Pulsed Laser Excitation}},\ }\href
  {https://doi.org/10.1103/PhysRevLett.114.137401} {\bibfield  {journal}
  {\bibinfo  {journal} {Physical Review Letters}\ }\textbf {\bibinfo {volume}
  {114}},\ \bibinfo {pages} {137401} (\bibinfo {year} {2015})}\BibitemShut
  {NoStop}%
\bibitem [{\citenamefont {Ulrich}\ \emph {et~al.}(2011)\citenamefont {Ulrich},
  \citenamefont {Ates}, \citenamefont {Reitzenstein}, \citenamefont
  {L{\ifmmode\ddot{o}\else\"{o}\fi}ffler}, \citenamefont {Forchel},\ and\
  \citenamefont {Michler}}]{Ulrich2011Jun}%
  \BibitemOpen
  \bibfield  {author} {\bibinfo {author} {\bibfnamefont {S.~M.}\ \bibnamefont
  {Ulrich}}, \bibinfo {author} {\bibfnamefont {S.}~\bibnamefont {Ates}},
  \bibinfo {author} {\bibfnamefont {S.}~\bibnamefont {Reitzenstein}}, \bibinfo
  {author} {\bibfnamefont {A.}~\bibnamefont
  {L{\ifmmode\ddot{o}\else\"{o}\fi}ffler}}, \bibinfo {author} {\bibfnamefont
  {A.}~\bibnamefont {Forchel}},\ and\ \bibinfo {author} {\bibfnamefont
  {P.}~\bibnamefont {Michler}},\ }\bibfield  {title} {\bibinfo {title}
  {{Dephasing of Triplet-Sideband Optical Emission of a Resonantly Driven
  InAs/GaAs Quantum Dot inside a Microcavity}},\ }\href
  {https://doi.org/10.1103/PhysRevLett.106.247402} {\bibfield  {journal}
  {\bibinfo  {journal} {Phys. Rev. Lett.}\ }\textbf {\bibinfo {volume} {106}},\
  \bibinfo {pages} {247402} (\bibinfo {year} {2011})}\BibitemShut {NoStop}%
\bibitem [{\citenamefont {{A. Nazir and D. P. S. McCutcheon}}(2016)}]{nazir16}%
  \BibitemOpen
  \bibfield  {author} {\bibinfo {author} {\bibnamefont {{A. Nazir and D. P. S.
  McCutcheon}}},\ }\bibfield  {title} {\bibinfo {title} {Modelling
  exciton-phonon interactions in optically driven quantum dots},\ }\href
  {https://doi.org/10.1088/0953-8984/28/10/103002} {\bibfield  {journal}
  {\bibinfo  {journal} {Journal of Physics: Condensed Matter}\ }\textbf
  {\bibinfo {volume} {28}},\ \bibinfo {pages} {103002} (\bibinfo {year}
  {2016})}\BibitemShut {NoStop}%
\bibitem [{\citenamefont {{G. D. Mahan}}(1990)}]{mahan}%
  \BibitemOpen
  \bibfield  {author} {\bibinfo {author} {\bibnamefont {{G. D. Mahan}}},\
  }\href@noop {} {\emph {\bibinfo {title} {Many-Particle Physics}}},\ \bibinfo
  {edition} {2nd}\ ed.\ (\bibinfo  {publisher} {Plenum Press, New York},\
  \bibinfo {year} {1990})\BibitemShut {NoStop}%
\bibitem [{\citenamefont {Wilson-Rae}\ and\ \citenamefont
  {Imamo{\ifmmode\breve{g}\else\u{g}\fi}lu}(2002)}]{Wilson-Rae2002May}%
  \BibitemOpen
  \bibfield  {author} {\bibinfo {author} {\bibfnamefont {I.}~\bibnamefont
  {Wilson-Rae}}\ and\ \bibinfo {author} {\bibfnamefont {A.}~\bibnamefont
  {Imamo{\ifmmode\breve{g}\else\u{g}\fi}lu}},\ }\bibfield  {title} {\bibinfo
  {title} {{Quantum dot cavity-QED in the presence of strong electron-phonon
  interactions}},\ }\href {https://doi.org/10.1103/PhysRevB.65.235311}
  {\bibfield  {journal} {\bibinfo  {journal} {Phys. Rev. B}\ }\textbf {\bibinfo
  {volume} {65}},\ \bibinfo {pages} {235311} (\bibinfo {year}
  {2002})}\BibitemShut {NoStop}%
\bibitem [{\citenamefont {Hargart}\ \emph {et~al.}(2016)\citenamefont
  {Hargart}, \citenamefont {M{\"u}ller}, \citenamefont {Roy-Choudhury},
  \citenamefont {Portalupi}, \citenamefont {Schneider}, \citenamefont
  {H{\"o}fling}, \citenamefont {Kamp}, \citenamefont {Hughes},\ and\
  \citenamefont {Michler}}]{hargart16}%
  \BibitemOpen
  \bibfield  {author} {\bibinfo {author} {\bibfnamefont {F.}~\bibnamefont
  {Hargart}}, \bibinfo {author} {\bibfnamefont {M.}~\bibnamefont {M{\"u}ller}},
  \bibinfo {author} {\bibfnamefont {K.}~\bibnamefont {Roy-Choudhury}}, \bibinfo
  {author} {\bibfnamefont {S.~L.}\ \bibnamefont {Portalupi}}, \bibinfo {author}
  {\bibfnamefont {C.}~\bibnamefont {Schneider}}, \bibinfo {author}
  {\bibfnamefont {S.}~\bibnamefont {H{\"o}fling}}, \bibinfo {author}
  {\bibfnamefont {M.}~\bibnamefont {Kamp}}, \bibinfo {author} {\bibfnamefont
  {S.}~\bibnamefont {Hughes}},\ and\ \bibinfo {author} {\bibfnamefont
  {P.}~\bibnamefont {Michler}},\ }\bibfield  {title} {\bibinfo {title} {Cavity
  enhanced simultaneous dressing of quantum dot exciton and biexciton states},\
  }\href {https://doi.org/10.1103/PhysRevB.93.115308} {\bibfield  {journal}
  {\bibinfo  {journal} {Physical Review B}\ }\textbf {\bibinfo {volume} {93}},\
  \bibinfo {pages} {115308} (\bibinfo {year} {2016})}\BibitemShut {NoStop}%
\bibitem [{\citenamefont {{R. Manson, K. Roy-Choudhury, and S.
  Hughes}}(2016)}]{ross16}%
  \BibitemOpen
  \bibfield  {author} {\bibinfo {author} {\bibnamefont {{R. Manson, K.
  Roy-Choudhury, and S. Hughes}}},\ }\bibfield  {title} {\bibinfo {title}
  {Polaron master equation theory of pulse-driven phonon phonon-assisted
  population inversion and single-photon emission from quantum-dot excitons},\
  }\href {https://doi.org/10.1103/PhysRevB.93.155423} {\bibfield  {journal}
  {\bibinfo  {journal} {Physical Review B}\ }\textbf {\bibinfo {volume} {93}},\
  \bibinfo {pages} {155423} (\bibinfo {year} {2016})}\BibitemShut {NoStop}%
\bibitem [{\citenamefont {Gustin}\ \emph {et~al.}(2021)\citenamefont {Gustin},
  \citenamefont {Hanschke}, \citenamefont {Boos}, \citenamefont
  {M{\ifmmode\ddot{u}\else\"{u}\fi}ller}, \citenamefont {Kremser},
  \citenamefont {Finley}, \citenamefont {Hughes},\ and\ \citenamefont
  {M{\ifmmode\ddot{u}\else\"{u}\fi}ller}}]{Gustin2021Jan}%
  \BibitemOpen
  \bibfield  {author} {\bibinfo {author} {\bibfnamefont {C.}~\bibnamefont
  {Gustin}}, \bibinfo {author} {\bibfnamefont {L.}~\bibnamefont {Hanschke}},
  \bibinfo {author} {\bibfnamefont {K.}~\bibnamefont {Boos}}, \bibinfo {author}
  {\bibfnamefont {J.~R.~A.}\ \bibnamefont
  {M{\ifmmode\ddot{u}\else\"{u}\fi}ller}}, \bibinfo {author} {\bibfnamefont
  {M.}~\bibnamefont {Kremser}}, \bibinfo {author} {\bibfnamefont {J.~J.}\
  \bibnamefont {Finley}}, \bibinfo {author} {\bibfnamefont {S.}~\bibnamefont
  {Hughes}},\ and\ \bibinfo {author} {\bibfnamefont {K.}~\bibnamefont
  {M{\ifmmode\ddot{u}\else\"{u}\fi}ller}},\ }\bibfield  {title} {\bibinfo
  {title} {{High-resolution spectroscopy of a quantum dot driven
  bichromatically by two strong coherent fields}},\ }\href
  {https://doi.org/10.1103/PhysRevResearch.3.013044} {\bibfield  {journal}
  {\bibinfo  {journal} {Phys. Rev. Res.}\ }\textbf {\bibinfo {volume} {3}},\
  \bibinfo {pages} {013044} (\bibinfo {year} {2021})}\BibitemShut {NoStop}%
\bibitem [{\citenamefont {Ota}\ \emph {et~al.}(2009)\citenamefont {Ota},
  \citenamefont {Iwamoto}, \citenamefont {Kumagai},\ and\ \citenamefont
  {Arakawa}}]{Ota2009Aug}%
  \BibitemOpen
  \bibfield  {author} {\bibinfo {author} {\bibfnamefont {Y.}~\bibnamefont
  {Ota}}, \bibinfo {author} {\bibfnamefont {S.}~\bibnamefont {Iwamoto}},
  \bibinfo {author} {\bibfnamefont {N.}~\bibnamefont {Kumagai}},\ and\ \bibinfo
  {author} {\bibfnamefont {Y.}~\bibnamefont {Arakawa}},\ }\bibfield  {title}
  {\bibinfo {title} {{Impact of electron-phonon interactions on quantum-dot
  cavity quantum electrodynamics}},\ }\href {https://arxiv.org/abs/0908.0788v1}
  {\bibfield  {journal} {\bibinfo  {journal} {arXiv}\ } (\bibinfo {year}
  {2009})},\ \Eprint {https://arxiv.org/abs/0908.0788} {0908.0788} \BibitemShut
  {NoStop}%
\bibitem [{\citenamefont {Ulhaq}\ \emph {et~al.}(2013)\citenamefont {Ulhaq},
  \citenamefont {Weiler}, \citenamefont {Roy}, \citenamefont {Ulrich},
  \citenamefont {Jetter}, \citenamefont {Hughes},\ and\ \citenamefont
  {Michler}}]{Ulhaq2013Feb}%
  \BibitemOpen
  \bibfield  {author} {\bibinfo {author} {\bibfnamefont {A.}~\bibnamefont
  {Ulhaq}}, \bibinfo {author} {\bibfnamefont {S.}~\bibnamefont {Weiler}},
  \bibinfo {author} {\bibfnamefont {C.}~\bibnamefont {Roy}}, \bibinfo {author}
  {\bibfnamefont {S.~M.}\ \bibnamefont {Ulrich}}, \bibinfo {author}
  {\bibfnamefont {M.}~\bibnamefont {Jetter}}, \bibinfo {author} {\bibfnamefont
  {S.}~\bibnamefont {Hughes}},\ and\ \bibinfo {author} {\bibfnamefont
  {P.}~\bibnamefont {Michler}},\ }\bibfield  {title} {\bibinfo {title}
  {{Detuning-dependent Mollow triplet of a coherently-driven single quantum
  dot}},\ }\href {https://doi.org/10.1364/OE.21.004382} {\bibfield  {journal}
  {\bibinfo  {journal} {Opt. Express}\ }\textbf {\bibinfo {volume} {21}},\
  \bibinfo {pages} {4382} (\bibinfo {year} {2013})}\BibitemShut {NoStop}%
\bibitem [{\citenamefont {Reigue}\ \emph {et~al.}(2017)\citenamefont {Reigue},
  \citenamefont {Iles-Smith}, \citenamefont {Lux}, \citenamefont {Monniello},
  \citenamefont {Bernard}, \citenamefont {Margaillan}, \citenamefont
  {Lemaitre}, \citenamefont {Martinez}, \citenamefont {McCutcheon},
  \citenamefont {M{\o}rk}, \citenamefont {Hostein},\ and\ \citenamefont
  {Voliotis}}]{Reigue2017Jun}%
  \BibitemOpen
  \bibfield  {author} {\bibinfo {author} {\bibfnamefont {A.}~\bibnamefont
  {Reigue}}, \bibinfo {author} {\bibfnamefont {J.}~\bibnamefont {Iles-Smith}},
  \bibinfo {author} {\bibfnamefont {F.}~\bibnamefont {Lux}}, \bibinfo {author}
  {\bibfnamefont {L.}~\bibnamefont {Monniello}}, \bibinfo {author}
  {\bibfnamefont {M.}~\bibnamefont {Bernard}}, \bibinfo {author} {\bibfnamefont
  {F.}~\bibnamefont {Margaillan}}, \bibinfo {author} {\bibfnamefont
  {A.}~\bibnamefont {Lemaitre}}, \bibinfo {author} {\bibfnamefont
  {A.}~\bibnamefont {Martinez}}, \bibinfo {author} {\bibfnamefont {D.~P.~S.}\
  \bibnamefont {McCutcheon}}, \bibinfo {author} {\bibfnamefont
  {J.}~\bibnamefont {M{\o}rk}}, \bibinfo {author} {\bibfnamefont
  {R.}~\bibnamefont {Hostein}},\ and\ \bibinfo {author} {\bibfnamefont
  {V.}~\bibnamefont {Voliotis}},\ }\bibfield  {title} {\bibinfo {title}
  {{Probing Electron-Phonon Interaction through Two-Photon Interference in
  Resonantly Driven Semiconductor Quantum Dots}},\ }\href
  {https://doi.org/10.1103/PhysRevLett.118.233602} {\bibfield  {journal}
  {\bibinfo  {journal} {Phys. Rev. Lett.}\ }\textbf {\bibinfo {volume} {118}},\
  \bibinfo {pages} {233602} (\bibinfo {year} {2017})}\BibitemShut {NoStop}%
\bibitem [{\citenamefont {Grange}\ \emph {et~al.}(2017)\citenamefont {Grange},
  \citenamefont {Somaschi}, \citenamefont
  {Ant{\ifmmode\acute{o}\else\'{o}\fi}n}, \citenamefont {De~Santis},
  \citenamefont {Coppola}, \citenamefont {Giesz}, \citenamefont
  {Lema{\ifmmode\hat{\imath}\else\^{\i}\fi}tre}, \citenamefont {Sagnes},
  \citenamefont {Auff{\ifmmode\grave{e}\else\`{e}\fi}ves},\ and\ \citenamefont
  {Senellart}}]{Grange2017Jun}%
  \BibitemOpen
  \bibfield  {author} {\bibinfo {author} {\bibfnamefont {T.}~\bibnamefont
  {Grange}}, \bibinfo {author} {\bibfnamefont {N.}~\bibnamefont {Somaschi}},
  \bibinfo {author} {\bibfnamefont {C.}~\bibnamefont
  {Ant{\ifmmode\acute{o}\else\'{o}\fi}n}}, \bibinfo {author} {\bibfnamefont
  {L.}~\bibnamefont {De~Santis}}, \bibinfo {author} {\bibfnamefont
  {G.}~\bibnamefont {Coppola}}, \bibinfo {author} {\bibfnamefont
  {V.}~\bibnamefont {Giesz}}, \bibinfo {author} {\bibfnamefont
  {A.}~\bibnamefont {Lema{\ifmmode\hat{\imath}\else\^{\i}\fi}tre}}, \bibinfo
  {author} {\bibfnamefont {I.}~\bibnamefont {Sagnes}}, \bibinfo {author}
  {\bibfnamefont {A.}~\bibnamefont {Auff{\ifmmode\grave{e}\else\`{e}\fi}ves}},\
  and\ \bibinfo {author} {\bibfnamefont {P.}~\bibnamefont {Senellart}},\
  }\bibfield  {title} {\bibinfo {title} {{Reducing Phonon-Induced Decoherence
  in Solid-State Single-Photon Sources with Cavity Quantum Electrodynamics}},\
  }\href {https://doi.org/10.1103/PhysRevLett.118.253602} {\bibfield  {journal}
  {\bibinfo  {journal} {Phys. Rev. Lett.}\ }\textbf {\bibinfo {volume} {118}},\
  \bibinfo {pages} {253602} (\bibinfo {year} {2017})}\BibitemShut {NoStop}%
\bibitem [{\citenamefont {Tighineanu}\ \emph {et~al.}(2018)\citenamefont
  {Tighineanu}, \citenamefont {Dree{\ss}en}, \citenamefont {Flindt},
  \citenamefont {Lodahl},\ and\ \citenamefont
  {S{\o}rensen}}]{Tighineanu2018Jun}%
  \BibitemOpen
  \bibfield  {author} {\bibinfo {author} {\bibfnamefont {P.}~\bibnamefont
  {Tighineanu}}, \bibinfo {author} {\bibfnamefont {C.~L.}\ \bibnamefont
  {Dree{\ss}en}}, \bibinfo {author} {\bibfnamefont {C.}~\bibnamefont {Flindt}},
  \bibinfo {author} {\bibfnamefont {P.}~\bibnamefont {Lodahl}},\ and\ \bibinfo
  {author} {\bibfnamefont {A.~S.}\ \bibnamefont {S{\o}rensen}},\ }\bibfield
  {title} {\bibinfo {title} {{Phonon Decoherence of Quantum Dots in Photonic
  Structures: Broadening of the Zero-Phonon Line and the Role of
  Dimensionality}},\ }\href {https://doi.org/10.1103/PhysRevLett.120.257401}
  {\bibfield  {journal} {\bibinfo  {journal} {Phys. Rev. Lett.}\ }\textbf
  {\bibinfo {volume} {120}},\ \bibinfo {pages} {257401} (\bibinfo {year}
  {2018})}\BibitemShut {NoStop}%
\bibitem [{\citenamefont {Kiraz}\ \emph {et~al.}(2004)\citenamefont {Kiraz},
  \citenamefont {Atat\"ure},\ and\ \citenamefont {Imamo\ifmmode~\breve{g}\else
  \u{g}\fi{}lu}}]{PhysRevA.69.032305}%
  \BibitemOpen
  \bibfield  {author} {\bibinfo {author} {\bibfnamefont {A.}~\bibnamefont
  {Kiraz}}, \bibinfo {author} {\bibfnamefont {M.}~\bibnamefont {Atat\"ure}},\
  and\ \bibinfo {author} {\bibfnamefont {A.}~\bibnamefont
  {Imamo\ifmmode~\breve{g}\else \u{g}\fi{}lu}},\ }\bibfield  {title} {\bibinfo
  {title} {Quantum-dot single-photon sources: Prospects for applications in
  linear optics quantum-information processing},\ }\href
  {https://doi.org/10.1103/PhysRevA.69.032305} {\bibfield  {journal} {\bibinfo
  {journal} {Phys. Rev. A}\ }\textbf {\bibinfo {volume} {69}},\ \bibinfo
  {pages} {032305} (\bibinfo {year} {2004})}\BibitemShut {NoStop}%
\bibitem [{\citenamefont {Woolley}\ \emph {et~al.}(2013)\citenamefont
  {Woolley}, \citenamefont {Lang}, \citenamefont {Eichler}, \citenamefont
  {Wallraff},\ and\ \citenamefont {Blais}}]{woolley13}%
  \BibitemOpen
  \bibfield  {author} {\bibinfo {author} {\bibfnamefont {M.~J.}\ \bibnamefont
  {Woolley}}, \bibinfo {author} {\bibfnamefont {C.}~\bibnamefont {Lang}},
  \bibinfo {author} {\bibfnamefont {C.}~\bibnamefont {Eichler}}, \bibinfo
  {author} {\bibfnamefont {A.}~\bibnamefont {Wallraff}},\ and\ \bibinfo
  {author} {\bibfnamefont {A.}~\bibnamefont {Blais}},\ }\bibfield  {title}
  {\bibinfo {title} {Signatures of {Hong–Ou–Mandel} interference at
  microwave frequencies},\ }\href
  {https://doi.org/10.1088/1367-2630/15/10/105025} {\bibfield  {journal}
  {\bibinfo  {journal} {New J. Phys.}\ }\textbf {\bibinfo {volume} {15}},\
  \bibinfo {pages} {105025} (\bibinfo {year} {2013})}\BibitemShut {NoStop}%
\bibitem [{\citenamefont {Wei}\ \emph {et~al.}(2014)\citenamefont {Wei},
  \citenamefont {He}, \citenamefont {Chen}, \citenamefont {Hu}, \citenamefont
  {He}, \citenamefont {Wu}, \citenamefont {Schneider}, \citenamefont {Kamp},
  \citenamefont {H{\ifmmode\ddot{o}\else\"{o}\fi}fling}, \citenamefont {Lu},\
  and\ \citenamefont {Pan}}]{Wei2014Oct}%
  \BibitemOpen
  \bibfield  {author} {\bibinfo {author} {\bibfnamefont {Y.-J.}\ \bibnamefont
  {Wei}}, \bibinfo {author} {\bibfnamefont {Y.-M.}\ \bibnamefont {He}},
  \bibinfo {author} {\bibfnamefont {M.-C.}\ \bibnamefont {Chen}}, \bibinfo
  {author} {\bibfnamefont {Y.-N.}\ \bibnamefont {Hu}}, \bibinfo {author}
  {\bibfnamefont {Y.}~\bibnamefont {He}}, \bibinfo {author} {\bibfnamefont
  {D.}~\bibnamefont {Wu}}, \bibinfo {author} {\bibfnamefont {C.}~\bibnamefont
  {Schneider}}, \bibinfo {author} {\bibfnamefont {M.}~\bibnamefont {Kamp}},
  \bibinfo {author} {\bibfnamefont {S.}~\bibnamefont
  {H{\ifmmode\ddot{o}\else\"{o}\fi}fling}}, \bibinfo {author} {\bibfnamefont
  {C.-Y.}\ \bibnamefont {Lu}},\ and\ \bibinfo {author} {\bibfnamefont {J.-W.}\
  \bibnamefont {Pan}},\ }\bibfield  {title} {\bibinfo {title} {{Deterministic
  and Robust Generation of Single Photons from a Single Quantum Dot with
  99.5{\%} Indistinguishability Using Adiabatic Rapid Passage}},\ }\bibfield
  {journal} {\bibinfo  {journal} {ACS Publications}\ }\href
  {https://doi.org/10.1021/nl503081n} {10.1021/nl503081n} (\bibinfo {year}
  {2014})\BibitemShut {NoStop}%
\bibitem [{\citenamefont {Fischer}\ \emph {et~al.}(2016)\citenamefont
  {Fischer}, \citenamefont {M{\" u}ller}, \citenamefont {Lagoudakis},\ and\
  \citenamefont {Vu{\v{c}}kovi{\'{c}}}}]{fischer16}%
  \BibitemOpen
  \bibfield  {author} {\bibinfo {author} {\bibfnamefont {K.~A.}\ \bibnamefont
  {Fischer}}, \bibinfo {author} {\bibfnamefont {K.}~\bibnamefont {M{\"
  u}ller}}, \bibinfo {author} {\bibfnamefont {K.~G.}\ \bibnamefont
  {Lagoudakis}},\ and\ \bibinfo {author} {\bibfnamefont {J.}~\bibnamefont
  {Vu{\v{c}}kovi{\'{c}}}},\ }\bibfield  {title} {\bibinfo {title} {Dynamical
  modeling of pulsed two-photon interference},\ }\href
  {https://doi.org/10.1088/1367-2630/18/11/113053} {\bibfield  {journal}
  {\bibinfo  {journal} {New J. Phys.}\ }\textbf {\bibinfo {volume} {18}},\
  \bibinfo {pages} {113053} (\bibinfo {year} {2016})}\BibitemShut {NoStop}%
\bibitem [{\citenamefont {Loredo}\ \emph {et~al.}(2016)\citenamefont {Loredo},
  \citenamefont {Zakaria}, \citenamefont {Somaschi}, \citenamefont {Anton},
  \citenamefont {de~Santis}, \citenamefont {Giesz}, \citenamefont {Grange},
  \citenamefont {Broome}, \citenamefont {Gazzano}, \citenamefont {Coppola},
  \citenamefont {Sagnes}, \citenamefont {Lemaitre}, \citenamefont {Auffeves},
  \citenamefont {Senellart}, \citenamefont {Almeida},\ and\ \citenamefont
  {White}}]{loredo2016}%
  \BibitemOpen
  \bibfield  {author} {\bibinfo {author} {\bibfnamefont {J.~C.}\ \bibnamefont
  {Loredo}}, \bibinfo {author} {\bibfnamefont {N.~A.}\ \bibnamefont {Zakaria}},
  \bibinfo {author} {\bibfnamefont {N.}~\bibnamefont {Somaschi}}, \bibinfo
  {author} {\bibfnamefont {C.}~\bibnamefont {Anton}}, \bibinfo {author}
  {\bibfnamefont {L.}~\bibnamefont {de~Santis}}, \bibinfo {author}
  {\bibfnamefont {V.}~\bibnamefont {Giesz}}, \bibinfo {author} {\bibfnamefont
  {T.}~\bibnamefont {Grange}}, \bibinfo {author} {\bibfnamefont {M.~A.}\
  \bibnamefont {Broome}}, \bibinfo {author} {\bibfnamefont {O.}~\bibnamefont
  {Gazzano}}, \bibinfo {author} {\bibfnamefont {G.}~\bibnamefont {Coppola}},
  \bibinfo {author} {\bibfnamefont {I.}~\bibnamefont {Sagnes}}, \bibinfo
  {author} {\bibfnamefont {A.}~\bibnamefont {Lemaitre}}, \bibinfo {author}
  {\bibfnamefont {A.}~\bibnamefont {Auffeves}}, \bibinfo {author}
  {\bibfnamefont {P.}~\bibnamefont {Senellart}}, \bibinfo {author}
  {\bibfnamefont {M.~P.}\ \bibnamefont {Almeida}},\ and\ \bibinfo {author}
  {\bibfnamefont {A.~G.}\ \bibnamefont {White}},\ }\bibfield  {title} {\bibinfo
  {title} {Scalable performance in solid-state single-photon sources},\ }\href
  {https://doi.org/10.1364/optica.3.000433} {\bibfield  {journal} {\bibinfo
  {journal} {Optica}\ }\textbf {\bibinfo {volume} {3}},\ \bibinfo {pages} {433}
  (\bibinfo {year} {2016})}\BibitemShut {NoStop}%
\bibitem [{\citenamefont {Schofield}\ \emph {et~al.}(2022)\citenamefont
  {Schofield}, \citenamefont {Clear}, \citenamefont {Hoggarth}, \citenamefont
  {Major}, \citenamefont {McCutcheon},\ and\ \citenamefont
  {Clark}}]{Schofield2022Jan}%
  \BibitemOpen
  \bibfield  {author} {\bibinfo {author} {\bibfnamefont {R.~C.}\ \bibnamefont
  {Schofield}}, \bibinfo {author} {\bibfnamefont {C.}~\bibnamefont {Clear}},
  \bibinfo {author} {\bibfnamefont {R.~A.}\ \bibnamefont {Hoggarth}}, \bibinfo
  {author} {\bibfnamefont {K.~D.}\ \bibnamefont {Major}}, \bibinfo {author}
  {\bibfnamefont {D.~P.~S.}\ \bibnamefont {McCutcheon}},\ and\ \bibinfo
  {author} {\bibfnamefont {A.~S.}\ \bibnamefont {Clark}},\ }\bibfield  {title}
  {\bibinfo {title} {{Photon indistinguishability measurements under pulsed and
  continuous excitation}},\ }\href
  {https://doi.org/10.1103/PhysRevResearch.4.013037} {\bibfield  {journal}
  {\bibinfo  {journal} {Phys. Rev. Res.}\ }\textbf {\bibinfo {volume} {4}},\
  \bibinfo {pages} {013037} (\bibinfo {year} {2022})}\BibitemShut {NoStop}%
\bibitem [{\citenamefont {Gardiner}\ \emph {et~al.}(2004)\citenamefont
  {Gardiner}, \citenamefont {Zoller},\ and\ \citenamefont
  {Zoller}}]{gardiner_quantum_2004}%
  \BibitemOpen
  \bibfield  {author} {\bibinfo {author} {\bibfnamefont {C.}~\bibnamefont
  {Gardiner}}, \bibinfo {author} {\bibfnamefont {P.}~\bibnamefont {Zoller}},\
  and\ \bibinfo {author} {\bibfnamefont {P.}~\bibnamefont {Zoller}},\
  }\href@noop {} {\emph {\bibinfo {title} {Quantum noise: a handbook of
  {Markovian} and non-{Markovian} quantum stochastic methods with applications
  to quantum optics}}},\ Vol.~\bibinfo {volume} {56}\ (\bibinfo  {publisher}
  {Springer Science \& Business Media},\ \bibinfo {year} {2004})\BibitemShut
  {NoStop}%
\bibitem [{\citenamefont {Hong}\ \emph {et~al.}(1987)\citenamefont {Hong},
  \citenamefont {Ou},\ and\ \citenamefont {Mandel}}]{Hong1987Nov}%
  \BibitemOpen
  \bibfield  {author} {\bibinfo {author} {\bibfnamefont {C.~K.}\ \bibnamefont
  {Hong}}, \bibinfo {author} {\bibfnamefont {Z.~Y.}\ \bibnamefont {Ou}},\ and\
  \bibinfo {author} {\bibfnamefont {L.}~\bibnamefont {Mandel}},\ }\bibfield
  {title} {\bibinfo {title} {{Measurement of subpicosecond time intervals
  between two photons by interference}},\ }\href
  {https://doi.org/10.1103/PhysRevLett.59.2044} {\bibfield  {journal} {\bibinfo
   {journal} {Phys. Rev. Lett.}\ }\textbf {\bibinfo {volume} {59}},\ \bibinfo
  {pages} {2044} (\bibinfo {year} {1987})}\BibitemShut {NoStop}%
\bibitem [{\citenamefont
  {Kir{\ifmmode\check{s}\else\v{s}\fi}ansk{\ifmmode\dot{e}\else\.{e}\fi}}\
  \emph {et~al.}(2017)\citenamefont
  {Kir{\ifmmode\check{s}\else\v{s}\fi}ansk{\ifmmode\dot{e}\else\.{e}\fi}},
  \citenamefont {Thyrrestrup}, \citenamefont {Daveau}, \citenamefont
  {Dree{\ss}en}, \citenamefont {Pregnolato}, \citenamefont {Midolo},
  \citenamefont {Tighineanu}, \citenamefont {Javadi}, \citenamefont {Stobbe},
  \citenamefont {Schott}, \citenamefont {Ludwig}, \citenamefont {Wieck},
  \citenamefont {Park}, \citenamefont {Song}, \citenamefont {Kuhlmann},
  \citenamefont {S{\ifmmode\ddot{o}\else\"{o}\fi}llner}, \citenamefont
  {L{\ifmmode\ddot{o}\else\"{o}\fi}bl}, \citenamefont {Warburton},\ and\
  \citenamefont {Lodahl}}]{Kirsanske2017Oct}%
  \BibitemOpen
  \bibfield  {author} {\bibinfo {author} {\bibfnamefont {G.}~\bibnamefont
  {Kir{\ifmmode\check{s}\else\v{s}\fi}ansk{\ifmmode\dot{e}\else\.{e}\fi}}},
  \bibinfo {author} {\bibfnamefont {H.}~\bibnamefont {Thyrrestrup}}, \bibinfo
  {author} {\bibfnamefont {R.~S.}\ \bibnamefont {Daveau}}, \bibinfo {author}
  {\bibfnamefont {C.~L.}\ \bibnamefont {Dree{\ss}en}}, \bibinfo {author}
  {\bibfnamefont {T.}~\bibnamefont {Pregnolato}}, \bibinfo {author}
  {\bibfnamefont {L.}~\bibnamefont {Midolo}}, \bibinfo {author} {\bibfnamefont
  {P.}~\bibnamefont {Tighineanu}}, \bibinfo {author} {\bibfnamefont
  {A.}~\bibnamefont {Javadi}}, \bibinfo {author} {\bibfnamefont
  {S.}~\bibnamefont {Stobbe}}, \bibinfo {author} {\bibfnamefont
  {R.}~\bibnamefont {Schott}}, \bibinfo {author} {\bibfnamefont
  {A.}~\bibnamefont {Ludwig}}, \bibinfo {author} {\bibfnamefont {A.~D.}\
  \bibnamefont {Wieck}}, \bibinfo {author} {\bibfnamefont {S.~I.}\ \bibnamefont
  {Park}}, \bibinfo {author} {\bibfnamefont {J.~D.}\ \bibnamefont {Song}},
  \bibinfo {author} {\bibfnamefont {A.~V.}\ \bibnamefont {Kuhlmann}}, \bibinfo
  {author} {\bibfnamefont {I.}~\bibnamefont
  {S{\ifmmode\ddot{o}\else\"{o}\fi}llner}}, \bibinfo {author} {\bibfnamefont
  {M.~C.}\ \bibnamefont {L{\ifmmode\ddot{o}\else\"{o}\fi}bl}}, \bibinfo
  {author} {\bibfnamefont {R.~J.}\ \bibnamefont {Warburton}},\ and\ \bibinfo
  {author} {\bibfnamefont {P.}~\bibnamefont {Lodahl}},\ }\bibfield  {title}
  {\bibinfo {title} {{Indistinguishable and efficient single photons from a
  quantum dot in a planar nanobeam waveguide}},\ }\href
  {https://doi.org/10.1103/PhysRevB.96.165306} {\bibfield  {journal} {\bibinfo
  {journal} {Phys. Rev. B}\ }\textbf {\bibinfo {volume} {96}},\ \bibinfo
  {pages} {165306} (\bibinfo {year} {2017})}\BibitemShut {NoStop}%
\bibitem [{\citenamefont {Gardiner}\ and\ \citenamefont
  {Collett}(1985)}]{PhysRevA.31.3761}%
  \BibitemOpen
  \bibfield  {author} {\bibinfo {author} {\bibfnamefont {C.~W.}\ \bibnamefont
  {Gardiner}}\ and\ \bibinfo {author} {\bibfnamefont {M.~J.}\ \bibnamefont
  {Collett}},\ }\bibfield  {title} {\bibinfo {title} {Input and output in
  damped quantum systems: Quantum stochastic differential equations and the
  master equation},\ }\href {https://doi.org/10.1103/PhysRevA.31.3761}
  {\bibfield  {journal} {\bibinfo  {journal} {Phys. Rev. A}\ }\textbf {\bibinfo
  {volume} {31}},\ \bibinfo {pages} {3761} (\bibinfo {year}
  {1985})}\BibitemShut {NoStop}%
\bibitem [{\citenamefont {Raghunathan}\ and\ \citenamefont
  {Brun}(2009)}]{Raghunathan2009Mar}%
  \BibitemOpen
  \bibfield  {author} {\bibinfo {author} {\bibfnamefont {S.}~\bibnamefont
  {Raghunathan}}\ and\ \bibinfo {author} {\bibfnamefont {T.}~\bibnamefont
  {Brun}},\ }\bibfield  {title} {\bibinfo {title} {{Continuous monitoring can
  improve indistinguishability of a single-photon source}},\ }\href
  {https://doi.org/10.1103/PhysRevA.79.033831} {\bibfield  {journal} {\bibinfo
  {journal} {Phys. Rev. A}\ }\textbf {\bibinfo {volume} {79}},\ \bibinfo
  {pages} {033831} (\bibinfo {year} {2009})}\BibitemShut {NoStop}%
\bibitem [{\citenamefont {Pathak}\ and\ \citenamefont
  {Hughes}(2010)}]{Pathak2010Jul}%
  \BibitemOpen
  \bibfield  {author} {\bibinfo {author} {\bibfnamefont {P.~K.}\ \bibnamefont
  {Pathak}}\ and\ \bibinfo {author} {\bibfnamefont {S.}~\bibnamefont
  {Hughes}},\ }\bibfield  {title} {\bibinfo {title} {{Coherently triggered
  single photons from a quantum-dot cavity system}},\ }\href
  {https://doi.org/10.1103/PhysRevB.82.045308} {\bibfield  {journal} {\bibinfo
  {journal} {Phys. Rev. B}\ }\textbf {\bibinfo {volume} {82}},\ \bibinfo
  {pages} {045308} (\bibinfo {year} {2010})}\BibitemShut {NoStop}%
\bibitem [{\citenamefont {Hughes}\ \emph {et~al.}(2019)\citenamefont {Hughes},
  \citenamefont {Franke}, \citenamefont {Gustin}, \citenamefont
  {Kamandar~Dezfouli}, \citenamefont {Knorr},\ and\ \citenamefont
  {Richter}}]{Hughes2019Aug}%
  \BibitemOpen
  \bibfield  {author} {\bibinfo {author} {\bibfnamefont {S.}~\bibnamefont
  {Hughes}}, \bibinfo {author} {\bibfnamefont {S.}~\bibnamefont {Franke}},
  \bibinfo {author} {\bibfnamefont {C.}~\bibnamefont {Gustin}}, \bibinfo
  {author} {\bibfnamefont {M.}~\bibnamefont {Kamandar~Dezfouli}}, \bibinfo
  {author} {\bibfnamefont {A.}~\bibnamefont {Knorr}},\ and\ \bibinfo {author}
  {\bibfnamefont {M.}~\bibnamefont {Richter}},\ }\bibfield  {title} {\bibinfo
  {title} {{Theory and Limits of On-Demand Single-Photon Sources Using
  Plasmonic Resonators: A Quantized Quasinormal Mode Approach}},\ }\href
  {https://doi.org/10.1021/acsphotonics.9b00849} {\bibfield  {journal}
  {\bibinfo  {journal} {ACS Photonics}\ }\textbf {\bibinfo {volume} {6}},\
  \bibinfo {pages} {2168} (\bibinfo {year} {2019})}\BibitemShut {NoStop}%
\bibitem [{\citenamefont {Barth}\ \emph {et~al.}(2016)\citenamefont {Barth},
  \citenamefont {L{\ifmmode\ddot{u}\else\"{u}\fi}ker}, \citenamefont {Vagov},
  \citenamefont {Reiter}, \citenamefont {Kuhn},\ and\ \citenamefont
  {Axt}}]{Barth2016Jul}%
  \BibitemOpen
  \bibfield  {author} {\bibinfo {author} {\bibfnamefont {A.~M.}\ \bibnamefont
  {Barth}}, \bibinfo {author} {\bibfnamefont {S.}~\bibnamefont
  {L{\ifmmode\ddot{u}\else\"{u}\fi}ker}}, \bibinfo {author} {\bibfnamefont
  {A.}~\bibnamefont {Vagov}}, \bibinfo {author} {\bibfnamefont {D.~E.}\
  \bibnamefont {Reiter}}, \bibinfo {author} {\bibfnamefont {T.}~\bibnamefont
  {Kuhn}},\ and\ \bibinfo {author} {\bibfnamefont {V.~M.}\ \bibnamefont
  {Axt}},\ }\bibfield  {title} {\bibinfo {title} {{Fast and selective
  phonon-assisted state preparation of a quantum dot by adiabatic
  undressing}},\ }\href {https://doi.org/10.1103/PhysRevB.94.045306} {\bibfield
   {journal} {\bibinfo  {journal} {Phys. Rev. B}\ }\textbf {\bibinfo {volume}
  {94}},\ \bibinfo {pages} {045306} (\bibinfo {year} {2016})}\BibitemShut
  {NoStop}%
\bibitem [{\citenamefont {Wu}\ \emph {et~al.}(2011)\citenamefont {Wu},
  \citenamefont {Piper}, \citenamefont {Ediger}, \citenamefont {Brereton},
  \citenamefont {Schmidgall}, \citenamefont {Eastham}, \citenamefont {Hugues},
  \citenamefont {Hopkinson},\ and\ \citenamefont {Phillips}}]{Wu2011Feb}%
  \BibitemOpen
  \bibfield  {author} {\bibinfo {author} {\bibfnamefont {Y.}~\bibnamefont
  {Wu}}, \bibinfo {author} {\bibfnamefont {I.~M.}\ \bibnamefont {Piper}},
  \bibinfo {author} {\bibfnamefont {M.}~\bibnamefont {Ediger}}, \bibinfo
  {author} {\bibfnamefont {P.}~\bibnamefont {Brereton}}, \bibinfo {author}
  {\bibfnamefont {E.~R.}\ \bibnamefont {Schmidgall}}, \bibinfo {author}
  {\bibfnamefont {P.~R.}\ \bibnamefont {Eastham}}, \bibinfo {author}
  {\bibfnamefont {M.}~\bibnamefont {Hugues}}, \bibinfo {author} {\bibfnamefont
  {M.}~\bibnamefont {Hopkinson}},\ and\ \bibinfo {author} {\bibfnamefont
  {R.~T.}\ \bibnamefont {Phillips}},\ }\bibfield  {title} {\bibinfo {title}
  {{Population Inversion in a Single InGaAs Quantum Dot Using the Method of
  Adiabatic Rapid Passage}},\ }\href
  {https://doi.org/10.1103/PhysRevLett.106.067401} {\bibfield  {journal}
  {\bibinfo  {journal} {Phys. Rev. Lett.}\ }\textbf {\bibinfo {volume} {106}},\
  \bibinfo {pages} {067401} (\bibinfo {year} {2011})}\BibitemShut {NoStop}%
\bibitem [{\citenamefont {Reiter}\ and\ \citenamefont
  {S{\o}rensen}(2012)}]{Reiter2012Mar}%
  \BibitemOpen
  \bibfield  {author} {\bibinfo {author} {\bibfnamefont {F.}~\bibnamefont
  {Reiter}}\ and\ \bibinfo {author} {\bibfnamefont {A.~S.}\ \bibnamefont
  {S{\o}rensen}},\ }\bibfield  {title} {\bibinfo {title} {{Effective operator
  formalism for open quantum systems}},\ }\href
  {https://doi.org/10.1103/PhysRevA.85.032111} {\bibfield  {journal} {\bibinfo
  {journal} {Phys. Rev. A}\ }\textbf {\bibinfo {volume} {85}},\ \bibinfo
  {pages} {032111} (\bibinfo {year} {2012})}\BibitemShut {NoStop}%
\bibitem [{\citenamefont {{H. J. Carmichael}}(1999)}]{carmichael}%
  \BibitemOpen
  \bibfield  {author} {\bibinfo {author} {\bibnamefont {{H. J. Carmichael}}},\
  }\href@noop {} {\emph {\bibinfo {title} {Statistical Methods in Quantum
  Optics 1: Master Equations and Fokker-Planck Equations}}}\ (\bibinfo
  {publisher} {Springer-Verlag, Berlin},\ \bibinfo {year} {1999})\BibitemShut
  {NoStop}%
\bibitem [{\citenamefont {Franke}\ \emph {et~al.}(2020)\citenamefont {Franke},
  \citenamefont {Richter}, \citenamefont {Ren}, \citenamefont {Knorr},\ and\
  \citenamefont {Hughes}}]{Franke2020Sep}%
  \BibitemOpen
  \bibfield  {author} {\bibinfo {author} {\bibfnamefont {S.}~\bibnamefont
  {Franke}}, \bibinfo {author} {\bibfnamefont {M.}~\bibnamefont {Richter}},
  \bibinfo {author} {\bibfnamefont {J.}~\bibnamefont {Ren}}, \bibinfo {author}
  {\bibfnamefont {A.}~\bibnamefont {Knorr}},\ and\ \bibinfo {author}
  {\bibfnamefont {S.}~\bibnamefont {Hughes}},\ }\bibfield  {title} {\bibinfo
  {title} {{Quantized quasinormal-mode description of nonlinear cavity-QED
  effects from coupled resonators with a Fano-like resonance}},\ }\href
  {https://doi.org/10.1103/PhysRevResearch.2.033456} {\bibfield  {journal}
  {\bibinfo  {journal} {Phys. Rev. Res.}\ }\textbf {\bibinfo {volume} {2}},\
  \bibinfo {pages} {033456} (\bibinfo {year} {2020})}\BibitemShut {NoStop}%
\bibitem [{\citenamefont {Gazzano}\ \emph {et~al.}(2013)\citenamefont
  {Gazzano}, \citenamefont {Michaelis~de Vasconcellos}, \citenamefont {Arnold},
  \citenamefont {Nowak}, \citenamefont {Galopin}, \citenamefont {Sagnes},
  \citenamefont {Lanco}, \citenamefont
  {Lema{\ifmmode\hat{\imath}\else\^{\i}\fi}tre},\ and\ \citenamefont
  {Senellart}}]{Gazzano2013Feb}%
  \BibitemOpen
  \bibfield  {author} {\bibinfo {author} {\bibfnamefont {O.}~\bibnamefont
  {Gazzano}}, \bibinfo {author} {\bibfnamefont {S.}~\bibnamefont {Michaelis~de
  Vasconcellos}}, \bibinfo {author} {\bibfnamefont {C.}~\bibnamefont {Arnold}},
  \bibinfo {author} {\bibfnamefont {A.}~\bibnamefont {Nowak}}, \bibinfo
  {author} {\bibfnamefont {E.}~\bibnamefont {Galopin}}, \bibinfo {author}
  {\bibfnamefont {I.}~\bibnamefont {Sagnes}}, \bibinfo {author} {\bibfnamefont
  {L.}~\bibnamefont {Lanco}}, \bibinfo {author} {\bibfnamefont
  {A.}~\bibnamefont {Lema{\ifmmode\hat{\imath}\else\^{\i}\fi}tre}},\ and\
  \bibinfo {author} {\bibfnamefont {P.}~\bibnamefont {Senellart}},\ }\bibfield
  {title} {\bibinfo {title} {{Bright solid-state sources of indistinguishable
  single photons - Nature Communications}},\ }\href
  {https://doi.org/10.1038/ncomms2434} {\bibfield  {journal} {\bibinfo
  {journal} {Nat. Commun.}\ }\textbf {\bibinfo {volume} {4}},\ \bibinfo {pages}
  {1} (\bibinfo {year} {2013})}\BibitemShut {NoStop}%
\bibitem [{\citenamefont {Ding}\ \emph {et~al.}(2016)\citenamefont {Ding},
  \citenamefont {He}, \citenamefont {Duan}, \citenamefont {Gregersen},
  \citenamefont {Chen}, \citenamefont {Unsleber}, \citenamefont {Maier},
  \citenamefont {Schneider}, \citenamefont {Kamp}, \citenamefont
  {H{\ifmmode\ddot{o}\else\"{o}\fi}fling}, \citenamefont {Lu},\ and\
  \citenamefont {Pan}}]{Ding2016Jan}%
  \BibitemOpen
  \bibfield  {author} {\bibinfo {author} {\bibfnamefont {X.}~\bibnamefont
  {Ding}}, \bibinfo {author} {\bibfnamefont {Y.}~\bibnamefont {He}}, \bibinfo
  {author} {\bibfnamefont {Z.-C.}\ \bibnamefont {Duan}}, \bibinfo {author}
  {\bibfnamefont {N.}~\bibnamefont {Gregersen}}, \bibinfo {author}
  {\bibfnamefont {M.-C.}\ \bibnamefont {Chen}}, \bibinfo {author}
  {\bibfnamefont {S.}~\bibnamefont {Unsleber}}, \bibinfo {author}
  {\bibfnamefont {S.}~\bibnamefont {Maier}}, \bibinfo {author} {\bibfnamefont
  {C.}~\bibnamefont {Schneider}}, \bibinfo {author} {\bibfnamefont
  {M.}~\bibnamefont {Kamp}}, \bibinfo {author} {\bibfnamefont {S.}~\bibnamefont
  {H{\ifmmode\ddot{o}\else\"{o}\fi}fling}}, \bibinfo {author} {\bibfnamefont
  {C.-Y.}\ \bibnamefont {Lu}},\ and\ \bibinfo {author} {\bibfnamefont {J.-W.}\
  \bibnamefont {Pan}},\ }\bibfield  {title} {\bibinfo {title} {{On-Demand
  Single Photons with High Extraction Efficiency and Near-Unity
  Indistinguishability from a Resonantly Driven Quantum Dot in a
  Micropillar}},\ }\href {https://doi.org/10.1103/PhysRevLett.116.020401}
  {\bibfield  {journal} {\bibinfo  {journal} {Phys. Rev. Lett.}\ }\textbf
  {\bibinfo {volume} {116}},\ \bibinfo {pages} {020401} (\bibinfo {year}
  {2016})}\BibitemShut {NoStop}%
\bibitem [{\citenamefont {Reitzenstein}\ and\ \citenamefont
  {Forchel}(2010)}]{Reitzenstein2010Jan}%
  \BibitemOpen
  \bibfield  {author} {\bibinfo {author} {\bibfnamefont {S.}~\bibnamefont
  {Reitzenstein}}\ and\ \bibinfo {author} {\bibfnamefont {A.}~\bibnamefont
  {Forchel}},\ }\bibfield  {title} {\bibinfo {title} {{Quantum dot
  micropillars}},\ }\href {https://doi.org/10.1088/0022-3727/43/3/033001}
  {\bibfield  {journal} {\bibinfo  {journal} {J. Phys. D: Appl. Phys.}\
  }\textbf {\bibinfo {volume} {43}},\ \bibinfo {pages} {033001} (\bibinfo
  {year} {2010})}\BibitemShut {NoStop}%
\bibitem [{\citenamefont {Kolatschek}\ \emph {et~al.}(2021)\citenamefont
  {Kolatschek}, \citenamefont {Nawrath}, \citenamefont {Bauer}, \citenamefont
  {Huang}, \citenamefont {Fischer}, \citenamefont {Sittig}, \citenamefont
  {Jetter}, \citenamefont {Portalupi},\ and\ \citenamefont
  {Michler}}]{Kolatschek2021Sep}%
  \BibitemOpen
  \bibfield  {author} {\bibinfo {author} {\bibfnamefont {S.}~\bibnamefont
  {Kolatschek}}, \bibinfo {author} {\bibfnamefont {C.}~\bibnamefont {Nawrath}},
  \bibinfo {author} {\bibfnamefont {S.}~\bibnamefont {Bauer}}, \bibinfo
  {author} {\bibfnamefont {J.}~\bibnamefont {Huang}}, \bibinfo {author}
  {\bibfnamefont {J.}~\bibnamefont {Fischer}}, \bibinfo {author} {\bibfnamefont
  {R.}~\bibnamefont {Sittig}}, \bibinfo {author} {\bibfnamefont
  {M.}~\bibnamefont {Jetter}}, \bibinfo {author} {\bibfnamefont {S.~L.}\
  \bibnamefont {Portalupi}},\ and\ \bibinfo {author} {\bibfnamefont
  {P.}~\bibnamefont {Michler}},\ }\bibfield  {title} {\bibinfo {title} {{Bright
  Purcell Enhanced Single-Photon Source in the Telecom O-Band Based on a
  Quantum Dot in a Circular Bragg Grating}},\ }\bibfield  {journal} {\bibinfo
  {journal} {Nano Lett.}\ }\href {https://doi.org/10.1021/acs.nanolett.1c02647}
  {10.1021/acs.nanolett.1c02647} (\bibinfo {year} {2021})\BibitemShut {NoStop}%
\bibitem [{\citenamefont {Najer}\ \emph {et~al.}(2019)\citenamefont {Najer},
  \citenamefont {S{\ifmmode\ddot{o}\else\"{o}\fi}llner}, \citenamefont
  {Sekatski}, \citenamefont {Dolique}, \citenamefont
  {L{\ifmmode\ddot{o}\else\"{o}\fi}bl}, \citenamefont {Riedel}, \citenamefont
  {Schott}, \citenamefont {Starosielec}, \citenamefont {Valentin},
  \citenamefont {Wieck}, \citenamefont {Sangouard}, \citenamefont {Ludwig},\
  and\ \citenamefont {Warburton}}]{Najer2019Nov}%
  \BibitemOpen
  \bibfield  {author} {\bibinfo {author} {\bibfnamefont {D.}~\bibnamefont
  {Najer}}, \bibinfo {author} {\bibfnamefont {I.}~\bibnamefont
  {S{\ifmmode\ddot{o}\else\"{o}\fi}llner}}, \bibinfo {author} {\bibfnamefont
  {P.}~\bibnamefont {Sekatski}}, \bibinfo {author} {\bibfnamefont
  {V.}~\bibnamefont {Dolique}}, \bibinfo {author} {\bibfnamefont {M.~C.}\
  \bibnamefont {L{\ifmmode\ddot{o}\else\"{o}\fi}bl}}, \bibinfo {author}
  {\bibfnamefont {D.}~\bibnamefont {Riedel}}, \bibinfo {author} {\bibfnamefont
  {R.}~\bibnamefont {Schott}}, \bibinfo {author} {\bibfnamefont
  {S.}~\bibnamefont {Starosielec}}, \bibinfo {author} {\bibfnamefont {S.~R.}\
  \bibnamefont {Valentin}}, \bibinfo {author} {\bibfnamefont {A.~D.}\
  \bibnamefont {Wieck}}, \bibinfo {author} {\bibfnamefont {N.}~\bibnamefont
  {Sangouard}}, \bibinfo {author} {\bibfnamefont {A.}~\bibnamefont {Ludwig}},\
  and\ \bibinfo {author} {\bibfnamefont {R.~J.}\ \bibnamefont {Warburton}},\
  }\bibfield  {title} {\bibinfo {title} {{A gated quantum dot strongly coupled
  to an optical microcavity}},\ }\href
  {https://doi.org/10.1038/s41586-019-1709-y} {\bibfield  {journal} {\bibinfo
  {journal} {Nature}\ }\textbf {\bibinfo {volume} {575}},\ \bibinfo {pages}
  {622} (\bibinfo {year} {2019})}\BibitemShut {NoStop}%
\bibitem [{\citenamefont {Liu}\ \emph {et~al.}(2018)\citenamefont {Liu},
  \citenamefont {Brash}, \citenamefont {O{'}Hara}, \citenamefont {Martins},
  \citenamefont {Phillips}, \citenamefont {Coles}, \citenamefont {Royall},
  \citenamefont {Clarke}, \citenamefont {Bentham}, \citenamefont {Prtljaga},
  \citenamefont {Itskevich}, \citenamefont {Wilson}, \citenamefont {Skolnick},\
  and\ \citenamefont {Fox}}]{Liu2018Sep}%
  \BibitemOpen
  \bibfield  {author} {\bibinfo {author} {\bibfnamefont {F.}~\bibnamefont
  {Liu}}, \bibinfo {author} {\bibfnamefont {A.~J.}\ \bibnamefont {Brash}},
  \bibinfo {author} {\bibfnamefont {J.}~\bibnamefont {O{'}Hara}}, \bibinfo
  {author} {\bibfnamefont {L.~M. P.~P.}\ \bibnamefont {Martins}}, \bibinfo
  {author} {\bibfnamefont {C.~L.}\ \bibnamefont {Phillips}}, \bibinfo {author}
  {\bibfnamefont {R.~J.}\ \bibnamefont {Coles}}, \bibinfo {author}
  {\bibfnamefont {B.}~\bibnamefont {Royall}}, \bibinfo {author} {\bibfnamefont
  {E.}~\bibnamefont {Clarke}}, \bibinfo {author} {\bibfnamefont
  {C.}~\bibnamefont {Bentham}}, \bibinfo {author} {\bibfnamefont
  {N.}~\bibnamefont {Prtljaga}}, \bibinfo {author} {\bibfnamefont {I.~E.}\
  \bibnamefont {Itskevich}}, \bibinfo {author} {\bibfnamefont {L.~R.}\
  \bibnamefont {Wilson}}, \bibinfo {author} {\bibfnamefont {M.~S.}\
  \bibnamefont {Skolnick}},\ and\ \bibinfo {author} {\bibfnamefont {A.~M.}\
  \bibnamefont {Fox}},\ }\bibfield  {title} {\bibinfo {title} {{High Purcell
  factor generation of indistinguishable on-chip single photons}},\ }\href
  {https://doi.org/10.1038/s41565-018-0188-x} {\bibfield  {journal} {\bibinfo
  {journal} {Nat. Nanotechnol.}\ }\textbf {\bibinfo {volume} {13}},\ \bibinfo
  {pages} {835} (\bibinfo {year} {2018})}\BibitemShut {NoStop}%
\bibitem [{\citenamefont {Hepp}\ \emph {et~al.}(2020)\citenamefont {Hepp},
  \citenamefont {Hornung}, \citenamefont {Bauer}, \citenamefont {Hesselmeier},
  \citenamefont {Yuan}, \citenamefont {Jetter}, \citenamefont {Portalupi},
  \citenamefont {Rastelli},\ and\ \citenamefont {Michler}}]{Hepp2020Dec}%
  \BibitemOpen
  \bibfield  {author} {\bibinfo {author} {\bibfnamefont {S.}~\bibnamefont
  {Hepp}}, \bibinfo {author} {\bibfnamefont {F.}~\bibnamefont {Hornung}},
  \bibinfo {author} {\bibfnamefont {S.}~\bibnamefont {Bauer}}, \bibinfo
  {author} {\bibfnamefont {E.}~\bibnamefont {Hesselmeier}}, \bibinfo {author}
  {\bibfnamefont {X.}~\bibnamefont {Yuan}}, \bibinfo {author} {\bibfnamefont
  {M.}~\bibnamefont {Jetter}}, \bibinfo {author} {\bibfnamefont {S.~L.}\
  \bibnamefont {Portalupi}}, \bibinfo {author} {\bibfnamefont {A.}~\bibnamefont
  {Rastelli}},\ and\ \bibinfo {author} {\bibfnamefont {P.}~\bibnamefont
  {Michler}},\ }\bibfield  {title} {\bibinfo {title} {{Purcell-enhanced
  single-photon emission from a strain-tunable quantum dot in a
  cavity-waveguide device}},\ }\href {https://doi.org/10.1063/5.0033213}
  {\bibfield  {journal} {\bibinfo  {journal} {Appl. Phys. Lett.}\ }\textbf
  {\bibinfo {volume} {117}},\ \bibinfo {pages} {254002} (\bibinfo {year}
  {2020})}\BibitemShut {NoStop}%
\bibitem [{\citenamefont {Dusanowski}\ \emph {et~al.}(2020)\citenamefont
  {Dusanowski}, \citenamefont {K{\ifmmode\ddot{o}\else\"{o}\fi}ck},
  \citenamefont {Shin}, \citenamefont {Kwon}, \citenamefont {Schneider},\ and\
  \citenamefont {H{\ifmmode\ddot{o}\else\"{o}\fi}fling}}]{Dusanowski2020Jul}%
  \BibitemOpen
  \bibfield  {author} {\bibinfo {author} {\bibfnamefont {{\L}.}~\bibnamefont
  {Dusanowski}}, \bibinfo {author} {\bibfnamefont {D.}~\bibnamefont
  {K{\ifmmode\ddot{o}\else\"{o}\fi}ck}}, \bibinfo {author} {\bibfnamefont
  {E.}~\bibnamefont {Shin}}, \bibinfo {author} {\bibfnamefont {S.-H.}\
  \bibnamefont {Kwon}}, \bibinfo {author} {\bibfnamefont {C.}~\bibnamefont
  {Schneider}},\ and\ \bibinfo {author} {\bibfnamefont {S.}~\bibnamefont
  {H{\ifmmode\ddot{o}\else\"{o}\fi}fling}},\ }\bibfield  {title} {\bibinfo
  {title} {{Purcell-Enhanced and Indistinguishable Single-Photon Generation
  from Quantum Dots Coupled to On-Chip Integrated Ring Resonators}},\
  }\bibfield  {journal} {\bibinfo  {journal} {Nano Lett.}\ }\href
  {https://doi.org/10.1021/acs.nanolett.0c01771} {10.1021/acs.nanolett.0c01771}
  (\bibinfo {year} {2020})\BibitemShut {NoStop}%
\bibitem [{\citenamefont {Mocza{\l}a-Dusanowska}\ \emph
  {et~al.}(2019)\citenamefont {Mocza{\l}a-Dusanowska}, \citenamefont
  {Dusanowski}, \citenamefont {Gerhardt}, \citenamefont {He}, \citenamefont
  {Reindl}, \citenamefont {Rastelli}, \citenamefont {Trotta}, \citenamefont
  {Gregersen}, \citenamefont {H{\ifmmode\ddot{o}\else\"{o}\fi}fling},\ and\
  \citenamefont {Schneider}}]{Moczala-Dusanowska2019Jun}%
  \BibitemOpen
  \bibfield  {author} {\bibinfo {author} {\bibfnamefont {M.}~\bibnamefont
  {Mocza{\l}a-Dusanowska}}, \bibinfo {author} {\bibfnamefont
  {{\L}.}~\bibnamefont {Dusanowski}}, \bibinfo {author} {\bibfnamefont
  {S.}~\bibnamefont {Gerhardt}}, \bibinfo {author} {\bibfnamefont {Y.~M.}\
  \bibnamefont {He}}, \bibinfo {author} {\bibfnamefont {M.}~\bibnamefont
  {Reindl}}, \bibinfo {author} {\bibfnamefont {A.}~\bibnamefont {Rastelli}},
  \bibinfo {author} {\bibfnamefont {R.}~\bibnamefont {Trotta}}, \bibinfo
  {author} {\bibfnamefont {N.}~\bibnamefont {Gregersen}}, \bibinfo {author}
  {\bibfnamefont {S.}~\bibnamefont {H{\ifmmode\ddot{o}\else\"{o}\fi}fling}},\
  and\ \bibinfo {author} {\bibfnamefont {C.}~\bibnamefont {Schneider}},\
  }\bibfield  {title} {\bibinfo {title} {{Strain-Tunable Single-Photon Source
  Based on a Quantum Dot{\textendash}Micropillar System}},\ }\bibfield
  {journal} {\bibinfo  {journal} {ACS Photonics}\ }\href
  {https://doi.org/10.1021/acsphotonics.9b00481} {10.1021/acsphotonics.9b00481}
  (\bibinfo {year} {2019})\BibitemShut {NoStop}%
\bibitem [{\citenamefont {Madsen}\ \emph {et~al.}(2014)\citenamefont {Madsen},
  \citenamefont {Ates}, \citenamefont {Liu}, \citenamefont {Javadi},
  \citenamefont {Albrecht}, \citenamefont {Yeo}, \citenamefont {Stobbe},\ and\
  \citenamefont {Lodahl}}]{Madsen2014Oct}%
  \BibitemOpen
  \bibfield  {author} {\bibinfo {author} {\bibfnamefont {K.~H.}\ \bibnamefont
  {Madsen}}, \bibinfo {author} {\bibfnamefont {S.}~\bibnamefont {Ates}},
  \bibinfo {author} {\bibfnamefont {J.}~\bibnamefont {Liu}}, \bibinfo {author}
  {\bibfnamefont {A.}~\bibnamefont {Javadi}}, \bibinfo {author} {\bibfnamefont
  {S.~M.}\ \bibnamefont {Albrecht}}, \bibinfo {author} {\bibfnamefont
  {I.}~\bibnamefont {Yeo}}, \bibinfo {author} {\bibfnamefont {S.}~\bibnamefont
  {Stobbe}},\ and\ \bibinfo {author} {\bibfnamefont {P.}~\bibnamefont
  {Lodahl}},\ }\bibfield  {title} {\bibinfo {title} {{Efficient out-coupling of
  high-purity single photons from a coherent quantum dot in a photonic-crystal
  cavity}},\ }\href {https://doi.org/10.1103/PhysRevB.90.155303} {\bibfield
  {journal} {\bibinfo  {journal} {Phys. Rev. B}\ }\textbf {\bibinfo {volume}
  {90}},\ \bibinfo {pages} {155303} (\bibinfo {year} {2014})}\BibitemShut
  {NoStop}%
\bibitem [{\citenamefont {Wang}\ \emph {et~al.}(2020)\citenamefont {Wang},
  \citenamefont {Denning}, \citenamefont {G{\ifmmode\ddot{u}\else\"{u}\fi}r},
  \citenamefont {Lu},\ and\ \citenamefont {Gregersen}}]{Wang2020Sep}%
  \BibitemOpen
  \bibfield  {author} {\bibinfo {author} {\bibfnamefont {B.-Y.}\ \bibnamefont
  {Wang}}, \bibinfo {author} {\bibfnamefont {E.~V.}\ \bibnamefont {Denning}},
  \bibinfo {author} {\bibfnamefont {U.~M.}\ \bibnamefont
  {G{\ifmmode\ddot{u}\else\"{u}\fi}r}}, \bibinfo {author} {\bibfnamefont
  {C.-Y.}\ \bibnamefont {Lu}},\ and\ \bibinfo {author} {\bibfnamefont
  {N.}~\bibnamefont {Gregersen}},\ }\bibfield  {title} {\bibinfo {title}
  {{Micropillar single-photon source design for simultaneous near-unity
  efficiency and indistinguishability}},\ }\href
  {https://doi.org/10.1103/PhysRevB.102.125301} {\bibfield  {journal} {\bibinfo
   {journal} {Phys. Rev. B}\ }\textbf {\bibinfo {volume} {102}},\ \bibinfo
  {pages} {125301} (\bibinfo {year} {2020})}\BibitemShut {NoStop}%
\bibitem [{\citenamefont {Fischer}\ \emph {et~al.}(2018)\citenamefont
  {Fischer}, \citenamefont {Hanschke}, \citenamefont {Kremser}, \citenamefont
  {Finley}, \citenamefont {M{\"{u}}ller},\ and\ \citenamefont
  {Vu{\v{c}}kovi{\'{c}}}}]{Fischer2018}%
  \BibitemOpen
  \bibfield  {author} {\bibinfo {author} {\bibfnamefont {K.~A.}\ \bibnamefont
  {Fischer}}, \bibinfo {author} {\bibfnamefont {L.}~\bibnamefont {Hanschke}},
  \bibinfo {author} {\bibfnamefont {M.}~\bibnamefont {Kremser}}, \bibinfo
  {author} {\bibfnamefont {J.~J.}\ \bibnamefont {Finley}}, \bibinfo {author}
  {\bibfnamefont {K.}~\bibnamefont {M{\"{u}}ller}},\ and\ \bibinfo {author}
  {\bibfnamefont {J.}~\bibnamefont {Vu{\v{c}}kovi{\'{c}}}},\ }\bibfield
  {title} {\bibinfo {title} {{Pulsed Rabi oscillations in quantum two-level
  systems: beyond the area theorem}},\ }\href
  {https://doi.org/10.1088/2058-9565/aa9269} {\bibfield  {journal} {\bibinfo
  {journal} {Quantum Science and Technology}\ }\textbf {\bibinfo {volume}
  {3}},\ \bibinfo {pages} {014006} (\bibinfo {year} {2018})}\BibitemShut
  {NoStop}%
\end{thebibliography}%
\end{document}